\documentclass[12pt]{article}
\usepackage{amssymb}
\usepackage{amsmath}
\usepackage{amstext}
\usepackage{graphicx,epsfig}
\usepackage{epsfig}
\usepackage{verbatim} 
\usepackage{fancybox}
\usepackage{color}
\usepackage{ulem}
\usepackage{enumitem}
\usepackage{subfigure}
\usepackage{bbm}
\usepackage{parskip}
\usepackage[utf8]{inputenc}

\newcommand{\Comment}[1]{{}}
\definecolor{MyDarkBlue}{rgb}{0.15,0.15,0.45}
\usepackage[linktocpage=true]{hyperref}
\hypersetup{
colorlinks=true,
citecolor=MyDarkBlue,
linkcolor=MyDarkBlue,
urlcolor=MyDarkBlue,
pdfauthor={Ercan Kilicarslan},
pdftitle={TOPOLOGICALLY MASSIVE GRAVITY: ANYON SCATTERING, WEYL-GAUGING AND CAUSALITY},
pdfsubject={hep-th}
}

\newcommand{\be}{\begin{equation}}
\newcommand{\ee}{\end{equation}}
\newcommand{\bea}{\begin{eqnarray}}
\newcommand{\eea}{\end{eqnarray}}
\newcommand{\beas}{\begin{eqnarray*}}
\newcommand{\eeas}{\end{eqnarray*}}

\def\({\left(}
\def\){\right)}

\title{TOPOLOGICALLY MASSIVE GRAVITY: ANYON SCATTERING, WEYL-GAUGING AND CAUSALITY}
\author{Ercan Kilicarslan\footnote{kercan@metu.edu.tr and physicsercan@gmail.com}}

\numberwithin{equation}{section}

\begin{document}


\begin{center}
{\LARGE \bf{\sc TOPOLOGICALLY MASSIVE GRAVITY: ANYON SCATTERING, WEYL-GAUGING AND CAUSALITY}}\footnote{This is a Ph.D. thesis defended in METU Physics Department on the $10^{th}$ of August 2017.}
\end{center} 
 \vspace{1truecm}
\thispagestyle{empty} \centerline{
{\large \bf {\sc Ercan kilicarslan${}^{a,}$}}\footnote{E-mail address: \Comment{\href{mailto:kercan@metu.edu.tr}}{\tt kercan@metu.edu.tr, physicsercan@gmail.com}}
                                                          }

\vspace{1cm}

\centerline{{\it ${}^a$ 
Department of Physics,}}
 \centerline{{\it Middle East Technical University,  06800, Ankara, Turkey.}}

\begin{abstract}
In this thesis, we studied the Topologically Massive Gravity (TMG) in two perspectives. Firstly, by using real scalar and abelian gauge fields, we built the Weyl-invariant extension of TMG which unifies cosmological TMG and Topologically Massive Electrodynamics (TME) with a Proca mass term. Here, we have demonstrated that the presence of (Anti)-de Sitter spaces as the background solution, spontaneously breaks the local Weyl symmetry, whereas the radiative corrections at two-loop level breaks the symmetry in flat vacuum. The breaking of Weyl symmetry fixes all the dimensionful parameters and provides masses to spin-2 and spin-1 particles as in the Higgs mechanism. Secondly, we calculated the tree-level scattering amplitude in the cosmological TMG plus the Fierz-Pauli mass term in (Anti)-de Sitter spaces and accordingly found the relevant weak field potential energies between two covariantly conserved localized point-like spinning sources. We have shown that in addition to spin-spin and mass-mass interactions, there also occurs a mass-spin interaction which is generated by the gravitational Chern-Simons term that changes the initial spin of particles converting them to gravitational anyons. In addition to these works concerning TMG, we have also discussed the issue of local causality in $2+1$ dimensional gravity theories and shown that Einstein's gravity, TMG and the new massive gravity are causal as long as the sign of the Newton's constant is set to negative. We study the causality discussion with the Shapiro time delay method.
\end{abstract}

\newpage

\tableofcontents
\newpage

\section{Introduction}
\parskip=5pt
\normalsize

General Relativity (GR) developed by Albert Einstein in 1916 is one of the most novel developments of the last century physics. As is well-known, GR is the simplest and a somewhat unique geometrical interpretation of gravitational interaction which describes a massless spin-2 particle (i.e., graviton) in the quantum field theory perspective. The predictions of GR have been approved by a large number of experiments, such as the perihelion precession of Mercury, deflection of light by massive objects, gravitational redshift of light and gravitational lensing etc. Recently, it has been observed by LIGO that the gravitational wave coming from two colliding black holes is in a complete and remarkable agreement with the one that is predicted by GR \cite{LIGO}.

Although Einstein's theory overwhelmingly provides very successful solutions whose strong and wide predictability have been approved by countless amount of observations, its bare form fails to be a complete foreseeing theory in the extreme regimes, namely in the UV and IR scales. Firstly, it is well-known that as one treats pure GR as a perturbative quantum field theory, due to the existence of the {\it{dimensionful}} parameter (that is, Newton's constant with mass dimension $-2$), the catastrophic infinities at the second loop-level resulting from the self-interaction of the gravitons cannot be handled and thereby the theory unavoidably turns out to be a perturbatively unacceptable model. Thus, the pure GR is valid only within a certain energy regime which means it is a low energy effective theory. As for the IR regimes, it has been  experimentally shown that GR in its bare form (without a dark sector) cannot explain the accelerated expansion of universe and the rotational curves of outer objects in galaxies that can be explained with by GR only if one anticipates a tremendous amount of dark matter comparing with ordinary matter. Moreover, GR has not been unified with the Standard model of particle physics. Thus, due to the noted problems in the both extreme scales, a proper modification of GR seems to be inescapable.

To this end, several alternative approaches have been introduced up to now in order to modify GR in such a way that the extended theory would particularly provide a well-defined gravity theory. Within these approaches, here we will specifically consider the massive gravity modifications of GR which suggest to give a small mass to the graviton: recall that providing a tiny mass to graviton will mediate Yukawa-like gravitational force as $\sim \frac{e^{-\lambda r}}{r^2}$ (where Yukawa length scale $\lambda$ is given in terms of graviton mass $m$ as $\lambda=\frac{1}{m}$). With this, gravitational interaction becomes less attractive in large distances which leads to screening effects on cosmological scales. Thus, this specific modification particularly has the potential to provide some understanding of the IR problems of GR.

In this respect, the first linearly consistent massive gravity theory was introduced in 1939 by Fierz and Pauli \cite{PF} which is called the Fierz-Pauli (FP) theory. In this theory, the FP mass term is added to Einstein theory with which graviton acquires a mass. With this modification, the FP theory describes 5 degrees of freedom (DOF) in four dimensions at the linear level while Einstein's gravity has 2 DOF. Thus, it is natural to expect the massive gravity theory to reduce to Einstein's gravity in the massless limit. However, as one tries to take the $m\rightarrow 0$ limit of FP theory, one will realize that this limit of FP theory does not reduce to Einstein's theory. This problem is called the van-Dam-Veltmann-Zakharov (vDVZ) discontinuity \cite{vdvz1,vdvz2}. To resolve this problem, although Vainshtein \cite{vainshtein} claimed that this discontinuity would appear only in the linear theory and thus disappear in the non-linear level, it has been shown that the massive gravity theory actually has an additional $6^{th}$ DOF which causes a ghost-like instability at the non-linear level. This extra DOF is called the Boulware-Deser ghost \cite{BD}. Thus, the massive modification of GR comes with the problems of vDVZ discontinuity and instability in $3+1$-dimensions.

To have some insights particularly in the idea of quantum gravity, due to its unique properties, it seems to be useful to study on $2+1$-dimensional toy models \cite{Deserjackiw1, Deserjackiw2}. The reasons of preferring $2+1$-dimensions as a platform is as follows: GR in $2+1$ dimensions does not possess any local DOF. This means that any vacuum solution of $2+1$-dimensional GR is locally equivalent to flat or maximally symmetric constant curvature backgrounds ($(A)dS$). That is, it is locally trivial. On the other hand, with a cosmological constant, global properties of the GR is non-trivial. Particularly, it then describes asymptotically $AdS_3$ black hole solutions \cite{BTZ} which naturally lead to the additional microscopic DOF giving rise to the celebrated Bekenstein-Hawking entropy. Additionally, Brown and Henneaux have shown that $2+1$-dimensional bulk $AdS$ gravity theory with asymptotically $AdS_3$ boundary conditions gives rise to a $2$-dimensional conformal field theory (CFT) on the boundary \cite{BH}. In this sense, cosmological Einstein's theory describes a healthy holographic description with globally non-trivial structure even if it does not have any local propagating DOF. Therefore, quantum version of the cosmological GR has the potential to provide a well-behaved $2+1$-dimensional quantum gravity theory in the AdS/CFT context. Here, the natural question is what kind of the modification in the cosmological GR will supply such a complete quantum gravity model which will also have local propagating DOF.

Topologically Massive Gravity (TMG) \cite{ DJT} seems to be the only viable lower dimensional massive extension of the three dimensional Einstein's theory with a dynamical DOF\footnote{There is also an other well-known massive gravity theory which is known as New Massive Gravity (NMG) \cite{BHT} that has quadratic curvature terms. The theory describes a massive spin-2 particle with two degrees of freedom.  NMG is a unitary theory for suitable choices of parameters but it is not renormalizable. This theory also suffers from the bulk-boundary unitary conflict, hence it is not suitable for holographic description.}. TMG is a unitary and renormalizable theory at the tree-level and it is a parity non-invariant theory which has a single massive spin-2 excitation about its flat or $AdS$ vacua. However, due to the emerging contradiction between the positivity of bulk excitations and the Brown-Henneaux's gravitational central charges for generic parameter $\mu$, TMG has the inevitable bulk-boundary unitarity clash in the AdS/CFT framework. Fortunately, a particular resolution of this conflict was suggested in \cite{Chiral} where it has been shown that the cosmological TMG is stable in the chiral limit. At the chiral limit, the theory has only right moving mode with positive central charge on the boundary. Thus, due to this fact, chiral limit of cosmological TMG\footnote{In \cite{GrumillerJohansson}, it has been claimed that Chiral gravity has problematic log mode solutions which lead the non-unitary. But, It was proposed in \cite{Maloney, Carlip} that there is a linearization instability and recently it was shown \cite{Altas} that these problematic modes are an artifact of first order perturbation theory and so they are not integrable to full solution.} has a notable potential to provide consistent quantum gravity theory in the AdS/CFT framework. Due to these appealing properties, many works have been devoted to understand the physical properties of TMG.

This thesis is based on the following papers:  
\begin{enumerate}
  \item S. Dengiz, E. Kilicarslan and B. Tekin, ``Weyl-gauging of Topologically Massive Gravity,`` Phys.\ Rev.\ D {\bf 86}, 104014 (2012)
   \cite{kilicarslan1}. 
  \item  S. Dengiz, E. Kilicarslan and B. Tekin, ``Scattering in Topologically Massive Gravity, Chiral Gravity and the corresponding Anyon-Anyon Potential Energy,`` Phys.\ Rev.\ D {\bf 89}, 024033 (2014)
   \cite{kilicarslan2}. 
  \item   J. D. Edelstein, G. Giribet, C. Gomez, E. Kilicarslan, M. Leoni and B. Tekin, ``Causality in 3D Massive Gravity Theories,`` Phys.\ Rev.\ D {\bf 95}, 104016 (2017) \cite{kilicarslan3}.
\end{enumerate}

In the first paper, the Weyl-invariant extension of TMG which unifies cosmological TMG and Topologically Massive Electrodynamics with a Proca mass term is constructed. Here, the tree-level perturbative unitarity of the Weyl-invariant TMG and its particle spectrum are studied in detail. It is shown that the theory does not have any dimensionful parameters; hence spin-$2$ and spin-$1$ particles acquire masses via the breaking of the Weyl's symmetry either spontaneously in $AdS$ vacuum as in the Higgs mechanism or radiatively in flat vacuum as a result of the Coleman-Weinberg mechanism.    

In the second paper, from the wave-type equation for cosmological TMG augmented with the FP mass term, the particle spectrum of the theory in various special limits in (A)dS backgrounds are studied. In doing so, we calculate the tree-level scattering amplitude and compute the associated non-relativistic potential energies between two locally spinning conserved point-like sources. It is shown that in addition to spin-spin and mass-mass interactions, there is also mass-spin interaction which is induced by the gravitational Chern-simons term which changes the initial spin of particles turning these into gravitational anyons. Here, we also briefly discuss the flat space chiral gravity limit of the scattering amplitude.    

In the last paper, the local causality problem in the $2+1$ dimensional massive gravity theories is studied in detail. Here, it has been shown that the causality and unitarity are not in contradiction in bare GR, TMG and NMG as long as the sign of Newton's constant is set to be negative. Furthermore, we also discuss the $2+1$-dimensional Born-Infeld gravity and show that the related causality and unitarity are compatible with each other. This is in sharp contrast to the Einstein-Gauss Bonnet and cubic theories in higher dimensions ($n\geq 4$) where causality and unitarity are in conflict. The causality issue has been studied in asymptotically Minkowski and $AdS$ spaces.  

\subsection{The Fierz-Pauli Theory}

The first modification of Einstein's gravity with a mass term was given by Fierz and Pauli and is described in the generic $D$-dimensional spacetime as follows \cite{PF}
\begin{equation}
 {I}=\frac{1}{ \kappa}\int d^Dx \sqrt{-g} \, \Big (  ( R -2 \Lambda)-\frac{m^2}{4}(h^2_{\mu\nu}-h^2)\bigg),
\label{pfaction}
\end{equation}
where $\kappa$ is the $D$ dimensional Newton's constant. It describes a free massive spin$-2$ particle. The second term in the action is the FP mass term and unavoidably breaks the gauge invariance of pure GR\footnote{That is, with the FP term, the modified GR is not invariant under the diffeomorphism defined as $\delta h_{\mu\nu}=2\nabla_{(\mu}\xi_{\nu)}$ any more.}. Because of the violation of gauge invariance, the theory has $D+1$ constraints and thus it defines $\frac{(D+1)(D-2)}{2}$ DOF at linearized level in $D$ dimensions. It is also important to note that FP mass term is a unique combination for ghost and tachyon-freedom. This is called the FP-tuning. More precisely, one should notice that if one chooses a different coefficient between $h_{\mu\nu}^2$ and $h^2$ instead of the FP tuning (that is, $-1$), then a scalar ghost appears as an extra DOF. To see this, let us consider the FP action with an arbitrary coefficient $a$ between $h_{\mu\nu}^2$ and $h^2$ instead of the FP tuning. For this purpose, we shall be interested in the weak field limit in which the linearized field equation about a $AdS$ background $ g_{\mu\nu}=\bar{g}_{\mu\nu}+h_{\mu\nu} $ reads
\begin{equation}
{\cal G}^L_{\mu \nu}+\frac{m^2}{2}(h_{\mu\nu}+a\bar{g}_{\mu\nu}h)=0,
\label{PFeom}
\end{equation}
where the related linearized tensors are given as follows \cite{Deser:2002jk}
\begin{equation}
\begin{aligned}
 {\cal G}^L_{\mu \nu}&=R^{L}_{\mu\nu}-\frac{1}{2} \bar{g}_{\mu\nu} R^{L}-\frac{2 \Lambda}{D-2} h_{\mu\nu}, \\
 R^{L}_{\mu\nu}&=\frac{1}{2} (\bar{\nabla}^\sigma  \bar{\nabla}_\mu h_{\nu\sigma}+\bar{\nabla}^\sigma \bar{\nabla}_\nu h_{\mu\sigma}-\bar{\square} h_{\mu\nu}-\bar{\nabla}_\mu \bar{\nabla}_\nu h ), \\
 R^{L}&=(\bar{g}^{\mu\nu}R_{\mu\nu})^{L}=-\bar{\square} h+\bar{\nabla}^\mu \bar{\nabla}^\nu h_{\mu\nu}-\frac{2 \Lambda}{D-2}  h.
 \label{curvaturet1}
 \end{aligned}
 \end{equation}
Here, $h=\bar{g}^{\mu\nu}h_{\mu\nu} $. Since we have $\bar{\nabla}^\mu{\cal G}^L_{\mu \nu}=0$, then the divergence of Eq.(\ref{PFeom}) yields (for $m\neq 0$)
\begin{equation}
\bar{\nabla}^\mu h_{\mu\nu}=-a\bar{\nabla}_\nu h.
\end{equation}
Using this, the linearized Einstein tensor ${\cal G}^L_{\mu \nu}$ can be recast as
\begin{equation}
\begin{aligned}
{\cal G}^L_{\mu \nu}=\frac{1}{2}\bigg(&-\bar{\square}h_{\mu\nu}+(a+1) \bar{g}_{\mu\nu}\bar{\square}h-(2a+1)
\bar{\nabla}_\mu \bar{\nabla}_\nu h\\&+\frac{4\Lambda}{(D-1)(D-2)} (h_{\mu\nu}+\frac{D-3}{2}\bar{g}_{\mu\nu}h)\bigg).
\label{ETPF}
\end{aligned}
\end{equation}
Subsequently, by inserting Eq.(\ref{ETPF}) into Eq.(\ref{PFeom}) and later taking the trace of it, one obtains
\begin{equation}
\bigg((D-2)(a+1)\bar{\square}+2\Lambda+m^2(1+aD)\bigg)h=0.
\label{FPghost1}
\end{equation}
Thus theory has a scalar ghost for $a\neq -1$, this can be seen explicitly by decomposing $h_{\mu\nu}$ as a transverse traceless helicity-$2$ tensor ($h^{TT}_{\mu\nu}$), a helicity-$1$ vector ($V^{\mu}$) and scalar field component ($\phi$). After plugging this decomposition into the FP action, one can see that scalar field comes with the wrong kinetic energy sign and verifies the Eq.(\ref{FPghost1}). Thus, one needs to select $ a =-1$ (that is, FP tuning) and $ m^2\neq \frac{2\Lambda}{(D-1)} $ for the ghost freedom. Here, the specific choice of $ m^2= \frac{2\Lambda}{(D-1)} $ is partially massless point at which $h$ is not fixed any more and a new symmetry arises.

As was mentioned above, one naturally expects that the massless limit of FP theory must yield the GR after working out some physical quantities. However, this has been shown to cause the problem of vDVZ discontinuity \cite{vdvz1,vdvz2}. As the corresponding weak field limits are analyzed, one will see that the massive gravity causes a deviation of 25 percent in the ordinary result for the light bending angle or a similar discretely different result for the static Newton's potential. The source of the discontinuity can be explicitly shown by using the Stuckelberg trick in which new gauge fields are introduced in such a way that the DOFs are intact without changing the dynamics of the theory \cite{stueckelberg,hinterbichler}. In order to see the origin of this, let us now consider the source coupled linearized FP action in a flat background \cite{hinterbichler}\footnote{We will do this analysis in flat space since it was shown that vDVZ discontinuity can be cured if one adds the cosmological constant to the theory, see for details \cite{Porrati:2000cp}.}

\begin{equation}
 {I}_{{\cal O}(h^2)}=\int d^Dx \sqrt{-g} \, \Big (  \frac{1}{2\kappa} h^{\mu\nu}{\cal G}^L_{\mu \nu}-\frac{m^2}{2\kappa}(h^2_{\mu\nu}-h^2)+h^{\mu\nu}T_{\mu\nu}\bigg),
\label{pfactionquadratic}
\end{equation}
where $T_{\mu\nu}$ is the energy-momentum tensor. To preserve gauge symmetry, let us now introduce the following transformations
\begin{equation}
h_{\mu\nu}\rightarrow h_{\mu\nu}+\partial_\mu A_{\nu}+\partial_\nu A_{\mu}, \hskip 1 cm A_{\mu}\rightarrow  A_{\mu}+\partial_\mu\phi,
\end{equation} 
where $A_{\mu}$ and $\phi$ are auxiliary Stukelberg vector and scalar fields, respectively. After scaling the fields as $\phi \rightarrow \frac{\phi}{m^2} $, $A_{\mu} \rightarrow \frac{A_{\mu}}{m} $ to bring the kinetic energies of fields to the canonical form and later taking massless limit ($m\rightarrow 0$), Eq.(\ref{pfactionquadratic}) takes the form
 \begin{equation}
 {I}_{{\cal O}(h^2)}=\int d^Dx \sqrt{-g} \, \Big (  \frac{1}{2\kappa} h^{\mu\nu} {\cal G}^L_{\mu \nu}(h)-\frac{1}{2}F_{\mu\nu}F^{\mu\nu}-\frac{2}{\kappa}(h_{\mu\nu}\partial^\mu\partial^\nu \phi-h\partial^2\phi)+h^{\mu\nu}T_{\mu\nu}\bigg).
\label{pfactionquadratic1}
\end{equation} 
Here, $F_{\mu\nu}=\partial_\mu A_\nu-\partial_\nu A_\mu$. Then, with the redefinition of graviton field as $h_{\mu\nu}\rightarrow \tilde{h}_{\mu\nu}+\frac{2}{D-2}\eta_{\mu\nu}\phi$, the scalar and spin-$2$ fields will decouple from each others and one will finally arrive at
 \begin{equation}
 \begin{aligned}
 {I}_{{\cal O}(\tilde{h}^2)}=\int d^Dx \sqrt{-g} \, \Big (&  \frac{1}{2\kappa} \tilde{h}^{\mu\nu} {\cal G}^L_{\mu \nu}(\tilde{h})-\frac{1}{2}F_{\mu\nu}F^{\mu\nu}-\frac{2}{\kappa}\frac{D-1}{D-2}\partial^\mu\partial^\nu \phi+\tilde{h}^{\mu\nu}T_{\mu\nu}\\&+\frac{2}{D-2}\phi T\bigg).
\label{pfactionquadratic12}
\end{aligned}
\end{equation} 
Observe that the gauge symmetries of the action are $\delta \tilde{h}_{\mu\nu}=2\nabla_{(\mu}\xi_{\nu)}$ and $\delta A_\mu=\partial_\mu\lambda$. It is also important to note that the source of vDVZ discontinuity is the coupling between scalar field $\phi$ and trace of energy momentum tensor $T$. On the other hand, one can realize that the theory still has $\frac{(D+1)(D-2)}{2}$ DOF within which $\frac{D(D-3)}{2}$ corresponds to massless spin-$2$ field, $D-2$ corresponds to massless vector field and $1$ corresponds to scalar field. Thus, the Stuckelberg mechanism does not alter the fundamental DOF. Furthermore, this analysis shows that the massless limit is not smooth since the extra scalar DOF which does not exist in GR survives in this limit.
 
Vainshtein \cite{vainshtein} claimed that the vDVZ discontinuity was an artifact of linear theory, at the non linear level, and thus it could be recovered by non-linear effects. However, it was shown that the massive gravity in four dimensions have 6 DOF at the non-linear level, while it has 5 DOF at the linear level. This extra sixth dynamical mode leads to a ghost-like instability and is known as Boulware-Deser ghost \cite{BD}.

\subsection{Topologically Massive Gravity}
Topologically Massive gravity (TMG) was developed by Deser, Jackiw and Templeton in 1982 \cite{DJT}. The theory is described by the following action
\begin{equation}
 S_{TMG}=\frac{1}{16\pi G}\int d^3 x \sqrt{-g} \bigg [ \sigma (R-2\Lambda) +\frac{1}{2\mu } \eta^{\lambda \mu \nu} \Big ( \Gamma^\rho_{\lambda \sigma} \partial_\mu \Gamma^\sigma_{\nu \rho}+\frac{2}{3} \Gamma^\rho_{\lambda \sigma} \Gamma^\sigma_{\mu \tau}\Gamma^\tau_{\nu \rho}  \Big ) \bigg ],
\label{TMG1}
\end{equation}
where $G$ is the usual $2+1$-dimensional Newton's constant, $ \sigma $ is a $\pm 1$ and $ \mu $ is the topological mass parameter. Here, $ \eta^{\mu \nu \alpha}$ is a rank-3 tensor described in terms of the Levi-Civita symbol as $\epsilon^{\mu \nu \alpha}/\sqrt{-g}$. We will work with the mostly plus signature. TMG is a parity non-invariant theory due to Chern-Simons term and it possesses more DOF than $3D$ Einstein's gravity since Chern-Simons term contains three derivatives of the metric and hence it propagates a massive spin-$2$. To see this, let us notice that the source-free field equations of TMG \footnote{See for details to Appendix \ref{chp:appendixa}.} are
\begin{equation}
 \sigma (R_{\mu \nu}-\frac{1}{2} g_{\mu \nu} R+ \Lambda g_{\mu \nu})+\frac{1}{\mu} C_{\mu \nu} =0,
\label{TMGfeq1}
 \end{equation}
 where $C_{\mu \nu}$ is the Cotton tensor that can be defined as
 \begin{equation}
 C^{\mu \nu}=\eta^{\mu \alpha \beta} \nabla_\alpha \Big ( R^\nu{_\beta}-\frac{1}{4} \delta^\nu{_\beta} R \Big ).
 \label{cotton3}
\end{equation}
It can be shown to have the following properties (divergence-free and traceless)
\begin{equation}
\nabla^\mu C_{\mu \nu}=0, \hskip .5 cm C^\mu_\mu=0.
\end{equation}
In three dimensions, it takes the role of the Weyl tensor which vanishes identically. If for a metric $g_{\mu \nu}$, $C^\mu{_{\nu}}=0$, then the metric is conformally flat. With this property of $C_{\mu \nu}$, all solutions of Einstein's theory in $3D$ also solve TMG. In particular $AdS$ is a solution to TMG. The linearization of Eq.(\ref{TMGfeq1}) about its $AdS$ background $ g_{\mu\nu}=\bar{g}_{\mu\nu}+h_{\mu\nu} $ reads
\begin{equation}
 \frac{1}{\kappa} {\cal G}^L_{\mu \nu}+\frac{1}{ \mu}C^L_{\mu\nu}=0,
\label{linerTMG1}
\end{equation}
where the related linearized tensors are given in Eq.(\ref{curvaturet1}).   The trace of Eq.(\ref{linerTMG1}) gives $g^{\mu\nu}{\cal G}^L_{\mu \nu}=-\frac{1}{2}R^{L}$ and this result requires $R^{L}$ to be zero. This allows one to choose the following transverse-traceless gauge
\begin{equation}
\bar{\nabla}^\mu h_{\mu\nu}=0, \hskip .5 cm h=0.
\end{equation} 
Under this gauge fixing, the linearized field equations in Eq.(\ref{linerTMG1}) take the form
\begin{equation}
\bigg(\sigma \delta^\beta_{\mu}+\frac{1}{ \mu} \eta_\mu\,^{ \alpha \beta} \bar{\nabla}_\alpha\bigg)\bigg(\bar{\square}-2\Lambda\bigg) h^{TT}_{\nu\beta}=0.
\label{TTeq}
\end{equation}
Acting on the left with the operator $\sigma \delta^\beta_{\mu}-\frac{1}{ \mu} \eta_\mu\,^{ \alpha \beta} \bar{\nabla}_\alpha $ to Eq.(\ref{TTeq}), one obtains 
\begin{equation}
\bigg(\bar{\square}-2\Lambda-(\mu^2\sigma^2+\Lambda)\bigg){\cal G}^L_{\mu \nu}=0.
\label{Speceq}
\end{equation}
Thus, by bearing in mind that the null-cone propagation for a massless spin-$2$ field in $(A)dS$ space is defined as $(\bar{\square}-2\Lambda)h_{\mu \nu}=0$, then one finds at that the theory actually describes a massive spin-$2$ graviton with a mass 
\begin{equation}
m_g^2=\mu^2\sigma^2+\Lambda.
\end{equation}
Observe that in the flat space limit ($\Lambda\rightarrow 0$), TMG still describes a single massive excitation with mass $m_g=\lvert \mu\sigma\lvert$ and $\sigma$ must be chosen to be negative for ghost freedom as opposed to bare GR. Note also that in the $\sigma \rightarrow 0$ limit, the theory reduces to the pure Chern-Simons theory which does not have any propagating DOF and cannot be deformed by a cosmological constant.

As is well-known, gauge/gravity duality (or more specifically $AdS/CFT$) is one of the promising approaches in constructing quantum gravity models. AdS/CFT states that each $d$ dimensional bulk $AdS$ gravity theory is holographically dual to the corresponding $d-1$ dimensional CFT on the boundary \cite{Malda}. In fact, such a duality was first introduced by Brown and Henneaux in 1986. Here, they have shown that three dimensional bulk $AdS$ gravity theory with asymptotically $AdS_3$ boundary conditions is equivalent to two dimensional CFT on the boundary \cite{BH}. Since TMG has asymptotical $AdS_3$ solutions, many studies have been devoted to apply $AdS/CFT$ correspondence to TMG. The left and right central charges in TMG are found to be
\begin{equation}
c_{L,R}=\frac{3\ell}{2G}(\sigma \mp \frac{1}{\mu\ell}),
\end{equation} 
where the $AdS_3$ radius is defined as $\Lambda=-\frac{1}{\ell^2}$. Although TMG seems to be an interesting candidate for a $2+1$-dimensional quantum gravity, it has been shown that the theory has an unstable mode causing negative energy in AdS/CFT context. More precisely, it has been demonstrated that the positivity of bulk excitations in TMG contradicts with the corresponding gravitational central charges for generic parameter $\mu$. Later, it has been shown that with the particular setting of parameters as $\sigma\mu\ell=1$, this trouble is being resolved and thus TMG then turns out to be stable at this chiral limit \cite{Chiral}. TMG at this critical point (i.e., Chiral gravity) has only the right moving mode with $c_R=\frac{3\ell}{2G}$ for $\sigma=1$.   

\subsubsection{Chiral Gravity}
Since it is an important candidate that has the potential to provide a complete quantum gravity toy model in AdS/CFT framework, let us now dwell on some fundamentals of the Chiral gravity by following \cite{Chiral}. To this end, let us note that Eq.(\ref{TTeq}) can be rewritten as follows \cite{Chiral}
\begin{equation}
\Big ( \mathcal{D}^L  \mathcal{D}^R   \mathcal{D}^{m_g}    h \Big )_{\mu \nu}=0,
\label{product}	
\end{equation}
where the three mutually commuting first order operators are
\begin{eqnarray}
 ( \mathcal{D}^{L/R})_\mu\,^\nu \equiv \delta_\mu^\nu \pm \ell  \,\eta_{\mu}\,^{\alpha \nu} \bar\nabla_\alpha, \nonumber 
\hskip .5 cm ( \mathcal{D}^{m_g})_\mu\,^\nu \equiv \delta_\mu^\nu + \frac{1}{\mu\sigma} \eta_{\mu}\,^{\alpha \nu} \bar\nabla_\alpha.
\end{eqnarray}
Hence, one can split the third-order field equations in Eq.(\ref{product}) into three isolated first derivative equations as follows
\begin{equation}
 ( \mathcal{D}^L h^L )_{\mu \nu}=0,\hskip .5 cm  ( \mathcal{D}^R h^R  )_{\mu \nu}=0,  \hskip .5 cm ( \mathcal{D}^{m_g} h^{m_g} )_{\mu \nu}=0.
\end{equation} 
Notice that the most general solution of Eq.(\ref{TTeq}) can be given as decomposing the fluctuation into left, right and massive moving modes 
\begin{equation}
h_{\mu\nu}=h^L_{\mu\nu}+h^R_{\mu\nu}+h^{m_g}_{\mu\nu}.
\end{equation} 
Observe that the left and right modes are also solution of linearized Einstein's gravity and massive mode is a solution of TMG. These solutions can indeed be found from the representations of the $AdS_3$ symmetries which is $SL(2,R)_L \times SL(2,R)_R$. For this purpose, one can choose the following $AdS_3$ metric which is a solution of TMG
\begin{equation}
ds^2=\ell^2(-\cosh^2\rho dt^2+\sinh^2\rho d\phi^2+d\rho^2).
\end{equation}
The most general solutions can be defined as follows
\begin{equation}
h_{\mu \nu} = e^{ -i t ( h + \bar h)}  e^{ -i \phi ( h - \bar h)} F_{\mu \nu}(\rho),
\end{equation}
where 
\begin{eqnarray}
F_{\mu\nu}(\rho)=f(\rho)\left(\begin{array}{ccc}
                                         1 & {h-\bar{h}\over2}& {2i\over\sinh(2\rho)} \\
                                          {h-\bar{h}\over2} & 1 & {i(h-\bar{h})\over\sinh(2\rho)} \\
                                         {2i\over\sinh(2 \rho)} &{i(h-\bar{h})\over \sinh(2 \rho)}   & - {4\over\sinh^2(2\rho)} \\
                                       \end{array}\right),
\end{eqnarray}
and 
\begin{equation}
f(\rho)=(\cosh{\rho})^{-(h+\bar{h})}\sinh^2{\rho}.
\end{equation}
Here, $(h, \bar  h)$ are the primary weights that can be found via the algebra. Without going into details, let us note that the primary weights for the left and the right modes which are the Einstein's gravity modes are $(2,0)$ and $(0,2)$, respectively. On the other side, the primary weights for the massive modes are
\begin{equation}
(h, \bar  h) = \Big ( \frac{ 3 + \mu \ell\sigma}{2},  \frac{ -1 + \mu \ell\sigma}{2} \Big ).
\label{massiveweghts}
\end{equation}
Note that at the chiral point, the massive mode operator $(\mathcal{D}^{m_g})$ is equal to the one of the left mode $(\mathcal{D}^L)$ that leads to a degeneration at which the massive mode weights Eq.(\ref{massiveweghts}) reduce to $(h, \bar  h)=(2,0)$.   

We are now ready to find the energies of excitations by constructing the Ostragradsky Hamiltonian. For this purpose, one needs to find the ${\cal{O}}(h^2)$ action yielding Eq.(\ref{linerTMG1}):
\begin{equation}
\begin{aligned}
S&=\frac{\sigma}{64\pi G\Lambda}\int d^3x \sqrt{-\bar{g}} h^{\mu\nu}\Big ( \mathcal{D}^L  \mathcal{D}^R   \mathcal{D}^{m_g}    h \Big )_{\mu \nu},\\
&=-\frac{1}{32\pi G}\int d^3x \sqrt{-\bar{g}} h^{\mu\nu}({\sigma\cal G}^{L}_{\mu\nu}+\frac{1}{\mu}C^{L}_{\mu\nu}),\\
&=\frac{1}{64\pi G}\int d^3x
 \sqrt{-\bar{g}}\{-\sigma\bar{\nabla}^\lambda h^{\mu\nu}\bar{\nabla}_\lambda
 h_{\mu\nu}+\frac{2\sigma}{
\ell^ 2}h^{\mu\nu}h_{\mu\nu}\\&
\hskip 3.9 cm-\frac{1}{\mu}\bar{\nabla}_\alpha
h^{\mu\nu}\eta_\mu\,^{\alpha\beta}(\bar{\nabla}^2+\frac{2}
{\ell^ 2})h_{\beta\nu}\}.
\end{aligned}
\end{equation}
 The conjugate momentum is
\begin{equation}
\Pi^{(1)\mu\nu}=-{\sqrt{-\bar{g}}\over64\pi G}
\left(\bar{\nabla}^0(2\sigma h^{\mu\nu}+{1\over\mu}\eta^{\mu\alpha}\,_{\beta}\bar{\nabla}_\alpha
h^{\beta \nu}) -{1\over\mu}\eta_\beta\,^{0\mu
}(\bar{\nabla}^2+{2\over \ell^ 2})h^{\beta\nu}\right).\end{equation}
With the help of the equations of motion, one obtains
\begin{equation}
\begin{aligned}
\Pi_M^{(1)\mu\nu}&=\frac{\sqrt{-\bar{g}}}{64\pi
G}(-\sigma\bar{\nabla}^0h^{\mu\nu}+\frac{1}{\mu}(\mu^2-\frac{1}{
\ell^ 2})\eta_\beta\,^{0\mu}h_M^{\beta\nu}),\\
\Pi_L^{(1)\mu\nu}&=-{\sqrt{-\bar{g}}\over64\pi G}(2\sigma-{1\over\mu
\ell}) \bar{\nabla}^0h_L^{\mu\nu},\\
\Pi_R^{(1)\mu\nu}&=-{\sqrt{-\bar{g}}\over64\pi G}(2\sigma+{1\over\mu
\ell})\bar{\nabla}^0h_R^{\mu\nu}.
\end{aligned}
\end{equation}
For the Ostrogradsky method, one needs to introduce another dynamical variable $K_{\mu\nu}\equiv\bar{\nabla}_0h_{\mu\nu}$
whose conjugate momentum reads
\begin{equation}\Pi^{(2)\mu\nu}=\frac{-\sqrt{-\bar{g}}g^{00}}{64\pi G\mu}\eta_\beta\,^{\lambda\mu}\bar{\nabla}_\lambda h^{\beta\nu}.
\end{equation}
Then, one gets
\begin{equation}
\begin{aligned}
\Pi^{(2)\mu\nu}_M&=\frac{-\sqrt{-\bar{g}}g^{00}}{64\pi G}\sigma h_M^{\mu\nu},\\
\Pi^{(2)\mu\nu}_R&=\frac{\sqrt{-\bar{g}}g^{00}}{64\pi G\mu
\ell}h_L^{\mu\nu},\\\Pi^{(2)\mu\nu}_L,&=\frac{-\sqrt{-\bar{g}}g^{00}}{64\pi
G\mu \ell}h_R^{\mu\nu},
\end{aligned}
\end{equation}
with which the Hamiltonian turns into
\begin{equation}H=\int d^2x\bigl(\dot{h}_{\mu\nu}\Pi^{(1)\mu\nu}+\dot{K}_{i\mu}\Pi^{(2)i\mu}-S\bigr).\end{equation}
Finally, the energies for the left and right excitations will become 
\begin{equation}
E_{L/R} = - \frac{c_{L/R}}{48\pi\ell}  \int d^2 x \, \sqrt{-\bar{g}} \,\bar{\nabla}^0  h_{L/R}^{\mu \nu} \, \partial_t h^{L/R}_{\mu \nu}, 
\label{energy11}
\end{equation}
while the energy for the massive mode is
\begin{equation}
E_{m_{g}^2} = \frac{m_{g}^2}{64\pi\mu G}   \int d^2 x \, \sqrt{-\bar{g}} \, \eta _\alpha\,^{0\mu} h_{m_g}^{\alpha \nu}\, \partial_t h^{m_g}_{\mu \nu}.\label{energy22}
\end{equation}
By using the solutions, one can easily show that all the energy integrals in Eq.(\ref{energy11})-Eq.(\ref{energy22}) above are actually negative. Let us now calculate the results of the integrals for the primary states. First of all, the energies for the left and right modes can be found as   
\begin{equation}
E_{L/R} =  \frac{c_{L/R}}{36  \ell}.  
\end{equation}
Also, the energy for the massive mode will read
\begin{equation}
E_{m_{g}^2} = \frac{m_{g}^2 \ell }{64\mu G} f(x) .
\label{energy23}
\end{equation}
Here, the dimensionless parameter $x$ is defined as
\begin{equation}
x \equiv  \frac{\sigma \mu \ell}{2},
\end{equation}
and the function $f(x)$ reads
\begin{eqnarray}
f(x) &=&-\frac{2^{4 x+5} (2 x+1)}{ 3+2x} \Bigg( \frac{(3 +2 x)}{2 (x+1)} \, _2F_1 [2 (x+1),4 x+5;2 x+3;-1 ] \\ \nonumber
&& \qquad\qquad\qquad\qquad- \,
   _2F_1 [2 x+3,4 x+5;2 (x+2);-1 ] \Bigg),
\end{eqnarray}
where the  $_2F_1$ is the hypergeometric function \cite{Nuevo}.  The important point here is that  for the physical regions $x \in [0,  \infty) $, the function $f(x)$ decreases and thus yields the values as  $ f(x) \in [-1, -2)$ which gives a negative energy for the massive mode. Therefore, to have positive energies for bulk excitations and positive or vanishing central charges, one must go to the chiral limit where the graviton mass ($m_g$) and left central charge ($c_L$) vanish. It is important to note that $\sigma$ must be chosen positive to have positive energy for BTZ black holes such that their energies are given as $E_{BTZ}\sim M\sigma$. Notice that this tells us that the bulk and boundary unitarity clash is recovered only in chiral limit with the choice of positive $\sigma$.  

On the other hand chiral gravity has problematic log mode solutions \cite{GrumillerJohansson} that are non-unitary modes. These modes do not satisfy the Brown-Henneaux (BH) boundary conditions \cite{BH} and they have linearly-growing time profile. It was also discussed in \cite{Altas} that there is a linearization instability, namely; these problematic modes are an artifact of perturbation theory and so they are not integrable to full solution. To say it in another way, there is no exact solution to chiral TMG, whose linearization about $AdS_3$ yields the problematic log mode.  

\subsection{Weyl Invariance}
Gauge theory is one of the most amazing developments in physics to describe the fundamental interactions in Nature. The concept of gauge theory goes back to Hermann Weyl's work in 1918\footnote{In fact, the word gauge (Eich) was first used by him}. In his work, Weyl tried to unify gravity with electromagnetism in a geometrical framework by starting with a quadratic curvature action \footnote{He started with a quadratic curvature action which is Weyl gauge invariant since Einstein theory is not gauge invariant.}. Weyl's idea is based on the rescaling of the metric:\footnote{See \cite{Weylhistory1} for details.}
	 \begin{equation}
 \mbox{g}_{\mu\nu} \rightarrow  \mbox{g}^{'}_{\mu\nu}= e^{\lambda \int {\cal W}_\alpha dx^\alpha} \mbox{g}_{\mu\nu},
 \label{weylmet1}
\end{equation}
where $\lambda$ is a real constant and ${\cal W}_\alpha$ is the vector field. Upon this idea, Einstein showed that there was a discrepancy between Weyl's idea and experimental evidence such that the spectral lines of atomic clocks would then depend on the causal history of the atoms. Thus, if one assumes that two hydrogen atoms have different location, then they could have different masses and frequencies. Therefore, Weyl's approach would ostensibly conflict with the physical principles. However, with the developing of wave mechanics by Erwin Schr\"odinger and Paul Dirac, London \cite{London} proposed that the Weyl's non-integrable scale factor should be replaced with a purely imaginary one in the case of coupling electromagnetism to charged fields with which it would correspond to the phase factor associated with the Schr\"odinger wave function. In the presence of the electromagnetic field, London elaborated Weyl's proposal to wave mechanics as follows
\begin{equation}
 \Psi(x) \rightarrow  \Psi^{'}(x)= e^{\frac{i}{\hbar} \int A_\alpha dx^\alpha} \Psi(x).
 \label{weylactfirstwww}
\end{equation} 
Note that in the absence of the electromagnetic field, the scale factor will be integrable and $A_\mu$ can be fixed by gauge choosing. Following London's work, Weyl proposed that gauge invariance incorporates matter into electromagnetism rather than gravity. Years later, Weyl's idea played the fundamental role in describing the microscopic interactions such as weak and strong interactions based on Yang-Mills gauge fields, a generalization of Weyl's ideas.

After giving a historical development of Weyl's method, let us give some basics of the Weyl transformations:
\subsubsection{Weyl Transformation}
To construct a Weyl invariant theory, let us consider the free scalar field action 
 \begin{equation}
S_{\Phi}=- \frac{1}{2}\int d^n x \sqrt{-\mbox{g}} \,\, \partial_\mu \Phi  \partial_\nu \Phi g^{\mu \nu},
\label{scafieldac}
\end{equation}
which is explicitly invariant under rigid Weyl transformations
\begin{equation}
 \mbox{g}_{\mu\nu} \rightarrow \mbox{g}^{'}_{\mu\nu}=e^{2 \lambda} \mbox{g}_{\mu\nu}, \hskip 1 cm \Phi \rightarrow \Phi^{'} =e^{-\frac{(n-2)}{2}\lambda} \Phi,
\end{equation}
where $\lambda$ is a constant. If one applies rigid Weyl invariance to a generic curved backgrounds, it could become unfruitful except conformally flat curved backgrounds. At this stage, one needs to replace rigid Weyl invariance by a local one. For this purpose, let us consider the local Weyl transformations with the following transformations of the metric and scalar field in $n$ dimensions
 \begin{equation}
 \mbox{g}_{\mu\nu} \rightarrow \mbox{g}^{'}_{\mu\nu}=e^{2 \lambda(x)} \mbox{g}_{\mu\nu}, \hskip 1 cm \Phi \rightarrow \Phi^{'} =e^{-\frac{(n-2)}{2}\lambda(x)} \Phi,
 \label{weylmetscalar}
\end{equation}
where $\lambda(x)$ is an arbitrary function of coordinates. Obviously, free scalar field action Eq.(\ref{scafieldac}) is not invariant under local Weyl transformations because $\partial_\mu \Phi $ transforms as
\begin{equation}
 \partial_\mu \Phi \rightarrow (\partial_\mu \Phi)^{'}=e^{-\frac{(n-2)}{2}\lambda(x)} \partial_\mu\Phi-\frac{(n-2)}{2}e^{-\frac{(n-2)}{2}\lambda(x)} \Phi\partial_\mu\lambda(x).
 \label{extratrm}
\end{equation}
To make the action Weyl-invariant, one needs to eliminate the extra terms induced by the transformations due to the partial derivatives. This can be achieved by replacing the ordinary partial derivatives $\partial_\mu$ in Eq.(\ref{scafieldac}) with the so-called gauge covariant derivative $\mathcal{D}_\mu$ which transforms according to
\begin{equation}
\mathcal{D}_\mu \Phi \rightarrow (\mathcal{D}_\mu \Phi)^{'}=e^{q\lambda(x)} \mathcal{D}_\mu \Phi,
\label{gaugecov}
\end{equation} 
where $q$ is the weight of the transformation. To ultimately reach a Weyl-invariant action, one needs to define $\mathcal{D}_\mu$ in such a way that its transformation cancels out the undesired extra term in Eq.(\ref{extratrm}). To this end, let us consider the following gauge covariant derivative of the scalar field
\begin{equation}
 \mathcal{D}_\mu \Phi =(\partial_\mu -q A_\mu)\Phi,
 \label{weytrametscal1}
\end{equation}
which transforms under the scale transformations Eq.(\ref{weylmetscalar})  
\begin{equation}
\mathcal{D}_\mu \Phi \rightarrow (\mathcal{D}_\mu \Phi)^{'}=e^{-\frac{(n-2)}{2}\lambda(x)} \bigg(\partial_\mu\Phi-\frac{(n-2)}{2} \Phi\partial_\mu\lambda(x)-qA'_\mu\phi\bigg).
\end{equation}
If we compare this with Eq.(\ref{extratrm}) and Eq.(\ref{gaugecov}), $q$ must be chosen as
$q=-\frac{(n-2)}{2}$ and the new gauge field, the so called Weyl's gauge field $ A_\mu $ must transform as
\begin{equation}
 A_\mu \rightarrow A^{'}_\mu = A_\mu - \partial_\mu \lambda(x).
\end{equation}
The local Weyl invariance requires introducing the Weyl's gauge field. Note also that Weyl's gauge field does not transform in the same way as the scalar field. Finally, transformed form of the gauge-covariant derivative of the scalar field will read 
 \begin{equation}
 (\mathcal{D}_\mu \Phi)^{'}=e^{-\frac{(n-2)}{2}\lambda(x)} \mathcal{D}_\mu \Phi.
 \label{weytrametscal12}
\end{equation}
Consequently, Weyl invariant form of the free scalar field action is obtained by replacing partial derivatives with gauge covariant derivative and introducing Weyl's gauge field. It follows from this that Eq.(\ref{scafieldac}) is invariant under local Weyl invariance. Moreover, one can add a Weyl invariant potential to the scalar field action as
\begin{equation}
S_\Phi=- \frac{1}{2}\int d^n x \sqrt{-\mbox{g}} \,\,\Big (   \mathcal{D}_\mu \Phi \mathcal{D}^\mu\Phi +\nu \, \Phi^{\frac{2n}{n-2}}\Big ) ,
\label{scalarwithpot1}
\end{equation}
where $\nu$ is a dimensionless positive coupling constant which provides a renormalizable theory in $n=3$ and $n=4$ dimensional flat backgrounds.

In the light of the above derivations, let us consider the free electromagnetic field action in an $n-$dimensional curved background. One can easily verify that field strength tensor associated with the gauge field $F_{\mu \nu}=\partial_\mu A_\nu-\partial_\nu A_\mu$ is invariant under local Weyl symmetry. Although the field strength tensor $F_{\mu \nu}$ is gauge invariant, $\sqrt{-\mbox{g}} F_{\mu \nu} F^{\mu \nu}$ term in the action is not invariant under Weyl transformations Eq.(\ref{weylmetscalar}). In order to have a Weyl-invariant Maxwell-type action, one needs a compensating scalar field with appropriate weight. In fact, it is straight forward to show that Weyl-gauged version of Maxwell-type action reads
 \begin{equation}
S_{A^\mu} =  - \frac{1}{2} \int d^n x \sqrt{-\mbox{g}}\,\, \Phi^{\frac{2(n-4)}{n-2}} F_{\mu \nu} F^{\mu \nu}.
\end{equation}
Note that the action does not require a compensating scalar field in four dimensions, as expected since in four dimensions, the Maxwell theory is already Weyl invariant.  
 
In the above discussions, we gave a general view to construct Weyl invariant actions such as a scalar field and electromagnetic field. Now, we want to apply the same framework to Einstein's gravity. To do so, one needs to introduce a Weyl invariant connection by using the Christoffel connection and then find the Weyl invariant form of Riemann and Ricci tensors and the curvature scalar. Before elaborating this, let us take a look at how gauge covariant derivative acts on the metric. It is simply given by
 \begin{equation}
 \mathcal{D}_\mu \mbox{g}_{\alpha \beta} =\partial_\mu \mbox{g}_{\alpha\beta} + 2 A_\mu \mbox{g}_{\alpha \beta},
\end{equation}   
which transforms according to 
\begin{equation}
 ( \mathcal{D}_\mu \mbox{g}_{\alpha \beta})^{'}=e^{2 \lambda(x)} \mathcal{D}_\mu \mbox{g}_{\alpha \beta}. 
\end{equation}
To insert Weyl-invariance into gravity, one needs to find the Weyl invariant version of Christoffel symbol in such a way that it remains invariant under local Weyl transformations and contains gauge covariant derivatives. Referring to \cite{DengizTekin} for details, let us notice that the following connection will fulfil the job  
 \begin{equation}
 \tilde{\Gamma}^\lambda_{\mu\nu}=\frac{1}{2}\mbox{g}^{\lambda\sigma} \Big ( \mathcal{D}_\mu \mbox{g}_{\sigma\nu}+\mathcal{D}_\nu \mbox{g}_{\mu\sigma}
-\mathcal{D}_\sigma \mbox{g}_{\mu\nu} \Big ).
\end{equation}
Then, the Weyl-invariant version of Riemann tensor can be obtained as
\begin{equation}
\begin{aligned}
  \tilde{R}^\mu{_{\nu\rho\sigma}} [\mbox{g},A]&=\partial_\rho \tilde{\Gamma}^\mu_{\nu\sigma}-\partial_\sigma \tilde{\Gamma}^\mu_{\nu\rho}
+ \tilde{\Gamma}^\mu_{\lambda\rho} \tilde{\Gamma}^\lambda_{\nu\sigma}-\tilde{\Gamma}^\mu_{\lambda\sigma} \tilde{\Gamma}^\lambda_{\nu\rho} \\
& =R^\mu{_{\nu\rho\sigma}}+\delta^\mu{_\nu}F_{\rho\sigma}+2 \delta^\mu{_[\sigma} \nabla_{\rho]} A_\nu 
+2 \mbox{g}_{\nu[\rho}\nabla_{\sigma]} A^\mu \\
& \quad +2 A_[\sigma   \delta_{\rho]}\,^\mu A_\nu  
+2 \mbox{g}_{\nu[\sigma}  A_{\rho]} A^\mu  +2 \mbox{g}_{\nu[\rho} \delta_{\sigma]}\,^\mu  A^2 , 
\label{weyinvriem1}
\end{aligned} 
\end{equation}
where $ 2 A_{[ \rho} B_{\sigma]} \equiv A_\rho B_\sigma -  A_\sigma B_\rho$; $\nabla_\mu A^\nu = \partial_\mu A^\nu+\Gamma^\nu_{\mu \rho}A^\rho$; $A^2= A_\mu A^\mu$.
After taking a contraction of Eq.(\ref{weyinvriem1}), the Weyl-invariant Ricci tensor can be computed as 
 \begin{equation}
\begin{aligned}
\tilde{R}_{\nu\sigma} [\mbox{g},A]&= \tilde{R}^\mu{_{\nu\mu\sigma}}[g,A] \\
&=R_{\nu\sigma}+F_{\nu\sigma}-(n-2)\Big [\nabla_\sigma A_\nu - A_\nu A_\sigma +A^2 \mbox{g}_{\nu\sigma} \Big ]-\mbox{g}_{\nu\sigma}\nabla \cdot A,
\label{weyinricciten1}
\end{aligned}
\end{equation}
where $\nabla \cdot A \equiv \nabla_\mu  A^\mu$. Finally taking the trace of the Ricci tensor, one gets the Weyl-extended the curvature scalar as 
\begin{equation}
 \tilde{R}[\mbox{g},A]=R-2(n-1)\nabla \cdot A-(n-1)(n-2) A^2,
 \label{ricciscalar1}
\end{equation}
which is {\it not} invariant under local Weyl transformations, but rather transforms like the inverse metric:
\begin{equation}
  \tilde{R}[\mbox{g},A] \rightarrow (\tilde{R}[\mbox{g}^{'},A^{'}])^{'} = e^{-2 \lambda (x) }  \tilde{R}[\mbox{g},A].
  \label{weyscalatrans1}
\end{equation}
Thus, unlike the Riemann and Ricci tensor, the curvature scalar is not invariant under Weyl symmetry. To construct the Weyl-invariant version of Einstein's gravity, one must resolve this problem by using a compensating scalar field \cite{DengizTekin}. Finally, the Weyl-gauged version of Einstein-Hilbert action can be written as
 \begin{equation}
 \begin{aligned}
S&= \int d^n x \sqrt{-\mbox{g}} \, \Phi^2  \tilde{R}[\mbox{g},A]\\
&= \int d^n x \sqrt{-\mbox{g}} \, \Phi^2\Big [R-2(n-1)\nabla \cdot A-(n-1)(n-2) A^2 \Big].
\label{weh1}
\end{aligned}
\end{equation}
Observe that the variation of Eq.(\ref{weh1}) with respect to Weyl's gauge field $A_\mu$ yields a constraint equation
\begin{equation}
A_{\mu }= \frac{2}{n-2}\partial_{\mu} \ln \Phi,
\label{wpuregauge1}
\end{equation}
which requires Weyl's gauge field $A_\mu$ to be pure gauge which is not dynamical. Hence, if one inserts Eq.(\ref{wpuregauge1}) into Eq.(\ref{weh1}), one can eliminate the gauge field and then obtains the "Conformally Coupled Scalar Tensor theory" described by the action
\begin{equation}
S=\int d^n x \sqrt{-\mbox{g}} \,\, \Big (\Phi^2 R+4\frac{(n-1)}{n-2} \partial_\mu \Phi \partial^\mu \Phi \Big).
\end{equation}
 
\subsection{Potential Energy From Tree-level Scattering Amplitude in Generic Gravity Theories}
\begin{figure}[h]
\centering
\includegraphics[width=0.5\textwidth]{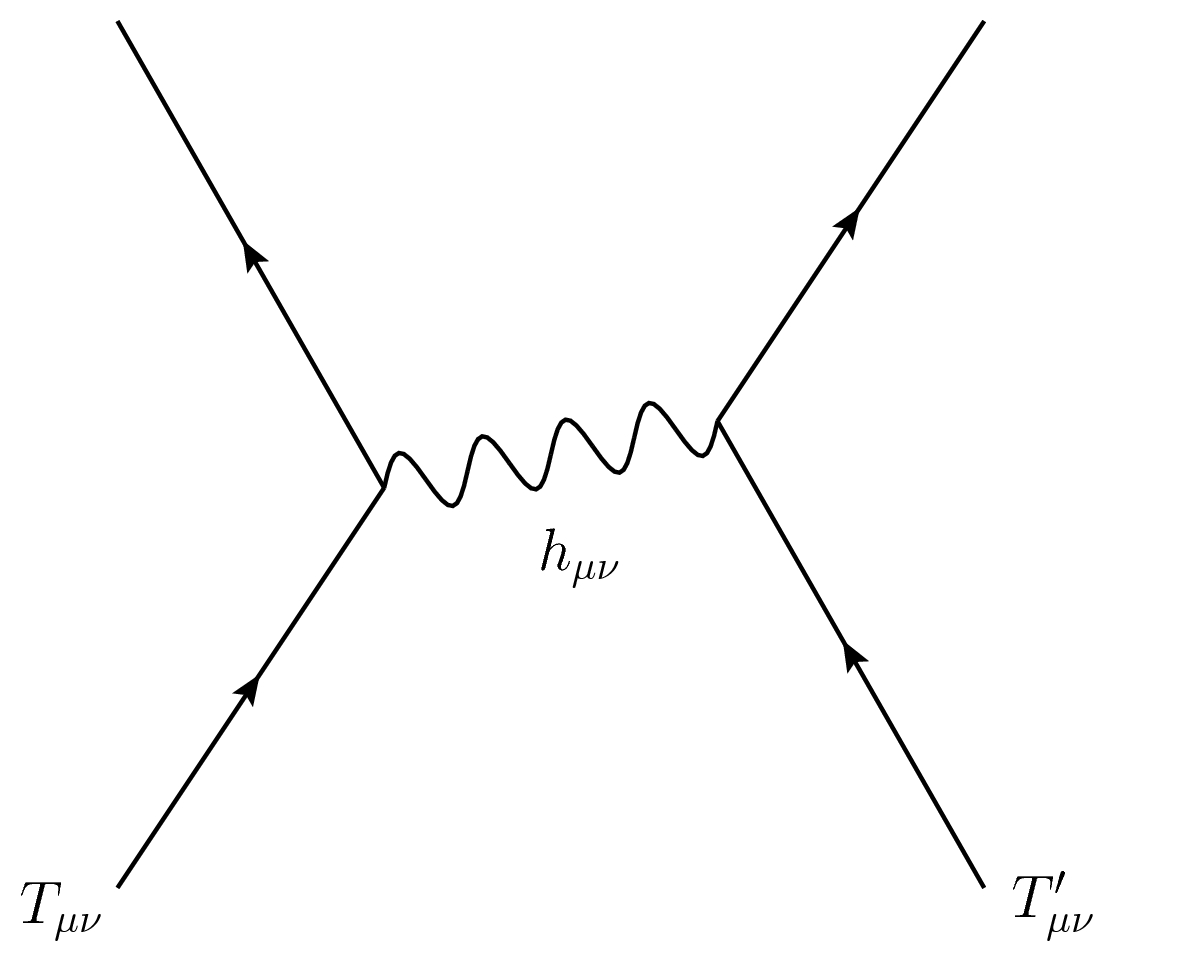}
\caption{Tree-level scattering amplitude between two conserved sources via the exchange of one graviton.}
\label{GravitonExchange}
\end{figure}

In this part, since we will consider the tree-level scattering amplitude which will provide the potential energy between two sources via the exchange of one graviton as in the Figure \ref{GravitonExchange}, let us briefly recapitulate some basics of the formulation. For this purpose, we will calculate the vacuum to vacuum transition amplitude between two covariantly conserved sources by using the path integral formalism which is defined as 
\begin{equation}
\langle 0| \exp^{-iHt}|0\rangle=\exp^{iUt}=W[T]=\int {\cal{D}}h_{\mu\nu}\exp^{iS[h,T]},
\label{vacuumtovacuum1}
\end{equation}
where $t$ is a large time and $S[h,T]$ is a linearized action about a generic background ($\bar{g}_{\mu\nu}$). For a source coupled generic gravity theory, $S[h,T]$ can be written in the following form in $n$ dimensions 
\begin{equation}
S[h,T]=-\int d^n x \sqrt{-\bar{g}} \,h^{\mu\nu}\bigg(\frac{1}{\kappa}E_{\mu\nu}(h)-T_{\mu\nu}\bigg),
\end{equation} 
whose linearized field equations are
\begin{equation}
E_{\mu\nu}(h)=\frac{\kappa}{2}T_{\mu\nu}.
\label{Treelvlfeq}
\end{equation}
Since we consider the covariantly conserved sources ($\bar{\nabla}^\mu T_{\mu\nu}=0$), it leads to $\bar{\nabla}^\mu E_{\mu\nu}=0$. To simplify the path integral further, one can apply the background field technique which is described by
\begin{equation}
h_{\mu\nu}\rightarrow h_{\mu\nu}+\bar{h}_{\mu\nu},
\label{Treelvlfr}
\end{equation}
where we suppose that $\bar{h}_{\mu\nu}$ is a solution of field equations Eq.(\ref{Treelvlfeq}). Note that path integral remains intact with respect to field redefinition Eq.(\ref{Treelvlfr}). It follows from  the field redefinition, linearized action $S[h,T]$ takes the form
\begin{equation}
S[h,T]=-\int d^n x \sqrt{-\bar{g}} \,\bigg(\frac{1}{\kappa}h^{\mu\nu}E_{\mu\nu}(h)-\frac{1}{2}\bar{h}^{\mu\nu} T_{\mu\nu}\bigg).
\label{Treelvlact}
\end{equation} 
Observe that field redefinition decouples $h_{\mu\nu}$ and $T_{\mu\nu}$. By plugging Eq.(\ref{Treelvlact}) into Eq.(\ref{vacuumtovacuum1}) and using the fact that the second term can be taken out of the path integral such that it does not have a $h_{\mu\nu}$ coupling and the first term leads to a normalization constant, then one arrives at
\begin{equation}
W[T]=W[0]\exp^{\frac{i}{2}\int  d^n x \sqrt{-\bar{g}}\bar{h}^{\mu\nu} T_{\mu\nu} }.
\label{Treelvlgenerator}
\end{equation}
On the other hand, the solution of $\bar{h}^{\mu\nu}$ can be found by Green's function technique. It is simply given by
\begin{equation}
\bar{h}^{\mu\nu}=\frac{\kappa}{2}\int d^n x' \sqrt{-\bar{g}}G^{\mu\nu}{_{\rho\sigma}}(x,x')T^{\rho\sigma}(x'),
\label{Treelvlsol}
\end{equation}
where $G_{\mu\nu\rho\sigma}(x,x')$ stands for the Green's function which obeys the following relation
\begin{equation}
{\cal{O}}_{\mu\nu\rho\sigma}G^{\rho\sigma}{_{\alpha\beta}}=\frac{1}{2}(\bar{g}_{\mu\alpha}\bar{g}_{\nu\beta}
+\bar{g}_{\mu\beta}\bar{g}_{\nu\alpha})\delta^{(n)}(x-x'),
\end{equation} 
with which the linearized field equation can be rewritten in the following form 
\begin{equation}
{\cal{O}}_{\mu\nu\rho\sigma}\bar{h}^{\rho\sigma}=\frac{\kappa}{2}T_{\mu\nu}.
\end{equation}
Thereupon, substituting Eq.(\ref{Treelvlsol}) back into the Eq.(\ref{Treelvlgenerator}), one gets
\begin{equation}
W[T]=W[0]\exp^{\frac{i\kappa}{4}\int  d^n x\,d^n x' \sqrt{-\bar{g}(x)}\,\sqrt{-\bar{g'}(x')} T^{\mu\nu}(x)\, G_{\mu\nu\rho\sigma}(x,x')\,T^{\rho\sigma}(x')}.
\end{equation}
Thus, one can obtain the potential energy in the following desired form \cite{gullu_spin} 
\begin{equation}
{\cal{U}}=-\frac{\kappa}{4t}\int  d^n x\,d^n x' \sqrt{-\bar{g}(x)}\,\sqrt{-\bar{g'}(x')} T^{\mu\nu}(x)\, G_{\mu\nu\rho\sigma}(x,x')\,T^{\rho\sigma}(x').
\end{equation}

\subsection{Shapiro Time Delay}
Shapiro's time delay is the fourth test of GR, which was proposed as an experiment in 1964 by Irwin I.Shapiro \cite{Shapiro:1964uw}, in the solar system. In his paper, he showed that when light rays pass near a massive object, they take longer time in the round trip due to gravitational potential of the massive object, compared to the time it takes when the massive object is absent. Shapiro suggested also an experimental set-up to measure the time delay; a radar signal is transmitted from the Earth to Venus or Mercury and back, in the presence of Sun. After a few years, the proposed experiment was carried out by him and observations verified the GR predictions. 

To calculate the time delay, Shapiro used the Schwarzschild geometry to describe gravitational field near the Sun. But we shall here give a derivation for the time delay by using a shock wave geometry since the shock wave solutions are more appropriate to study on higher order gravity theories \footnote{Shock waves are exact solution of any theory of gravity whose action consist of Riemann tensor and its contractions \cite{Horowitz89}.}. For this purpose, let us consider an $n$- dimensional shock wave metric  generated by a high-energy massless particle moving in the $+x$ direction with momentum $p^\mu=\lvert p\rvert(\delta^\mu_0+\delta^\mu_x) $:
\begin{equation}
  ds^2=-dudv+H(u,{\bf{x}})du^2+\sum_{i=1}^{n-2} (dy^i)^2.
  \label{shapirometric1}
\end{equation}
where $u=t-x$ and $v=t+x$ null coordinates. The corresponding energy momentum tensor can be given as 
\begin{equation}
  T_{uu}=\lvert p\rvert \delta(u)\prod_{i=1}^{n-2}\delta(y^i).
\end{equation}
For the shock wave ansatz Eq.(\ref{shapirometric1}), the Einstein's field equations reduce to 
\begin{equation}
\begin{aligned}
\sum_{i=1}^{n-2}\partial_{y_{i}}^2H(u,{\bf{x}})=-16\pi G\lvert p\rvert\delta(u)\prod_{i=1}^{n-2}\delta(y^i)  ~. 
\label{shapiroeqn1}
\end{aligned}
\end{equation}
To solve this equation, assume that the solution is in the form $H(u,{\bf{x}})=\delta(u)g({\bf{x}})$. If one puts this into Eq.(\ref{shapiroeqn1}), one obtains  
\begin{equation}
\nabla_{\perp}^2 g({\bf{x}})=-16G\lvert p\rvert\prod_{i=1}^{n-2}\delta(y^i),
\end{equation}
where $\nabla_{\perp}^2=\sum_{i=1}^{n-2}\partial_{y_{i}}^2$. One can also realize that $g(r)$ is a Green's function of the $n-2$ dimensional Laplace operator. The easiest way to solve this equation is by going to spherical coordinates:
\begin{equation}
\begin{aligned}
&\int_{V}\nabla_{\perp}^2 g(r)dV=-16\pi G\lvert p\rvert\int_{V}\prod_{i=1}^{n-2}\delta(y^i)dV,\\&
\hskip 0.6 cm \Omega_{n-2}r^{n-3}\frac{dg(r)}{dr} =-16\pi G\lvert p\rvert,
\label{shapeq22}
\end{aligned}
\end{equation}
where we have used the Gauss' theorem and $\Omega_{n-2}$ is the $n-2$ dimensional solid angle which is given as $\Omega_{n-2}=\frac{2\pi^{\frac{n-2}{2}}}{\Gamma(\frac{n-2}{2})}$. Then from the Eq.(\ref{shapeq22}), the solution reads
\begin{equation}
g(r)=\frac{16\pi G\lvert p\rvert}{(n-4)\,\Omega_{n-2}r^{n-4}},
\end{equation}
\begin{figure}[h]
\centering
\includegraphics[width=0.5\textwidth]{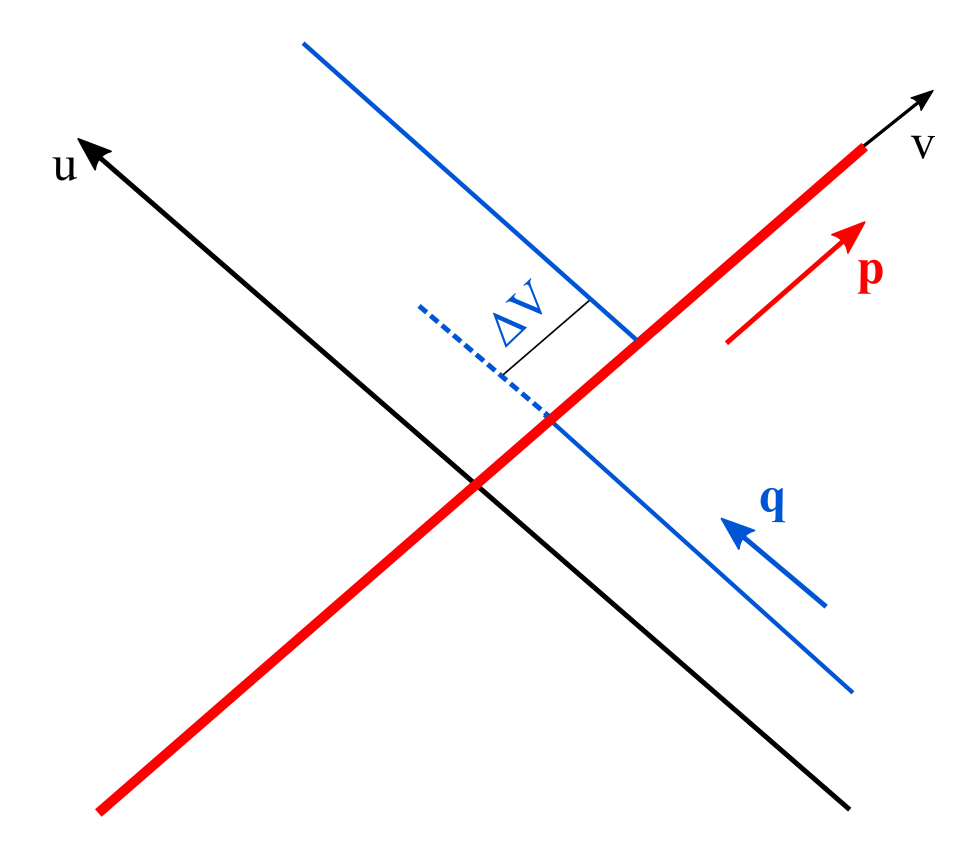}
\caption{A shock wave generated by a massless particle at $u=0$ (red line). When another test particle with momentum $q$ (blue line) crosses the shock-wave, there will be a shifting in the null cone coordinate, $\Delta v$. }
\label{shapirogeodesics}
\end{figure}
here we dropped the integral constant to have zero profile function ($H(u,{\bf{x}})$) in the asymptotic limit. Finally, the most general solution can be found as
\begin{equation}
H(u,{\bf{x}})=\frac{4 \Gamma(\frac{n-4}{2})}{\pi^{\frac{n-4}{2}}}\frac{G\lvert p\rvert}{r^{n-4}}\delta(u).
\end{equation}
Let us now consider massless test particle with momentum $q$ crossing the shock wave geometry which is generated by another massless particle, with an impact parameter $r=b$ as shown in the Figure \ref{shapirogeodesics}. In this case, shock wave metric takes the following form
\begin{equation}
 ds^2=-du\bigg(dv-\frac{4 \Gamma(\frac{n-4}{2})}{\pi^{\frac{n-4}{2}}}\frac{G\lvert p\rvert}{r^{n-4}}\delta(u)du\bigg)+\sum_{i=1}^{n-2} (dy^i)^2.
\end{equation}
Observe that there is a discontinuity in the $u$ coordinate and it can be cancelled out at $r=b$ by re-defining the other null coordinate as
\begin{equation}
 v \equiv v_{new}+\frac{4 \Gamma(\frac{n-4}{2})}{\pi^{\frac{n-4}{2}}}\frac{G\lvert p\rvert}{b^{n-4}}\theta(u).
\end{equation}
This tells us that when a particle crosses the shock wave, it experiences a time delay given by
\begin{equation}
\Delta v=\frac{4 \Gamma(\frac{n-4}{2})}{\pi^{\frac{n-4}{2}}}\frac{G\lvert p\rvert}{b^{n-4}},
\label{shapfinaleq}
\end{equation}
which is positive definite as expected. If $\Delta v$ is positive, the particle suffers from a time delay; otherwise, it leads to causality violation. Note that Eq.(\ref{shapfinaleq}) ostensibly diverges in four dimensions due to gamma function, but one can easily show that $g(r)$ is given in the logarithmic form like $\log(r)$ in four dimensions, which yields
 \begin{equation}
\Delta v=-8G\lvert p\rvert\log(b)+f,
\end{equation}
where $f$ is regular function which satisfies $\nabla_{\perp}^2 f=0$.

\section[Weyl gauging of topologically massive gravity]{Weyl gauging of topologically massive gravity\footnote{The results of this chapter are published in \cite{kilicarslan1}. }}
Pure GR cannot provide viable explanations for some gedanken or real phenomena at both small (UV) and large (IR) scales. To be more precise, recall that as one approaches GR from the perturbative quantum field theory context, one sees that due to the existing dimensionful coupling constant (that is, Newton's constant with mass dimension $-2$), the infinities coming from the self-interaction of gravitons cannot be regulated to a finite value and thus the theory unavoidably turns out to be a non-renormalizable one. As for the IR regime, it is known that pure GR breaks down to give explanations to the accelerating expansion of universe and rotational curves of spiral galaxies. Thus, GR is valid only in a certain energy regime and thus a viable modification of GR seems to be essential in order to have a full theory. Particularly in the UV scale, one has to somehow do it for the sake of the long-lived idea of quantum gravity theory. Even though there are several perturbative or non-perturbative alternative approaches for a complete GR in the UV scale in the literature, one can in fact address the higher curvature modifications to heel the undesired propagator structure and the scattering potentials etc. In this regard, by bearing in mind the superficial degrees of GR in the perturbative aspect, one can attempt to amend the pure GR by adding scalar higher order curvature as follows
\begin{equation}
I=\int d^4x \sqrt{-g} (\sigma R+\alpha R^2+\beta R_{\mu\nu}^2).
\label{firsteq}
\end{equation}
However, this comes with an unexpected problem albeit several intriguing remedies, particularly the renormalization. More precisely, with the above modification, Eq.(\ref{firsteq}) acquires additional extra DOF of massive spin-2 and also a massless spin-2 excitations in addition to the regular massless spin-2 belonging to the pure GR \cite{stelle}. Unfortunately, although the above approach remarkably gets over the renormalization obstacle, the unitarity of massive and massless spin-2 excitations are contradiction and thus the theory unavoidably turns out be a non-unitary one. As to the large scale modification, as a distinct method to the common approaches, one may attempt to supply an adequately small amount of mass to graviton which, with the emerging extra DOF, will apparently be valuable candidates to dark matter and energy. In this perspective, although there are several alternative proposals, the following two models undoubtedly are the only ones which stand out and deserve to be considered deeply: the first one is the so-called Fierz-Pauli (FP) massive gravity theory which has been constructed in 1939 \cite{PF}. This model recasts the bare GR via an appropriate mass term in such a way that the graviton gains an acceptable mass at the end. However, this modification possesses the problems of vDVZ discontinuity \cite{vdvz1,vdvz2} at the linearized level and at the non-linear level the Boulware-Deser ghost mode \cite{BD} as well as the breaking of gauge-invariance. As to the second one and as compared with the other foremost models, the $2+1$-dimensional Topologically Massive Gravity (TMG) \cite{ DJT} and its particular limits seem to be the only viable lower dimensional massive theory that deserves to be elaborated even if they also contain some unavoidable loop-holes. Needless to say that the priority of addressing the $2+1$-dimensions is nothing but merely to get some insights in the idea of quantum gravity. TMG is a renormalizable \cite{DZ} and unitary theory and describes a single massive spin-2 particle. Particularly, due to having asymptotically $AdS_3$ solutions and several other forthcoming reasons, the Cosmological TMG has the potential to provide a well-behaved quantum gravity in AdS/CFT framework \cite{Chiral}.  

Contemplating on the discussions so far and the promising properties of TMG as well as the \cite{DengizTekin, Tanhayi:2011aa, Tanhayi:2012nn, Dengiz1} in which it is shown that the corresponding masses of the particles in the Weyl-invariant New Massive Gravity can be obtained via the breaking of the Weyl's symmetry as in the Higgs mechanism, here we would like to answer the question of whether the mass of spin-$2$ particle in TMG can also be produced in the same way or not. To do so, let us now jump to the construction of the Weyl-invariant TMG:\\

\subsection{Weyl-gauging of TMG}

Recall that neither pure GR nor its conformally invariant version (that is, conformally coupled scalar-tensor theory)  
\begin{equation}
S=\int d^3 x \sqrt{-g} \Big (\Phi^2 R +8 \partial_\mu \Phi \partial^\mu \Phi-\frac{\nu \Phi^6}{2} \Big),
\label{cieha}
\end{equation}
do not possess any local physical DOF in $2+1$ dimensions. Notice that this can be seen either by expanding the action up to second order in fluctuations around its vacuum or by going to the Einstein frame in which the related DOF will be manifest \cite{Tanhayi:2011aa}. On the other hand, once one augments Einstein's theory with a gravitational Chern-Simons term one ends up with the TMG as follows
\begin{equation}
 S_{TMG}=\int d^3 x \sqrt{-g} \bigg [ \sigma m R +\frac{k}{2 } \eta^{\lambda \mu \nu} \Big ( \Gamma^\rho_{\lambda \sigma} \partial_\mu \Gamma^\sigma_{\nu \rho}+\frac{2}{3} \Gamma^\rho_{\lambda \sigma} \Gamma^\sigma_{\mu \tau}\Gamma^\tau_{\nu \rho}  \Big ) \bigg ].
\label{tmg1-21}
\end{equation}
The theory defines a dynamical parity-violating massive spin-2 graviton with mass $ M_{graviton}=\lvert\frac{\sigma m}{k}\rvert$ about the flat vacua. Here, $ \sigma,k $ are dimensionless parameters and $ \eta^{\mu \nu \alpha}$ is a rank-3 tensor described in terms of the Levi-Civita symbol as $\epsilon^{\mu \nu \alpha}/\sqrt{-g}$. Notice that, one must pick $\sigma<0$ and $m>0$ in order to avoid having a ghost in the theory. Moreover, by adding a cosmological constant to TMG, one gets the cosmological TMG with mass as $M^2_{graviton}=\frac{\sigma^2 m^2}{k^2 }+\Lambda $ \cite{Carlip:2008jk, Gurses:2011fv}. Here, if the cosmological constant is particularly set to the following certain value 
\begin{equation}
 \frac{\sigma^2 m^2\ell^2}{k^2}=1,\hskip .5 cm\Lambda=-\frac{1}{\ell^2},
\end{equation}
the theory is then called the chiral limit of TMG (namely, Chiral gravity \cite{Chiral}) which interestingly satisfies the bulk-boundary unitarity conditions to some degree. (See also \cite{Bagchi:2012yk} in which it is proposed that the flat-space Chiral gravity may provide a holographic correspondence between an asymptotically flat limit of TMG and a $1+1$ dimensional CFT.)

The Chern-Simons term is diffeomorphism and conformally invariant up to a boundary term.\footnote{If one takes $ \sigma=0 $ which corresponds to pure gravitational Chern-Simons theory, it was shown that theory is equivalent to gauge theory \cite{Horne:1988jf} or in \cite{Afshar:2011qw} in the AdS/CFT context.} However, due to Einstein sector, TMG as a whole does not remain intact under conformal transformations. Furthermore, by using Eq.(\ref{cieha}), the conformally invariant version of TMG, up to a boundary, will read as follows \cite{Deser:2004wd}
\begin{equation}
\begin{aligned}
 S_{CTMG}=\int d^3 x \sqrt{-g} \bigg [&\sigma \Phi^2 R +8 \partial_\mu \Phi \partial^\mu \Phi-\frac{\nu \Phi^6}{2}\\
& + \frac{k}{2 } \eta^{\lambda \mu \nu} \Big ( \Gamma^\rho_{\lambda \sigma} \partial_\mu \Gamma^\sigma_{\nu \rho}+\frac{2}{3} \Gamma^\rho_{\lambda \sigma} \Gamma^\sigma_{\mu \tau} \Gamma^\tau_{\nu \rho}  \Big ) \bigg ].  
\label{ctmg}
\end{aligned}
\end{equation}
Observe that as one sets the scalar field to a non-zero vacuum expectation (VEV) value as $ \langle \Phi \rangle=m^{1/2}$, Eq.(\ref{ctmg}) reduces to ordinary TMG in Eq.(\ref{tmg1-21}) with an effective Newton's constant generated from the VEV of scalar field. This cursory analysis actually brings up an interesting point of if this particular limit comes about as the vacuum solution of the conformal TMG or not. Put it other words, whether the conformal symmetry is broken by the vacua or not. Indeed, as is shown in \cite{DengizTekin,Tanhayi:2011aa,Tanhayi:2012nn}, the situation is so: firstly, the local Weyl's symmetry is radiatively broken at two and one loop-levels in $2+1$ and $3+1$ dimensions in flat vacuum in an analogy with the Coleman-Weinberg mechanism \cite{tantekin, coleman}. Secondly, the conformal symmetry is spontaneously broken by the existence of $(A)dS$ vacua as in the Standard model Higgs mechanism. Thus, the masses of the particles are generated by the virtue of legitimate symmetry breaking mechanisms. In this section, we will construct the Weyl-invariant extension of TMG and accordingly show that the similar symmetry-breaking mechanisms create masses of the particles here. In doing so, we will see that the Weyl-gauged TMG reconciles regular TMG with the Topologically Massive Electrodynamics (TME) with a Proca mass term. Recall that TME
\begin{equation}
S_{TME}=\int d^3x \sqrt{-g} \Big [ -\frac{1}{4}F_{\mu\nu}^2+\frac{\mu}{4} \eta^{ \mu \nu \lambda} F_{\mu\nu} A_\lambda \Big ], 
\end{equation}
admits a single spin-1 particle with the mass $ M_{gauge}=\lvert \mu \rvert $, whereas TME plus Proca theory has two massive spin-1 particles with different masses in flat space. (See below for the masses of the gauge field in TME-Proca theory in (A)dS backgrounds). As was studied in \cite{Weylhistory1, DengizTekin,Tanhayi:2011aa,Tanhayi:2012nn, Dengiz1, Dengiz2, Dengiz3}, the rigid global scale symmetry\footnote{Setting $x^\mu \rightarrow \lambda x^\mu $ and $ \Phi \rightarrow \lambda^d \Phi $ where $ d $ is the scaling dimension of the field and $\lambda$ is a constant.} can be promoted to a local Weyl's symmetry in order to attain Poincar\'e invariant models in \emph{arbitrarily} curved backgrounds. This procedure is performed by Weyl-gauging which holds the following relations in $2+1$ dimensions 
\begin{equation}
\begin{aligned}
 &g_{\mu\nu} \rightarrow g^{'}_{\mu\nu}=e^{2 \zeta(x)} g_{\mu\nu}, \hskip 1 cm \Phi \rightarrow \Phi^{'} =e^{-\frac{\zeta(x)}{2}} \Phi, \\
&{\cal{D}}_\mu \Phi =\partial_\mu\Phi -\frac{1}{2} A_\mu \Phi, \hskip 1 cm {\cal{D}}_\mu g_{\alpha \beta}=\partial_\mu g_{\alpha\beta}+ 2 A_\mu g_{\alpha \beta}, \\
&A_\mu \rightarrow A^{'}_\mu = A_\mu - \partial_\mu \zeta(x),
\label{transform1}
\end{aligned}
\end{equation}
where $ {\cal{D}}_\mu $ is gauge covariant derivative. To find the Weyl-invariant version of Eq.(\ref{tmg1-21}), one needs to find Weyl-gauged Christoffel connection:
\begin{equation}
 \widetilde{\Gamma}^\lambda_{\mu\nu}=\frac{1}{2}g^{\lambda\sigma} \Big ({\cal{D}}_\mu g_{\sigma\nu}+{\cal{D}}_\nu g_{\mu\sigma}
-{\cal{D}}_\sigma g_{\mu\nu} \Big),
\label{christofel}
\end{equation}
or equivalently
\begin{equation}
 \widetilde{\Gamma}^\lambda_{\mu \nu} =\Gamma^\lambda_{\mu \nu}+\delta^\lambda_\nu A_\mu+\delta^\lambda_\mu A_\nu-g_{\mu \nu} A^\lambda.
\label{gammm1}
\end{equation}
By using of Eq.(\ref{gammm1}), the Weyl-invariant Riemann tensor can be obtained as follows
 \begin{equation}
 \begin{aligned}
\widetilde{R}^\mu{_{\nu\rho\sigma}} [g,A]
=&R^\mu{_{\nu\rho\sigma}}+\delta^\mu{_\nu}F_{\rho\sigma}+2
\delta^\mu_{[\sigma} \nabla_{\rho]} A_\nu
+2 g_{\nu[\rho}\nabla_{\sigma]} A^\mu 
 +2 A_{[\sigma} \delta_{\rho]}^\mu A_\nu \\&+2 g_{\nu[\sigma}
A_{\rho]} A^\mu +2 g_{\nu[\rho} \delta_{\sigma]}^\mu  A^2,
\label{wiriem}
\end{aligned}
\end{equation}
and followingly the Weyl-invariant Ricci tensor becomes
\begin{equation}
\begin{aligned}
\widetilde{R}_{\nu\sigma} [g,A]&= \widetilde{R}^\mu{_{\nu\mu\sigma}}[g,A] \\
&=R_{\nu\sigma}+F_{\nu\sigma}-(n-2)\Big [\nabla_\sigma A_\nu - A_\nu A_\sigma +A^2  g_{\nu\sigma} \Big ]-g_{\nu\sigma}\nabla \cdot A,
\label{wiricc}
\end{aligned}
\end{equation}
here $\nabla\cdot A\equiv \nabla_\mu A^\mu$. Taking one more contraction from Eq.(\ref{wiricc}), one finally gets Weyl-extended scalar curvature tensor 
\begin{equation}
\widetilde{R}[g,A]=R-2(n-1)\nabla \cdot A-(n-1)(n-2) A^2,
\label{wisclcr}
\end{equation}
which is {\it {not}} invariant under Weyl transformations but rather transforms according to $(\widetilde{R}[g,A])'=e^{-2 \zeta(x)}\widetilde{R}[g,A]$. To get a Weyl-invariant Einstein theory, one can resolve this obstacle by using a compensating scalar field.
 
Consequently, by using all these set-ups and taking care of the contributions coming from the volume parts, one will finally get the Weyl-invariant extension of TMG as follows
\begin{equation}
 \begin{aligned}
S_{WTMG}=&\int d^3x  \sqrt{-g}\,\, \sigma \Phi^2 [R-4 \nabla . A-2 A^2] \\
&+ \frac{k}{2}\int d^3 x \sqrt{-g} \,\, \epsilon^{\lambda \mu \nu} \Big ( \tilde{\Gamma}^\rho_{\lambda \sigma} \partial_\mu \tilde{\Gamma}^\sigma_{\nu \rho}+\frac{2}{3} \tilde{\Gamma}^\rho_{\lambda \sigma} \tilde{\Gamma}^\sigma_{\mu \tau} \tilde{\Gamma}^\tau_{\nu \rho}  \bigg ). 
\label{witmg1} 
\end{aligned}
\end{equation}
Here, denoting the Weyl-gauged gravitational Chern-Simons term as $ CS(\widetilde{\Gamma}) $, one can easily show that 
\begin{equation}
 \begin{aligned}
  CS(\widetilde{\Gamma})=CS(\Gamma)+ \frac{k}{4}\epsilon^{\lambda \mu \nu} A_\lambda F_{\mu\nu}  -\partial_\mu \Big [ \frac{k}{2}\epsilon^{\lambda \mu \nu} g^{\alpha \sigma} (\partial_\lambda g_{\nu \sigma}) A_\alpha- \frac{k}{2} \epsilon^{\lambda \mu \nu} \Gamma^\rho_{\lambda \rho} A_\nu  \Big ],
 \end{aligned}
\end{equation}
with which the explicit form of the Weyl-invariant version of TMG, up to a boundary term, will turn into 
\begin{equation}
 \begin{aligned}
S_{WTMG}=&\int d^3x  \sqrt{-g} \,\,\sigma \Phi^2 [R-4 \nabla . A-2 A^2]+ \frac{k}{4}\int d^3 x \sqrt{-g} \,\, \eta^{\lambda \mu \nu} A_\lambda F_{\mu \nu}\\
&+ \frac{k}{2}\int d^3 x \sqrt{-g} \,\, \eta^{\lambda \mu \nu} \bigg ( \Gamma^\rho_{\lambda \sigma} \partial_\mu \Gamma^\sigma_{\nu \rho}+\frac{2}{3} \Gamma^\rho_{\lambda \sigma} \Gamma^\sigma_{\mu \tau} \Gamma^\tau_{\nu \rho}  \bigg ).  
\label{witmgac} 
\end{aligned}
\end{equation}
Here, it is worth pointing out a generic blunder: that is, the Weyl-invariance and the conformal-invariance are generally confused to each others in literature even if the conformal invariance is a subgroup of Weyl invariance. To see this, let us notice that as the Weyl gauge is assumed to be a pure-gauge in Eq.(\ref{witmgac})
\begin{equation}
 A_\mu=2\partial_\mu \ln \Phi,
\end{equation}
then, up to the scalar potential, the Weyl-invariant TMG turns out to be the renowned conformally-invariant TMG in Eq.(\ref{ctmg}). Note also that the Weyl-invariant version of gravitational Chern-Simons term incorporates the abelian Chern-Simons term and thus unlike the conformal invariant TMG, it is only Weyl-gauging method that yields the abelian Chern-Simons term.

To have a full dynamical theory, one naturally needs to also consider the Weyl-invariant version of scalar and Maxwell-type theories which respectively are 
\begin{equation}
 S_{\Phi}=-\frac{\alpha}{2} \int d^3x \sqrt{-g}\,\,(D_\mu \Phi D^\mu \Phi+\nu \Phi^6) \,,  \hskip .5 cm S_{A^\mu}=-\frac{\beta}{4}\int d^3x \sqrt{-g}\,\, \Phi^{-2} F_{\mu\nu}F^{\mu \nu}.
\end{equation}
where $\alpha, \nu$ and $\beta$ are dimensionless parameters that are necessary in the Weyl invariance. Observe that the Weyl-invariant scalar potential is also taken into account in order to get the cosmological TMG in the vacua. Note also that the Maxwell-type action is achieved to be Weyl-invariant with the help of a specifically tunned compensating Weyl scalar field. Here, the dimensions of the fundamental fields can be given in terms of mass-dimensions as follows 
\begin{equation}
 [g_{\mu\nu}]=M^0=1 \hskip 1cm ; \hskip 1cm [\Phi]=M^{1/2} \hskip 1cm ; \hskip 1cm [A_\mu]=M.
\end{equation}
Thus, collecting all the stuff, the full Lagrangian density of the Weyl extension of TMG will read
\begin{equation}
 \begin{aligned}
{\cal L}_{WTMG}&=\sigma \Phi^2 [R-4 \nabla . A-2 A^2]+ \frac{k}{2} \epsilon^{\lambda \mu \nu} \bigg ( \Gamma^\rho_{\lambda \sigma} \partial_\mu \Gamma^\sigma_{\nu \rho}+\frac{2}{3} \Gamma^\rho_{\lambda \sigma} \Gamma^\sigma_{\mu \tau} \Gamma^\tau_{\nu \rho}  \bigg )\\
&+ \frac{k}{4} \epsilon^{\lambda \mu \nu} A_\lambda F_{\mu\nu} -\frac{\alpha}{2}(D_\mu \Phi D^\mu \Phi+\nu \Phi^6)-\frac{\beta}{4}\Phi^{-2} F_{\mu\nu}F^{\mu \nu}.
\label{witmg2} 
\end{aligned}
\end{equation}
To study the existing symmetry mechanism and other fundamental features of the model, one naturally needs the arising field equations. Therefore, by skipping the detailed calculations, let us first notice that the variation of Eq.(\ref{witmg2}) with respect to $g_{\mu\nu}$ will become
 \begin{equation}
\begin{aligned}
 & \sigma \Phi^2 G_{\mu \nu}+(\sigma-\frac{\alpha}{8}) g_{\mu \nu} \Box \Phi^2-(\sigma-\frac{\alpha}{4})\nabla_\mu \nabla_\nu \Phi^2-(4\sigma+\frac{\alpha}{4})\Phi^2 \nabla_\mu A_\nu \\
&+(2\sigma+\frac{\alpha}{8})g_{\mu \nu} \Phi^2\nabla . A -(2\sigma+\frac{\alpha}{8})\Phi^2 A_\mu A_\nu+(\sigma+\frac{\alpha}{16})g_{\mu \nu} \Phi^2 A^2 +\frac{\alpha}{4} g_{\mu \nu}(\nabla_\alpha \Phi)^2\\
&+\frac{\alpha \nu}{4} g_{\mu \nu} \Phi^6-\frac{\alpha}{2} (\nabla_\mu \Phi)(\nabla_\nu \Phi) 
+\frac{\beta}{8} g_{\mu \nu}\Phi^{-2} F^2_{\alpha \beta}+\frac{\beta}{2} \Phi^{-2} F_{\mu \alpha} F^\alpha{_\nu} +k C_{\mu \nu}=0.
\label{jkhl1}
\end{aligned}
\end{equation}
Subsequently, the variation with respect to $A_{\mu}$ will yield  
\begin{equation}
 (4 \sigma+\frac{\alpha}{4}) \nabla_\mu \Phi^2-(4 \sigma+\frac{\alpha}{4}) \Phi^2 A_\mu+\frac{k}{2} \eta_\mu{^{\lambda \nu}} \nabla_\lambda A_\nu - \beta \nabla^\nu (\Phi^{-2} F_{\mu\nu})=0. 
\end{equation}
Finally, the variation with respect to scalar field $ \Phi $ will read
\begin{equation}
 2 \sigma \Phi \Big [R-4 \nabla .A-2 A^2 \Big]+\alpha \Big [\Box \Phi-\frac{1}{2} \Phi \nabla .A-\frac{1}{4} \Phi A^2-3\nu \Phi^5 \Big]+\frac{\beta}{2} \Phi^{-3} F^2_{\mu \nu}=0.
\end{equation}

Let us now consider the symmetric and non-symmetric (broken phase) vacua behaviours of the theory. First of all, in the symmetric vacuum, $  \langle \Phi \rangle=0 $, Eq.(\ref{witmg2}) reduces to a pure gravitational Chern-Simons term without a propagating DOF. The theory is diffeomorphism and conformally invariant and the Weyl gauge field must be vanish because of the Maxwell term. As for the the broken phase with  $  \langle \Phi \rangle=m^{1/2} $, Eq.(\ref{witmg2}) boils down to
\begin{equation}
 \begin{aligned}
{\cal L}_{WTMG}&= \sigma m R-\frac{\alpha \nu}{2}m^3+ \frac{k}{2} \eta^{\lambda \mu \nu} \bigg ( \Gamma^\rho_{\lambda \sigma} \partial_\mu \Gamma^\sigma_{\nu \rho}+\frac{2}{3} \Gamma^\rho_{\lambda \sigma} \Gamma^\sigma_{\mu \tau} \Gamma^\tau_{\nu \rho}  \bigg ) \\
&-\frac{\beta}{4m} F_{\mu\nu}F^{\mu \nu}+ \frac{k}{4} \eta^{\lambda \mu \nu} A_\lambda F_{\mu\nu}-\frac{m}{2} \Big (4 \sigma +\frac{\alpha}{4} \Big ) A^2,
\label{vwitmg} 
\end{aligned}
\end{equation}
which clearly shows that TMG with a cosmological constant is generically coupled to TME-Proca theory. The theory describes a single massive spin-$2$ graviton and two massive spin-$1$ helicity modes around its flat and $(A)dS$ vacua. From the earlier works \cite{Carlip:2008jk, Gurses:2011fv}, the mass of spin-2 excitation in the $(A)dS$ background will be evaluated as follows
\begin{equation}
 M^2_{graviton}=\frac{m^2 \sigma^2}{k^2}+\Lambda \hskip 0.5cm \mbox{where} \hskip 0.5cm \Lambda=\frac{\alpha \nu m^2}{4 \sigma}.
\label{grav_mass}
\end{equation}
On the other hand, the spin-$1$ helicity modes propagated in the TME-Proca theory have the same mass \cite{tantekin, Dunne:1998qy} 
\begin{equation}
 M^\pm_{gauge}(\Lambda=0)= \frac{1}{2} \bigg \{\sqrt{\frac{k^2m^2}{\beta^2}+\frac{4m}{\beta} \Big (4 \sigma+\frac{\alpha}{4}\Big)}\pm \frac{m \lvert k \rvert}{\beta}  \bigg \},
\label{mass_gauge_3}
\end{equation}
about flat backgrounds. Note that in the case of vanishing of the Proca mass term, (that is, $ 16 \sigma+\alpha=0 $), there is a single massive helicity-1 mode with $ M^\pm_{gauge}= \frac{m \lvert k \rvert}{\beta} $ such that one of the propagating DOF inevitably becomes a pure gauge. As for the $(A)dS$ vacua, since it is a bit subtle, we will find the particle spectrum of the TME-Proca theory in the next section. But, for the sake of completeness, here let us just quote result
\begin{equation}
M^2_{gauge_\pm}(\Lambda\ne0)=\frac{15 \Lambda}{4}+M^2_{gauge_\pm}(\Lambda = 0).
\label{gauge_kutle}
\end{equation}
Now that we have given the masses, we  need to check the physical consistency of the particle spectrum with the constraints which is coming from tree-level unitarity of the theory that is  absence of tachyons and ghosts around a constant curvature backgrounds. First of all, for flat spaces ($\Lambda=0$), the theory is unitary as long as
\begin{equation}
 \sigma <0, \hskip 1cm \beta>0 \hskip 0.5cm \mbox{and} \hskip 0.5cm \alpha+\frac{k^2 m}{\beta} \ge -16 \sigma.
\end{equation}
On the other side, for AdS spaces ($ \Lambda<0 $), there are more possibilities. As for spin-2 graviton, mass square of the particles has to obey Breitenlohner-Freedmann (BF) bound \cite{bf,waldron}, $ M^2_{graviton} \ge \Lambda $, which it does in our case. For the gauge field, $ M^2 _{gauge} (\Lambda) \ge 0$ condition must be satisfied for non-tachyonic excitations, this brings an constraint on cosmological constant $ \Lambda $ as follows
\begin{equation}
 \Lambda \ge -\frac{4}{15} M^2_{gauge}(\Lambda=0).
\end{equation}
Finally, for dS spaces ($ \Lambda>0 $),  Eq.(\ref{grav_mass}) must satisfy Higuchi bound \cite{higuchi} $ M^2_{graviton}\ge\Lambda>0 $. However, this condition does not impose any extra constraint except the existence of a dS vacuum forces to sign of Einstein term to be negative as $ \sigma>0 $ (assuming $\alpha>0, \nu>0$).

\subsection{Unitarity of Weyl-invariant Topologically Massive Gravity}
In the previous section, we have found the particle spectrum of the Weyl-invariant TMG by freezing the scalar field to the vacuum value. One could ask that this method may not be a conclusive way for studying the unitarity and the stability of the theory. To search this issue at least at tree-level, we will study perturbative unitarity of the model by expanding the action up to the second order in fluctuations which will ultimately provide the basic oscillators around the vacua \cite{Tanhayi:2011aa,Tanhayi:2012nn,Gullu:2010em}. To do so, let us now assume that the fundamental fields fluctuate about their vacuum values as follows 
\begin{equation}
\begin{aligned}
 &\Phi\equiv \sqrt{m}+\tau \Phi^L,\hskip 0.8cm g_{\mu\nu} \equiv \bar{g}_{\mu\nu}+\tau h_{\mu\nu}, \hskip 0.8cm A_\mu \equiv \tau A^L_\mu, \\
& g^{\mu\nu}=\bar{g}^{\mu\nu}-\tau h^{\mu\nu}+\tau^2 h^{\mu\rho}h^\nu_\rho, \hskip 0.8cm \sqrt{-g}=\sqrt{-\bar{g}}\,[1+\frac{\tau}{2} h+\frac{\tau^2}{8}(h^2-2h^2_{\mu\nu})], \\
&\nabla_\mu A_\alpha =\tau \bar{\nabla}_\mu A^L_\alpha-\tau^2 (\Gamma^\gamma_{\mu\alpha})_L A^L_\gamma-\tau^2 h^\gamma_\beta (\Gamma^\beta_{\mu\alpha})_L A^L_\gamma, 
\label{fluct}
\end{aligned}
\end{equation}
where a small dimensionless parameter $ \tau $ is introduced to follow the expansion. Using the specified fluctuations of the field in Eq.(\ref{fluct}) as well as the vacuum field equation $ \Lambda=\frac{\alpha \nu m^2}{4 \sigma} $, one will finally get the quadratic order expansion of Weyl-invariant TMG as follows
\begin{equation}
\begin{aligned}
 {I}_{WTMG}^{(2)}=\int d^3x\sqrt{-\bar{g}}\bigg \{& -\frac{\alpha}{2}(\partial_\mu \Phi^L)^2 -6 \alpha \nu m^2 \Phi^2_L-\sqrt{m} \Big ( 8 \sigma +\frac{\alpha}{2} \Big ) \Phi^L \bar{\nabla}\cdot A^L \\
& -\frac{\beta}{4m}(F^L_{\mu\nu})^2+\frac{k}{4} \eta^{\lambda \mu \nu} A^L_\lambda F^L_{\mu\nu}- m \Big ( 2 \sigma +\frac{\alpha}{8} \Big ) A^2_L \\
&-\frac{\sigma m}{2} h^{\mu \nu} {\cal G}^{L}_{\mu \nu}+\frac{k}{2}h^{\mu \nu} C^{L}_{\mu \nu} + 2 \sigma \sqrt{m} \Phi^L R^L \bigg \},
\label{linearform}
\end{aligned}
\end{equation}
where we have discarded the irrelevant boundary terms. Here, the $2+1$ dimensional linearized curvature tensors are \cite{deser_tekin_en}
\begin{equation}
\begin{aligned}
 C_{L}^{\mu \nu}&=\frac{\eta^{\mu \alpha\beta}}{\sqrt{-\bar{g}}} \bar{g}_{\beta \sigma} \nabla_\alpha \Big ( R^{\sigma\nu}_L-2 \Lambda h^{\sigma \nu}- \frac{1}{4} \bar{g}_{\beta \nu}  R_L \Big ),\hskip 0.3cm {\cal G}_{\mu\nu}^L=R_{\mu\nu}^L-\frac{1}{2}\bar{g}_{\mu\nu}R^L-2\Lambda h_{\mu\nu}, \\
R^{L}_{\nu \sigma}&=\frac{1}{2} \Big (\bar{\nabla}_\mu \bar{\nabla}_\sigma h^\mu_\nu+\bar{\nabla}_\mu \bar{\nabla}_\nu
 h^\mu_\sigma- \bar{\Box}h_{\sigma \nu}-\bar{\nabla}_\sigma \bar{\nabla}_\nu h \Big),  R^{L}=\bar{\nabla}_\mu \bar{\nabla}_\nu h^{\mu \nu}-\bar{\Box}h-2 \Lambda h.
\end{aligned}
\end{equation}
Note that Eq.(\ref{linearform}) still involves coupled terms which have to be decoupled from each others in order to get the particle spectrum. For this purpose, let us first consider the following redefinition of the fluctuations 
\begin{equation}
h_{\mu\nu} \equiv\widetilde{h}_{\mu\nu}-\frac{4}{\sqrt{m}}\bar{g}_{\mu\nu}\Phi_L \hskip 1cm \mbox{and} \hskip 1cm A^L \equiv \widetilde{A}_\mu+\frac{2}{\sqrt{m}} \partial_\mu \Phi_L ,
\label{rede}
\end{equation}
and then plug them into Eq.(\ref{linearform}). In doing so, one will get 
\begin{equation}
\begin{aligned}
 {I}_{WTMG}^{(2)}=\int d^3x\sqrt{-\bar{g}}\bigg \{& -\frac{\beta}{4m}(\widetilde{F}^L_{\mu\nu})^2+\frac{k}{4} \eta^{\lambda \mu \nu} \widetilde{A}^L_\lambda \widetilde{F}^L_{\mu\nu} - m \Big ( 2 \sigma +\frac{\alpha}{8} \Big ) \widetilde{A}^2_L \\
& -\frac{\sigma m}{2} \widetilde{h}^{\mu \nu} \Big [ \widetilde{{\cal G}^{L}}_{\mu \nu}- \frac{k}{\sigma m} \widetilde{C}^{L}_{\mu \nu} \Big ] \bigg \}, 
\label{linearform1}
\end{aligned}
\end{equation}
 where the relevant redefined tensors are
\begin{equation}
 \begin{aligned}
(R_{\mu\nu})_L=&(\widetilde{R}_{\mu\nu})_L+\frac{2}{\sqrt{m}}(\bar{\nabla}_\mu\partial_\nu\Phi_L+\bar{g}_{\mu\nu}\bar{\Box}\Phi_L),
\hskip 0.1cm  R_L=\widetilde{R}_L+\frac{8}{\sqrt{m}}(\bar{\Box}\Phi_L+3\Lambda\Phi_L),\\
  {\cal G}_{\mu\nu}^L=&\widetilde{{\cal
G}}^L_{\mu\nu}+\frac{2}{\sqrt{m}}\Big(\bar{\nabla}_\mu\partial_\nu\Phi_L-\bar{g}_{\mu\nu}\bar{\Box}\Phi_L-2\Lambda
\bar{g}_{\mu\nu}\Phi_L\Big), \hskip 0.7cm \widetilde{h}^{\mu \nu} \widetilde{C}^{L}_{\mu \nu}=h^{\mu \nu} C^{L}_{\mu \nu},\\
h^{\mu\nu}{\cal G}^L_{\mu\nu}=&\widetilde{h}^{\mu\nu}{\cal
\widetilde{G}}^L_{\mu\nu}+\frac{4}{\sqrt{m}}\widetilde{R}_L\Phi_L+\frac{16}{m}\Phi_L\bar{\Box}\Phi_L+\frac{48}{m}\Lambda\Phi_L^2.
 \end{aligned}
\end{equation}
As it was studied in the previous part, the first line of the Eq.(\ref{linearform1}) is TME-Proca theory which propagates unitary massive spin-1 DOF with the mass in Eq.(\ref{gauge_kutle}). Notice that the second line is the action for the parity-noninvariant TMG theory which has single unitary massive spin-2 graviton with the mass in Eq.(\ref{grav_mass}). To read the mass of the spin-1 particle, one needs a further study. Therefore, let us proceed accordingly.  

\subsection{Topologically Massive Electrodynamics-Proca theory  in $ (A)dS$}

This section is devoted to derive the masses of the gauge field given by Eq.(\ref{gauge_kutle}). For this purpose, let us consider the field equation of the TME with a Proca mass term about arbitrary background 
\begin{equation}
 a \nabla_\nu F^{\nu \mu}+b \eta^{\lambda \nu \mu} F_{\lambda \nu}+c A^\mu=0,
\label{mcspe1}
\end{equation}
where $a,b$ and $c$ are
\begin{equation}
 a=\frac{\beta}{m},\hskip .5 cm b=\frac{k}{2},\hskip .5 cm c=-\chi=-m (4\sigma+\alpha/4). 
\end{equation}
To analyze particle spectrum of the theory, one needs to transform Eq.(\ref{mcspe1}) into a source-free wave type equation. For that reason, let us first take the divergence of Eq.(\ref{mcspe1}). In doing so, one will get
\begin{equation}
c\,\nabla_\mu A^\mu=0.
\end{equation}
Thus, for $c \ne 0$, the Lorenz gauge is dictated by the model and thus one of the DOF drops out. To go further, let us now define
\begin{equation}
 \widetilde{F}^\mu=\frac{1}{2} \eta^{\mu \lambda \nu } F_{\lambda \nu} , \hskip 1cm {\cal B}^\mu=\eta^{ \mu \lambda \nu} \nabla_\lambda \widetilde{F}_\nu,
\end{equation}
 with $ {\cal B}^\mu=\nabla_\alpha F^{\alpha \mu} $. Accordingly, exerting the operator $\eta^{\alpha \nu \mu}  \nabla_\nu $ to Eq.(\ref{mcspe1}) yields
\begin{equation}
 a \eta^{\alpha \nu \mu} \nabla_\nu {\cal B}_\mu+2b\eta^{\alpha \nu \mu}\nabla_\nu \widetilde{F}_\mu+c\eta^{\alpha \nu \mu}\nabla_\nu A_\mu=0.
\label{mcspe3}
\end{equation}
By using of 
\begin{equation}
 \eta^{\alpha \nu \mu} \nabla_\nu {\cal B}_\mu=\Box \widetilde{F}^\alpha-R^\alpha{_\beta} \widetilde{F}^\beta,
\label{mcspe5}
\end{equation}
one can obtain
\begin{equation}
 {\cal B}^\alpha=-\frac{1}{2 b} \Big[a (\Box \widetilde{F}^\alpha-R^\alpha{_\beta} \widetilde{F}^\beta)+ c \widetilde{F}^\alpha \Big].
\label{bdenklemi}
\end{equation}
Performing the operation  $\eta^{\sigma \lambda}{_\alpha}  \nabla_\lambda $ to  Eq.(\ref{mcspe3}) as well as using $\nabla_\alpha {\cal B}^\alpha=0$, which follows from the Bianchi identity, 
one gets
\begin{equation}
a \Box {\cal B}^\sigma-a R^\sigma{_\alpha}{\cal B}^\alpha+2 b \eta^{\sigma \lambda \alpha} \nabla_\lambda {\cal B}_\alpha+c {\cal B}^\sigma=0.
\label{mcspe7}
\end{equation}
Thus, substituting Eq.(\ref{bdenklemi}) into Eq.(\ref{mcspe7}) ends up with a fourth-order equation for TME-Proca theory in a arbitrary background 
\begin{equation}
\begin{aligned}
 &\bigg [-\frac{a^2}{2b} \delta^\sigma_\beta \Box^2+ \Big (\frac{a^2}{b} R^\sigma{_\beta}-\frac{ac}{b} \delta^\sigma_\beta+2b \delta^\sigma_\beta \Big)\Box \\
&+\frac{a^2}{2b} (\Box  R^\sigma{_\beta} ) 
- \frac{a^2}{2b}R^\sigma{_\alpha} R^\alpha{_\beta}+\Big (\frac{ac}{b}-2b \Big) R^\sigma{_\beta}-\frac{c^2}{2b}\delta^\sigma_\beta \bigg] \widetilde{F}^\beta=0,
\label{mcspe8}
\end{aligned}
 \end{equation}
which with $ R^\alpha{_\beta}=2 \Lambda \delta^\alpha{_\beta} $, turns out to be 
\begin{equation}
 \bigg [-\frac{a^2}{2b} \Box^2+\Big (\frac{2 \Lambda a^2}{b}-\frac{ac}{b}+2b \Big) \Box+\Big(-\frac{2a^2 \Lambda^2}{b}+\frac{2ac\Lambda}{b}-4b\Lambda-\frac{c^2}{2b} \Big) \bigg ] \widetilde{F}^\sigma=0,
\label{mcspe9}
\end{equation}
in $(A)dS$ spacetimes. Thereupon, by fixing $\Lambda =0$, the flat space limit will become 
\begin{equation}
 \bigg [ -\frac{a^2}{2b} \partial^4+ \Big(2b-\frac{ac}{b} \Big) \partial^2-\frac{c^2}{2b} \bigg]\widetilde{F}^\beta=0,
\end{equation}
whose the masses are 
\begin{equation}
 M^\pm_{gauge}(\Lambda=0)= \frac{1}{2} \bigg \{\sqrt{\frac{k^2m^2}{\beta^2}+\frac{4m}{\beta} \Big (4 \sigma+\frac{\alpha}{4}\Big)}\pm \frac{m \lvert k \rvert}{\beta}  \bigg \},
\end{equation}
where we have made use of $ a=\frac{\beta}{m};b=\frac{k}{2};c=-\chi=-m (4\sigma+\alpha/4) $.

For $(A)dS$, the Eq.(\ref{mcspe9}) can alternatively be recast as follows
\begin{equation}
 \frac{\beta^2}{m^2}(\Box-\xi^2_+)(\Box-\xi^2_-) \widetilde{F}^\sigma=0,
\label{mcspe11}
\end{equation}
in which one has
\begin{equation}
 \xi^2_\pm \equiv 2 \Lambda +M^2_{gauge_\pm}(\Lambda=0).
\end{equation}
Observe that this parity-non-invariant gauge theory has \emph{two} propagating DOF with distinct masses. To find the masses of the fluctuations, one should take care of the tilted null-cone propagation for massless spin-1 field in $2+1 $-dimensional $AdS$ space \cite{Deser:1983mm}
\begin{equation}
 \Big (\Box +\frac{7}{4} \Lambda \Big) A^\mu=0,
\end{equation}
where we have used the Lorenz gauge $ \nabla_\mu A^\mu=0 $. Hence from Eq.(\ref{mcspe11}), masses for helicity $ \pm1 $ components of the gauge field can be easily obtained to be in Eq.(\ref{gauge_kutle}).\\

\section[Scattering in topologically massive gravity, chiral gravity, and the corresponding anyon-anyon potential energy]{Scattering in topologically massive gravity, chiral gravity, and the corresponding anyon-anyon potential energy\footnote{The results of this chapter are published in \cite{kilicarslan2}.}}
\label{chp:introduction}

As is well-known, Einstein's gravity in $2+1$ dimensions  does not possess only any local degrees of freedom (DOF) but also any black holes and gravitational waves solution around its flat backgrounds. As for the $(A)dS$ spaces, the theory however describes black hole solutions \cite{BTZ} which naturally lead to the additional microscopic DOF leading to the celebrated Bekenstein-Hawking entropy. Recall that the statistical mechanics roots the macroscopic quantities such as entropy, temperature etc. to the microscopic states. Therefore, the emergent of these extra DOF has led experts to contemplate on a well-behaved $2+1$-dimensional quantum gravity theory.  Here, the main question is that what sort of the modification in the pure theory will provide such a complete quantum model. In this respect, there is no doubt the renowned Topologically Massive Gravity (TMG) \cite{DJT} is the only viable candidate to fulfil this pivotal job even if there have been proposed many alternative models hitherto. However, it is known that the model comprises shortcomings in AdS/CFT perspective. That is, it has the bulk/boundary unitary clash. Fortunately, this controversial issue has been resolved in the  chiral limit of TMG \cite{Chiral}. Thus, due to this fact,  Chiral gravity has a notable potential to supply a complete quantum gravity theory in the AdS/CFT paradigm. 

In the light of the above discussion, one can ask the following crucial question: how can one find the tree level scattering amplitude and the associated  Newtonian gravitational potential energy between two covariantly conserved sources in cosmological TMG integrated with a Fierz-Pauli mass term and its chiral limit? Here, it is worth mentioning that the Fierz-Pauli is assumed in order to obviate the arising the zero-modes during calculation of retarded Green functions. For this purpose, we will calculate the corresponding scattering amplitude of the theory. On the other hand, one can realize that, in the existence of Chern-Simons term, topological mass $\mu$ induces an additional spin with $\frac{\kappa m}{\mu}$, where $\kappa$ is $3D$ Newton's constant and $m$ is the mass of the particle. Due to gravitational Chern-Simons term, particles behave like gravitational anyons \cite{DeserAnyon} which are exotic particles with having different statistics. These gravitational anyons show the same behaviour with their electromagnetic counterpart where Abelian Chern-Simons term changes the statistics of charged particles and turns them into an anyon \cite{Wilczek}. In this chapter, we will try to extend the gravitational anyons discussion and construct an analogy between them and their Abelian counterparts.

\subsection{Cosmological TMG with a Fierz-Pauli Term in (A)dS Backgrounds}

The action of TMG with a Fierz-Pauli mass term is given by 
 \begin{equation}
 \begin{aligned}
 {I}=\int d^3x \sqrt{-g} \, \Big \{& \frac{1}{ \kappa} ( R -2 \Lambda)-\frac{m^2}{4 \kappa}(h^2_{\mu\nu}-h^2)\\&+\frac{1}{2 \mu} \, \eta^{\mu \nu \alpha} \Gamma^\beta{_{\mu \sigma}} \Big (\partial_\nu \Gamma^\sigma{_{\alpha \beta}}+\frac{2}{3} \Gamma^\sigma{_{\nu \lambda}}  \Gamma^\lambda{_{\alpha \beta}} \Big )+ {\cal L}_{matter} \Bigg \},
\label{wtics1}
 \end{aligned}
\end{equation}
where $ \kappa $ is the usual $3D$ Newton's constant and $ \mu $ is dimensionless coupling constant and the tensor $ \eta^{\mu \nu \alpha}$ is a rank-3 tensor described in terms of the Levi-Civita symbol as $\epsilon^{\mu \nu \alpha}/\sqrt{-g}$. Generically, theory has three modes about its flat and (A)dS vacua. 
Observe that when Fierz-Pauli mass term vanishes, which is the TMG  theory, there is a single massive spin-2 graviton and taking the $ \mu \to \infty $ limit, which yields the Fierz-Pauli massive gravity theory, there are two massive spin-2 excitations.
In the full theory, in flat space, the unitarity analysis and particle spectrum of the theory were given in \cite{DeserTekinMode}. In this part, we extend this result to the maximally symmetric curved backgrounds.

To be able to obtain the fluctuations propagated around the (A)dS vacua (see for details Appendix \ref{chp:appendixb}), let us first recall that the variation of Eq.(\ref{wtics1}) yields the following field equation 
\begin{equation}
 \frac{1}{\kappa} (R_{\mu \nu}-\frac{1}{2} g_{\mu \nu} R+ \Lambda g_{\mu \nu})+\frac{1}{\mu} C_{\mu \nu}+\frac{m^2}{2 \kappa}(h_{\mu\nu}-g_{\mu\nu}h) =\tau_{\mu \nu}.
\label{tmgfeq}
 \end{equation}
Here, $ C_{\mu \nu} $ is the well-known Cotton-York tensor that is described as follows 
\begin{equation}
 C^{\mu \nu}=\eta^{\mu \alpha \beta} \nabla_\alpha \Big ( R^\nu{_\beta}-\frac{1}{4} \delta^\nu{_\beta} R \Big ).
 \label{cotton3}
\end{equation}
Remember that the Cotton-York tensor is symmetric, divergence free and traceless and is also known as the $2+1$ dimensional cousin of Weyl tensor. For our main aim, let us now recall that Eq.(\ref{cotton3}) can alternatively be recast in the following explicitly symmetric version
\begin{equation}
 C^{\mu \nu}=\frac{1}{2} \eta^{\mu \alpha \beta} \nabla_\alpha G^\nu{_\beta}+\frac{1}{2} \eta^{\nu \alpha \beta} \nabla_\alpha G^\mu{_\beta},
 \label{coteins}
\end{equation}
with which the linearization of Eq.(\ref{tmgfeq}) about a generic background $ g_{\mu\nu}=\bar{g}_{\mu\nu}+h_{\mu\nu} $ reads
\begin{equation}
 \frac{1}{\kappa} {\cal G}^L_{\mu \nu}+\frac{1}{2 \mu} \eta_{\mu \alpha \beta} \bar{\nabla}^\alpha {\cal G}^L_{\nu}{^\beta}+\frac{1}{2 \mu} \eta_{\nu \alpha \beta} \bar{\nabla}^\alpha {\cal G}^L_{\mu}{^\beta}+\frac{m^2}{2 \kappa}(h_{\mu\nu}-\bar{g}_{\mu\nu}h)=T_{\mu \nu}.
\label{linertmg}
\end{equation}
Notice that the background covariantly conserved  energy momentum tensor $ T_{\mu\nu}$ is perturbatively defined as $ T_{\mu\nu}=\tau_{\mu\nu}+\Theta(h^2,h^3,...) $. Moreover, the relevant linearized curvature tensors are given in Eq.(\ref{curvaturet1}).

Before going into further details, let us now emphasise  an ambiguity associated to the sign of Einstein term and unitarity: recall that, in the usual TMG, although the theory is tachyon-free, the ghost-freedom about the flat vacuum requires the sign of Einstein term to be opposite \cite{DJT}. However, with Fierz-pauli mass term, the unitarity of theory compels the sign to be same as the usual one. Otherwise, both of the excitations inevitably turns into imaginary \cite{Deser:2002jk} which is undesired issue.

 Subsequently,  let us notice that one needs to somehow convert Eq.(\ref{linertmg}) into a Poison-type wave equation of the form
 \begin{equation}
  (\bar{\square}-2 \Lambda-m^2_1) (\bar{\square}-2\Lambda-m^2_2) (\bar{\square}-2 \Lambda-m^2_3)h_{\mu\nu}=\tilde{T}_{\mu\nu},
  \label{waveeqn}
 \end{equation}
in order to find the excitations. Note that as the right hand side of Eq.(\ref{waveeqn}) vanishes, the terms $ m_i $ become the masses of the particles. To get the accurate masses, one needs to keep in mind that unlike the flat case (that is, $ \bar{\square}h_{\mu\nu}=0$), the null-cone propagation for a massless spin-2 field in $2+1$ dimensional (A)dS backgrounds is described by  $ (\bar{\square}-2 \Lambda)h_{\mu\nu}=0$. 

To rewrite all the curvature tensors in terms of the source terms, one needs to firstly find the divergence of Eq.(\ref{linertmg}). In doing so, one will get
 \begin{equation}
  m^2( \bar{\nabla}^\mu h_{\mu\nu}-\bar{\nabla}_\nu )h=0,
  \label{condtion1}
\end{equation}
that gives $R^L=-2 \Lambda h$. Followingly, by using of Eq.(\ref{condtion1}) as well as the trace of Eq.(\ref{linertmg}), one eventually obtains 
 \begin{equation}
 \quad  h=\frac{\kappa}{\Lambda-m^2} T, \quad {\cal G}^L\equiv \bar{g}^{\mu\nu} {\cal G}^L_{\mu\nu} =\frac{\Lambda \kappa}{\Lambda-m^2} T. 
  \label{scaleinsstre}
 \end{equation} 
 
Now that we have rewritten the curvature tensors in terms of sources, we can proceed further in order to convert the genuine equation into the desired form. For this purpose, by performing the operation $ \eta^{\mu \sigma \rho} \bar{\nabla}_\sigma $ to Eq.(\ref{linertmg}), one arrives at
\begin{equation}
\begin{aligned}
  \frac{1}{\kappa} \eta^{\mu \sigma \rho} \bar{\nabla}_\sigma {\cal G}^{L}_{\mu\nu}-&\frac{1}{\mu} \bar{\square}{\cal G}^{L}_\nu{^\rho}+\frac{3 \Lambda}{\mu}{\cal G}^{L}_\nu{^\rho} +\frac{m^2}{2 \kappa} \eta^{\mu \sigma \rho} \bar{\nabla}_\sigma (h_{\mu\nu}-\bar{g}_{\mu\nu}h)\\
  &= \eta^{\mu \sigma \rho} \bar{\nabla}_\sigma T_{\mu\nu}+\frac{\Lambda}{\mu} \delta^\rho{_\nu} {\cal G}^{L}
  +\frac{1}{2 \mu} \bar{\nabla}_\nu \bar{\nabla}^\rho {\cal G}^{L}-\frac{1}{2 \mu} \delta^\rho{_\nu} \bar{\square} {\cal G}^{L},
\label{ghdgh}
  \end{aligned}
  \end{equation}
which is traceless. Here, the following identity is used
\begin{equation}
 \begin{aligned}
\eta^{\mu \sigma \rho} \eta_{\nu \alpha \beta} = \bigg [& -\delta^\mu{_\nu} \Big ( \delta^\sigma{_\alpha} \delta^\rho{_\beta}-\delta^\sigma{_\beta} \delta^\rho{_\alpha} \Big ) 
+\delta^\mu{_\alpha} \Big ( \delta^\sigma{_\nu} \delta^\rho{_\beta}-\delta^\sigma{_\beta} \delta^\rho{_\nu} \Big ) \\
 & -\delta^\mu{_\beta} \Big ( \delta^\sigma{_\nu} \delta^\rho{_\alpha}-\delta^\sigma{_\alpha} \delta^\rho{_\nu} \Big ) \bigg ],
 \label{identity}
 \end{aligned}
\end{equation}
during the derivation of Eq.(\ref{ghdgh}). Moreover, with the help of Eq.(\ref{coteins}) and Eq.(\ref{tmgfeq}), one can also recast the first term as follows  
\begin{equation}
\eta_{ \rho\sigma\mu } \bar{\nabla}^\sigma  {\cal G}^{L}_\nu {^\mu}=\mu T_{\rho\nu}- \eta_{ \rho\sigma\nu } \bar{\nabla}^\sigma R^L-\frac{\mu}{\kappa}{\cal G}^{L}_{\rho\nu}-\frac{\mu m^2}{2\kappa}(h_{\rho\nu}-\bar{g}_{\rho\nu}h),
\label{fteq}
\end{equation}
with which Eq.(\ref{ghdgh}) turns into
 \begin{equation}
 \begin{aligned}
 &\Big (\bar{\square}-3 \Lambda-\frac{\mu^2}{\kappa^2} \Big ) {\cal G}^{L}{_{\rho \nu}}-\frac{\mu^2 m^2}{2 \kappa^2} (h_{\rho\nu}-\bar{g}_{\rho\nu}h)-\frac{\mu m^2}{2 \kappa} \eta^{\mu \sigma}{_\rho} \bar{\nabla}_\sigma (h_{\mu\nu}-\bar{g}_{\mu\nu}h)\\
 &=\frac{\mu}{2} \, \eta_\rho{^{\mu\sigma}} \bar{\nabla}_\mu T_{\sigma\nu}+ \frac{\mu}{2} \, \eta_\nu{^{\mu\sigma}} \bar{\nabla}_\mu T_{\sigma\rho}-\frac{\mu^2}{\kappa} T_{\rho \nu} -\Lambda \bar{g}_{\rho \nu} {\cal G}^{L}-\frac{1}{2} \bar{\nabla}_\nu \bar{\nabla}_\rho {\cal G}^{L}+\frac{1}{2} \bar{g}_{\rho\nu} \bar{\square} {\cal G}^{L}.
 \label{klmapt2}
 \end{aligned}
\end{equation}
Observe that the term $ \eta_{ \rho\sigma\nu } \bar{\nabla}^\sigma R^L$ drops out due to existing symmetry property. To convert Eq.(\ref{klmapt2})
into a wave-type equation, one needs to also rewrite the Fierz-Pauli term in terms of $ {\cal G}^L_{\mu\nu} $ and its contractions. For this purpose, let us define
\begin{equation}
 \eta_{\mu\alpha\beta} \bar{\nabla}^\alpha {\cal G}^L_{\nu}{^\beta}={\cal B}_{\mu\nu},
 \label{redfphg}
\end{equation}
which satisfies $\bar{g}^{\mu\nu}{\cal B}_{\mu\nu}={\cal B}=0$ and $ \bar{\nabla}^\mu {\cal B}_{\mu\nu}=0 $.
Then, by substituting Eq.(\ref{redfphg}) into Eq.(\ref{klmapt2}) and applying  $ \eta_{\rho \alpha \beta} \bar{\nabla}^\alpha $, one ultimately gets
\begin{equation}
\begin{aligned}
& -\frac{1}{\kappa} \eta_{\rho \alpha \beta}\bar{\nabla}^\alpha {\cal B}^\rho{_\nu}+\frac{1}{2 \mu} \eta_{\rho \alpha \beta} \eta^{\mu \sigma \rho}\bar{\nabla}^\alpha\bar{\nabla}_\sigma {\cal B}_{\mu \nu}+\frac{1}{2 \mu} \eta_{\rho \alpha \beta} \eta^{\mu \sigma \rho}\bar{\nabla}^\alpha \bar{\nabla}_\sigma {\cal B}_{\nu \mu} \\
 &+\frac{m^2}{2 \kappa}\eta_{\rho \alpha \beta} \eta^{\mu \sigma \rho} \bar{\nabla}^\alpha \bar{\nabla}_\sigma (h_{\mu\nu}-\bar{g}_{\mu\nu}h)= \eta_{\rho \alpha \beta}\eta^{\mu \sigma \rho} \bar{\nabla}^\alpha  \bar{\nabla}_\sigma T_{\mu \nu}.
\label{newdef2}
 \end{aligned}
  \end{equation}
Accordingly, by using the above-developed tools, one will recast the Fierz-Pauli term as follows
\begin{equation}
\begin{aligned}
 \frac{m^2}{2 \kappa} ( h_{\beta \nu}- \bar{g}_{\beta\nu} h)&=-\frac{m^2}{ \kappa}(\bar{\square}-2 \Lambda)^{-1} {\cal G}^L_{\beta \nu}+\frac{\Lambda}{\kappa} (\bar{\square}-2 \Lambda)^{-1} \bar{g}_{\beta \nu} {\cal G}^L \\
 &-\frac{m^2}{2 \kappa}(\bar{\square}-2 \Lambda)^{-1} \Big( \bar{g}_{\beta\nu } \bar{\square}-\bar{\nabla}_\beta \bar{\nabla}_\nu \Big)h-\Lambda (\bar{\square}-2 \Lambda)^{-1} \bar{g}_{\beta\nu} T.
\label{explfierzpaul}
 \end{aligned}
 \end{equation}
Here, the inverse stands for the related Green's function. Accordingly, plugging Eq.(\ref{explfierzpaul}) into Eq.(\ref{klmapt2}) as well as using Eq.(\ref{scaleinsstre}), one finally arrives at 
   \begin{equation}
 \begin{aligned}
   &\Bigg ((\bar{\square}-3 \Lambda-\frac{\mu^2}{\kappa^2})+\frac{2 \mu^2 m^2}{ \kappa^2}(\bar{\square}-2 \Lambda)^{-1}-\frac{\mu^2 m^4}{ \kappa^2}(\bar{\square}-2 \Lambda)^{-2} \Bigg) {\cal G}^{L}{_{\rho \nu}} \\
&=\frac{\mu}{2} \, \eta_\rho{^{\mu\sigma}} \bar{\nabla}_\mu T_{\sigma\nu}+ \frac{\mu}{2} \, \eta_\nu{^{\mu\sigma}} \bar{\nabla}_\mu T_{\sigma\rho}-\frac{\mu^2}{\kappa} T_{\rho \nu}+\frac{\mu^2 m^2}{ \kappa} (\bar{\square}-2 \Lambda)^{-1} T_{\rho\nu}\\
&-\frac{\mu^2 m^2}{2 \kappa \Lambda(1-\frac{m^2}{\Lambda})}\Bigg \{(\bar{\square}-2 \Lambda)^{-1} \Big (1-m^2 (\bar{\square}-2 \Lambda)^{-1}\Big)-\frac{\kappa^2 \Lambda}{\mu^2 m^2} \Bigg \}\\&\times \Big(\bar{g}_{\rho\nu }(\bar{\square}-2 \Lambda)-\bar{\nabla}_\rho \bar{\nabla}_\nu \Big)T.
\label{res1223}
\end{aligned}
\end{equation}
Here, one has 
\begin{equation}
 {\cal G}^L_{\rho\nu}= -\frac{1}{2} (\bar{\square}-2 \Lambda)h_{\rho\nu}+\frac{1}{2} \bar{\nabla}_\rho \bar{\nabla}_\nu h.
  \end{equation}
Note that by exploting $ h=\frac{\kappa}{\Lambda-m^2} T $, one can rewrite Eq.(\ref{res1223}) in the following desired form 
\begin{equation}
{\cal O} h_{\rho\nu}=\tilde{T}_{\rho\nu}.
\label{fineqdsn}
 \end{equation}
Once Eq.(\ref{fineqdsn}) is obtained, we can now proceed to read the masses of the excitations around flat and (A)dS backgrounds: observe that the most economical way to count the DOF and thereby examine the particle spectrum is working in the source-free regions  (i.e., $T_{\rho\nu}=0$). In this regard, the linearized Einstein tensor becomes ${\cal G}^L_{\rho\nu}= -\frac{1}{2} (\bar{\square}-2 \Lambda)h_{\rho\nu}$ and the right-hand side of Eq.(\ref{res1223}) vanishes in the vacuum and thus one eventually attains
\begin{equation}
 \Bigg [ \Big(\bar{\square}-3 \Lambda-\frac{\mu^2}{\kappa^2} \Big )(\bar{\square}-2 \Lambda)^2+\frac{2 \mu^2 m^2}{ \kappa^2}(\bar{\square}-2 \Lambda)-\frac{\mu^2 m^4}{ \kappa^2} \Bigg]h_{\rho \nu}=0,
\label{vacuum3}
 \end{equation}
which boils down to
 \begin{equation}
 (\partial^2)^3-\frac{\mu^2}{\kappa^2}(\partial^2)^2+\frac{2 \mu^2 m^2}{ \kappa^2}\partial^2 -\frac{\mu^2 m^4}{ \kappa^2}=0,
 \label{massflat}
\end{equation}
in flat background. Observe that it has three massive modes. Furthermore, as $ m^2=0 $, TMG possesses a single massive spin-2 excitations with $ M_{graviton}=-\mu/\kappa $. Thus, to have a unitarity in flat vacuum, $\kappa $ must be negative. As to the generic case, the model can admit imaginary roots of Eq.(\ref{massflat}) which is a catastrophic possibility. But, as is given in \cite{DeserTekinMode}, for the particular choice $ \mu^2/m^2 \geq 27/4 $, all the roots turns into real.  Notice that for the lowest limit $ \mu^2/m^2 = 27/4 $, one reads the masses as follows
 \begin{equation}
  m_1^2=m^2_2=3 m^2, \qquad m^2_3=\frac{3 m^2}{4},
 \end{equation}
which are actually the same as the ones given in \cite{DeserTekinMode}.

As for the spectrum about (A)dS spaces, by bearing in mind that the null-cone propagation for spin-2 field satisfies $ (\bar{\square}-2 \Lambda)h_{\mu\nu}=0 $, hence Eq.(\ref{vacuum3}) turns into
\begin{equation}
 (\bar{\square}-2 \Lambda-m^2_1) (\bar{\square}-2\Lambda-m^2_2) (\bar{\square}-2 \Lambda-m^2_3)h_{\mu\nu}=0,
\end{equation}
where the relevant roots obey 
\begin{equation}
 \begin{aligned}
m^2_1+m^2_2+m^2_3&= \Lambda+\frac{\mu^2}{\kappa^2},  \\
m^2_1m^2_2m^2_3&=\frac{\mu^2 m^4}{\kappa^2}, \\
m^2_1m^2_2+m^2_1m^2_3+m^2_2m^2_3 &=\frac{2 \mu^2 m^2}{\kappa^2}. 
\label{massads1}
 \end{aligned}
\end{equation}
Giving the explicit form of the masses obtained from Eq.(\ref{massads1}) is rather cumbersome. However, here, we need to emphasize that it might provide special limits. For example, taking the $\mu\rightarrow \infty$ limit, which gives the Fierz-Pauli theory with two excitations with mass $m$. On the other hand,
at the chiral point $ (\mu^2/\kappa^2=-\Lambda)$, there occur three roots such that two of them become tachyon. That is to say, contrary to Einstein gravity, the theory strictly rejects any Fierz-Pauli mass deformation about its chiral point
and so there is no such chiral gravity extension of Fierz-Pauli theory.

\subsubsection*{Scattering Amplitudes}

Hereafter, we will consider the tree-level scattering amplitude between two locally spinning conserved point-like sources. To do so, one needs to first single out the non-propagating DOF from the model. For this purpose, let us assume the following decomposition of graviton field
\begin{equation}
  h_{\mu\nu} \equiv  h^{TT}_{\mu\nu}+\bar{\nabla}_{(\mu}V_{\nu)}+\bar{\nabla}_\mu \bar{\nabla}_\nu \phi+\bar{g}_{\mu\nu} \psi,
  \label{dectth}
 \end{equation}
 where $ h^{TT}_{\mu\nu} $ is the transverse-traceless, $ V_\mu $ is the divergence-free vector and $ \phi $ and $ \psi $ are scalar components of $ h_{\mu\nu} $. To get rid of $\phi $ and thus rewrite the graviton field in terms of the source terms, one needs to find the trace of Eq.(\ref{dectth}) and the divergence of Eq.(\ref{condtion1}). In doing so, one arrives at 
\begin{equation}
 h=\frac{1}{\Lambda} (\bar{\square}+3 \Lambda) \psi.
\end{equation}
Thereupon, by using Eq.(\ref{scaleinsstre}), one gets
 \begin{equation}
  \psi = \frac{\kappa}{1-\frac{m^2}{\Lambda}} (\bar{\square}+3 \Lambda)^{-1} T.
  \label{psittt}
 \end{equation}
On the other side, to link $ h^{TT}_{\mu\nu} $ to the source, one has to utilize the Lichnerowicz operator $ \triangle^{(2)}_L $ that acts on the graviton field according to \cite{Porrati:2000cp} 
  \begin{equation}
  \triangle^{(2)}_L h_{\mu\nu} =-\bar{\square} h_{\mu\nu}-2 \bar{R}_{\mu\rho\nu\sigma} h^{\rho\sigma}+2\bar{R}^\rho{_{(\mu}} h_ {\nu)\rho}.
 \end{equation}
Here, one has 
 \begin{equation}
  \begin{aligned}
    \triangle^{(2)}_L \nabla_{(\mu}V_{\nu)}&= \nabla_{(\mu} \triangle^{(1)}_L V_{\nu)} , \qquad  \quad \,\,\, \triangle^{(1)}_L V_{\mu}=(-\square+\Lambda)V_\mu, \\
    \nabla^\mu \triangle^{(2)}_L h_{\mu\nu} &=\triangle^{(1)}_L \nabla^\mu h_{\mu\nu}, \qquad  \,\, \nabla^\mu \triangle^{(1)}_L V_{\mu}= \triangle^{(0)}_L \nabla^\mu V_\mu \\
    \triangle^{(2)}_L g_{\mu\nu} \phi &= g_{\mu\nu}\triangle^{(0)}_L \phi, \qquad  \qquad \quad \triangle^{(0)}_L \phi=-\square \phi,
  \label{lichng}
  \end{aligned}
 \end{equation}
which supplies to recast ${\cal G}^{TTL}{_{\rho \nu}}$ in terms of Lichnerowicz operator as follows 
 \begin{equation}
  {\cal G}^{TTL}{_{\rho \nu}}=\frac{1}{2} \Big ( \triangle^{(2)}_L-4\Lambda \Big ) h^{TT}_{\rho\nu}.
  \label{einstett2}
 \end{equation}
Hence, plugging Eq.(\ref{einstett2}) into Eq.(\ref{res1223}) yields
 \begin{equation}
  \begin{aligned}
    h^{TT}_{\rho\nu}&=\mu \, {\cal O}^{-1} (\bar{\square}-2 \Lambda)^{2} \eta_\rho{^{\mu\sigma}} \bar{\nabla}_\mu T^{TT}_{\sigma\nu}+ \mu \,{\cal O}^{-1} (\bar{\square}-2 \Lambda)^{2} \eta_\nu{^{\mu\sigma}} \bar{\nabla}_\mu T^{TT}_{\sigma\rho} \\
&-\frac{2 \mu^2}{\kappa} {\cal O}^{-1} (\bar{\square}-2 \Lambda)^{2}T^{TT}_{\rho \nu}+\frac{2 \mu^2 m^2}{ \kappa} {\cal O}^{-1} (\bar{\square}-2 \Lambda) T^{TT}_{\rho\nu}.
 \label{htt}
 \end{aligned}
 \end{equation}
Here, the corresponding Green's function is
\begin{equation}
  {\cal O}^{-1} \equiv \Bigg \{ \Big [ (\bar{\square}-2 \Lambda)^{2} \Big (\bar{\square}-3 \Lambda-\frac{\mu^2}{\kappa^2} \Big)+\frac{2 \mu^2 m^2}{ \kappa^2}(\bar{\square}-2 \Lambda)-\frac{\mu^2 m^4}{ \kappa^2}\Big ] \times  \Big ( \triangle^{(2)}_L-4\Lambda \Big ) \Bigg \}^{-1}. 
\end{equation}
Similarly, the decomposition of $T_{\rho\nu} $ will read \cite{Gullu:2009vy, Gullutez}
\begin{equation}
 T^{TT}_{\rho\nu}=T_{\rho\nu}-\frac{1}{2} \bar{g}_{\mu\nu}T+\frac{1}{2} \Big (\bar{\nabla}_\mu \bar{\nabla}_\nu + \Lambda \bar{g}_{\mu\nu} \Big ) \times (\bar{\square}+3 \Lambda )^{-1} T.
\label{ttstrener}
 \end{equation}
Recall that the tree-level scattering amplitude between two localized conserved spinning point-like sources a la one graviton exchange is described by
\begin{equation}
\begin{aligned}
 {\cal A}&=\frac{1}{4} \int d^3 x \sqrt{-\bar{g}} T^{'}_{\rho\nu}(x) h^{\rho\nu}(x) \\
 &=\frac{1}{4} \int d^3 x \sqrt{-\bar{g}} (T^{'}_{\rho\nu} h^{TT\rho\nu}+T^{'} \psi ).
 \label{scatdef}
\end{aligned}
 \end{equation}
Thus, by inserting Eq.(\ref{psittt}), Eq.(\ref{einstett2}) and Eq.(\ref{ttstrener}) into Eq.(\ref{scatdef}), one will finally get the related scattering amplitude in (A)dS background as follows
\begin{equation}
 \begin{aligned}
  4{\cal A}&=2\mu T^{'}_{\rho\nu} {\cal O}^{-1} (\bar{\square}-2 \Lambda)^{2} \eta^{\rho\mu\sigma}\bar{\nabla}_\mu T_\sigma{^\nu}-\frac{2 \mu^2}{\kappa}T^{'}_{\rho\nu} {\cal O}^{-1}(\bar{\square}-2 \Lambda)(\bar{\square}-2 \Lambda-m^2) T^{\rho\nu} \\
  &-\frac{\mu^2}{\kappa}T^{'}_{\rho\nu} {\cal O}^{-1} (\bar{\square}-2 \Lambda)(\bar{\square}-2 \Lambda-m^2) (\bar{\nabla}^\rho \bar{\nabla}^\nu + \Lambda \bar{g}^{\rho\nu}) \times \Big (\bar{\square}+3 \Lambda \Big )^{-1} T \\
&+\frac{\mu^2}{\kappa} T^{'} {\cal O}^{-1}(\bar{\square}-2 \Lambda)(\bar{\square}-2 \Lambda-m^2) T + \frac{\kappa}{1-\frac{m^2}{\Lambda}} T^{'} (\bar{\square}+3 \Lambda)^{-1} T.
\label{mainressct}
\end{aligned}
\end{equation}
Here, the integral signs are suppressed for the sake of simplicity. As is manifest, scattering amplitude Eq.(\ref{mainressct}) for generic constant curvature spaces intricate. The pole structures of a theory generally indicates the existence of particles. Thus, here one can do the unitarity analysis by using the pole structure. In doing so, it is obvious that Eq.(\ref{mainressct}) has four poles. One of them is $ \bar{\square}_1 = -3 \Lambda $ and the others are nothing but the roots of cubic Eq.(\ref{massads1}). Here, determining the physical poles is a bit unwieldy so, for simplicity, let us set Fierz-Pauli term to zero and examine its chiral limit. Therefore, by setting $ m^2=0 $, $ h=0 $ and $\mu^2/\kappa^2 =- \Lambda $, one will get the scattering amplitude in the chiral limit 
\begin{equation}
 \begin{aligned}
  4{\cal A}&=2 \sqrt{-\Lambda} T^{'}_{\rho\nu} \Bigg \{\Big (\bar{\square}-2 \Lambda \Big) \times  \Big (\triangle^{(2)}_L-4\Lambda \Big ) \Bigg \}^{-1} \eta^{\rho\mu\sigma}\bar{\nabla}_\mu T_\sigma{^\nu}\\
  &+2 \kappa \Lambda T^{'}_{\rho\nu} \Bigg \{\Big (\bar{\square}-2 \Lambda \Big) \times  \Big ( \triangle^{(2)}_L-4\Lambda\Big ) \Bigg \}^{-1} T^{\rho\nu}. 
\label{anyon-anyon2}
\end{aligned}
\end{equation}
Observe that there are two poles 
\begin{equation}
  \bar{\square}_1=2 \Lambda, \qquad \bar{\square}_2=-4 \Lambda.
\end{equation}
But since the residue of the second pole is zero, it is not a physical pole. Thus, we finally have one physical poles at $ \bar{\square}_1=2 \Lambda$.
Observe that we have massless graviton which satisfy Breitenlohner-Freedmann bound $(M^2 \geq \Lambda)$.

\subsection{Flat Space Considerations}

In this part, we will calculate the three-level scattering amplitude for various theories in flat spaces which will provide the desired non-relativistic gravitational potential energies $ {\cal U} $ between two covariantly point-like spinning sources. To do so,  let us consider the following energy-momentum tensors 
\begin{equation}
T_{00}= m_a \delta^{(2)}({\bf x}-{\bf x}_a),\qquad T^{i}{_0}=-1/2 J_a \epsilon^{ij} \partial_j \delta^{(2)}({\bf x}-{\bf x}_a).
\end{equation}
Here, $m_a $ are mass and $ J_a $ are the spin of sources where $a=1,2$.

\subsubsection{Scattering of anyons in TMG with Fierz-Pauli Term}
 
 This part is devoted to the anyon-anyon scattering and also the corresponding Newtonian potential energies in the TMG augmented by a Fierz-Pauli mass term. Therefore, let us notice that by applying the flat-space limit (i.e., $ \Lambda \to 0 $) in Eq.(\ref{mainressct}), one gets
 \begin{equation}
 \begin{aligned}
  4{\cal A}&=-2 \mu T^{'}_{\rho\nu} \frac{\partial^2}{\partial^4 (\partial^2-\frac{\mu^2}{\kappa^2})+\frac{2 \mu^2 m^2}{\kappa^2} \partial^2-\frac{\mu^2 m^4}{\kappa^2}} \eta^{\rho\mu\sigma}\partial_\mu T_\sigma{^\nu}\\
  &+\frac{2 \mu^2}{\kappa}T^{'}_{\rho\nu}  \frac{\partial^2-m^2}{\partial^4 (\partial^2-\frac{\mu^2}{\kappa^2})+\frac{2 \mu^2 m^2}{\kappa^2} \partial^2-\frac{\mu^2 m^4}{\kappa^2}} T^{\rho\nu}
   \\&-\frac{\mu^2}{\kappa} T^{'}\frac{\partial^2-m^2}{\partial^4 (\partial^2-\frac{\mu^2}{\kappa^2})+\frac{2 \mu^2 m^2}{\kappa^2} \partial^2-\frac{\mu^2 m^4}{\kappa^2}} T. 
\label{scatflat2}
  \end{aligned}
\end{equation}
In general, the explicit form of propagators can be rather cumbersome. In such cases, one can alternatively decompose them in terms of the known ones. 
In this respect, one can recast the relevant main propagator as  follows
\begin{equation}
\frac{\partial^2-m^2}{\partial^4 (\partial^2-\frac{\mu^2}{\kappa^2})+\frac{2 \mu^2 m^2}{\kappa^2} \partial^2-\frac{\mu^2 m^4}{\kappa^2}} 
\equiv \sum_{k=1}^{3} \prod_{\substack{r=1 \\ r \neq k}}^3 \frac{(M^2_k-m^2)}{(M^2_k-M^2_r)} G_k({\bf x}, {\bf x}^{'},t,t^{'}).
 \label{decgreen1}
  \end{equation}
Here, $G_k({\bf x}, {\bf x}^{'}, t, t^{'})=(\partial^2-M_k^2)^{-1}$ where $M_k =M_k (\kappa^2, \mu^2,m^2) $, $ k=1,2,3 $, are the roots of the cubic equation. Thus, by plugging Eq.(\ref{decgreen1}) into Eq.(\ref{scatflat2}) and ensuingly evaluating the time integrals, one will eventually obtain
\begin{equation}
 \begin{aligned}
4\,{\cal U}=&\sum_{k=1}^{3} \prod_{\substack{r=1 \\ k \neq r}}^3(M^2_k-M^2_r)^{-1} \Bigg \{\frac{\mu^2 M^2_k}{\kappa} \Big (\frac{\kappa m_2 }{\mu} J_1 +\frac{\kappa m_1 }{\mu} J_2  + J_1 J_2(1-\frac{ m^2 }{M^2_k}) \Big ) \\
& \times \int d^2 x \int d^2 x^{'}\,\,\,\delta^{(2)}({\bf x}^{'}-{\bf x}_2) \partial_i \partial^i \hat{G}_k({\bf x}, {\bf x}^{'}) \delta^{(2)}({\bf x}-{\bf x}_1) \\
& +  \frac{\mu^2 m_1 m_2 }{\kappa} ( M^2_k-m^2) \int d^2 x \int d^2 x^{'}\,\,\, \delta^{(2)}({\bf x}^{'}-{\bf x}_2) \hat{G}_k({\bf x}, {\bf x}^{'}) \delta^{(2)}({\bf x}-{\bf x}_1) \Bigg \}.
  \end{aligned}
\end{equation}
Here, the potential energy is $ {\cal U}={\cal A}/t $ \cite{gullu_spin} and the time-integrated scalar Green's function $G_k({\bf x}, {\bf x}^{'})$ is
\begin{equation}
  \hat{G}_k ({\bf x},{\bf x}^{'})=\int d t^{'} \, G_k ({\bf x},{\bf x}^{'},t,t^{'})=\frac{1}{2 \pi}  K_{0}\, (M_k r),
\end{equation}
where $ r = \lvert {\bf x}_1-{\bf x}_2 \rvert$ and $K_{0}(M_k r) $ is the modified Bessel function of the second kind. Hence, by virtue of the identity between the Bessel functions,
\begin{equation}
  \vec{\nabla}^2 K_0 (M_k r)= \frac{M^2_k}{2} \Big (K_0 (M_k r)+K_2 (M_k r) \Big ),
\label{recur}
\end{equation} 
 the Newtonian potential energy will read
\begin{equation}
 \begin{aligned}
{\cal U}&=\sum_{k=1}^{3} \prod_{\substack{r=1 \\ k \neq r}}^3(M^2_k-M^2_r)^{-1} \Bigg \{\frac{\mu^2 M^4_k}{16 \pi \kappa} \Big (J^{tot}_1 J^{tot}_2-\frac{\kappa^2 m_1 m_2}{\mu^2}- \frac{ m^2 J_1 J_2 }{M^2_k} \Big )  K_{2}(M_k r) \\
&\qquad \qquad\qquad\qquad\qquad +  \frac{\mu^2 M^4_k}{16 \pi \kappa}\Big [\frac{2 m_1 m_2 }{ M^2_k}(1-\frac{m^2 }{M^2_k} )+ \Big (J^{tot}_1 J^{tot}_2-\frac{ \kappa^2 m_1 m_2}{\mu^2}\\&\qquad \qquad\qquad\qquad\qquad- \frac{ m^2 J_1 J_2 }{M^2_k} \Big )\Big ] \times K_{0}\, (M_k r) \Bigg \}.
  \label{newtpot1}
  \end{aligned}
\end{equation}
 
Before going further, let us now quote one of the final and very important result: unlike the ordinary Einstein theory, topological mass $ \mu $ in TMG induces an additional spins with $  J^{ind}_a=\kappa m_a/\mu \,\,\,; a=1,2 $ which change the initial spin of the particles and turns them into an anyon \cite{DeserAnyon} by
\begin{equation}
 J^{tot}_a =J_a+ J^{ind}_a.
 \label{totspin1}
\end{equation}
 Observe that, in addition to total spin-spin and mass-mass interactions (which is consistent with the result when $ m^2=0 $), there also occurs the Fierz-Pauli mass and initial spin interactions.  Notice that as is seen in Eq.(\ref{newtpot1}), the Fierz-Pauli mass term couples only to the initial spins of particles. Note also that depending on the choice of $(J_i, m_i, m^2)$, $ {\cal U} $ can be either negative or positive. 

Let us now turn our attention to the extreme distance limits of potential energy. Firstly, it is straightforward to show that the potential turns into
   \begin{equation}
 \begin{aligned}
{\cal U}& \sim \sum_{k=1}^{3} \prod_{\substack{r=1 \\ k \neq r}}^3(M^2_k-M^2_r)^{-1} \Bigg \{\frac{\mu^2 M^2_k}{8 \pi \kappa} \Big (J^{tot}_1 J^{tot}_2-\frac{\kappa^2 m_1 m_2}{\mu^2}- \frac{ m^2 J_1 J_2 }{M^2_k} \Big ) \times \frac{1}{r^2} \\
&\qquad \qquad\qquad\qquad\qquad - \frac{\mu^2 M^4_k}{16 \pi \kappa}\Big [\frac{2 m_1 m_2 }{ M^2_k}(1-\frac{m^2 }{M^2_k} )+ \Big (J^{tot}_1 J^{tot}_2-\frac{\kappa^2 m_1 m_2}{\mu^2}\\
&\qquad \qquad\qquad\qquad\qquad- \frac{ m^2 J_1 J_2 }{M^2_k} \Big )\Big ] \times \Big (\ln(M_k r)+\gamma_{E} \Big) \Bigg \},
  \label{newtpot3}
  \end{aligned}
\end{equation}
at the short range. Here, $ \gamma_E $ is the Euler-Mascheroni constant. It is necessary to stress that for the induced angular momenta for the positive $ \kappa$ at the critical value 
\begin{equation}
 J_1=-\frac{\kappa m_1 J_2}{(\mu J_2 +\kappa m_2 )- \frac{\mu m^2 J_2}{M^2_k} },  \qquad \frac{m^2}{M^2_k} < 0,
  \label{potcond2}
\end{equation}
Eq.(\ref{newtpot3}) acts like Newton's potential energy. As for the large regimes, by keeping in mind that the modified Bessel functions behave like
\begin{equation}
  K_n(M_k r) \sim \sqrt{\frac{\pi}{ 2 M_k r}} \, e^{-M_k r},
  \label{Besselfp2}
 \end{equation}
at these scales, one will get the corresponding potential energy as follows
\begin{equation}
 \begin{aligned}
{\cal U}& \sim \sum_{k=1}^{3} \prod_{\substack{r=1 \\ k \neq r}}^3(M^2_k-M^2_r)^{-1} \frac{\mu^2 M^4_k}{8 \pi \kappa}\Big [\frac{ m_1 m_2 }{ M^2_k}(1-\frac{m^2 }{M^2_k} )+ \Big (J^{tot}_1 J^{tot}_2-\frac{\kappa^2 m_1 m_2}{\mu^2}\\
&\qquad \qquad\qquad\qquad\qquad\qquad\quad- \frac{ m^2 J_1 J_2 }{M^2_k} \Big )\Big ] \times \sqrt{\frac{\pi}{ 2 M_k r}} \, e^{-M_k r},
 \label{newtpot5}
\end{aligned}
\end{equation}
which asymptotically approaches zero. 
  
\subsubsection{Scattering of anyons in TMG}
In this section, we focus on the scattering amplitude of anyons and hence the corresponding non-relativistic gravitational potential energy for the usual TMG (see Appendix \ref{chp:appendixb1} for details). Accordingly, applying the $ m^2 \to 0 $ limit in Eq.(\ref{mainressct}) yields 
\begin{equation}
 \begin{aligned}
  4{\cal A}&=-2\mu T^{'}_{\rho\nu} \frac{1}{(\partial^2-\frac{\mu^2}{\kappa^2})\partial^2} \eta^{\rho\mu\sigma}\partial_\mu T_\sigma{^\nu}+\frac{2 \mu^2}{\kappa}T^{'}_{\rho\nu} \frac{1}{(\partial^2-\frac{\mu^2}{\kappa^2})\partial^2}   T^{\rho\nu} \\
  &-\frac{\mu^2}{\kappa} T^{'} \frac{1}{(\partial^2-\frac{\mu^2}{\kappa^2})\partial^2} T + \kappa T^{'} \frac{1}{\partial^2} T.
\label{anyon-anyon1}
\end{aligned}
\end{equation}
Note that here one admits a massive and a massless modes. Observe that, in the $ \mu \to \infty $ limit, Eq.(\ref{anyon-anyon1}) reduces to the Einstein's theory which implies that van Dam-Veltman-Zakharov (vDVZ) discontinuity disappears \cite{Porrati:2000cp}. Additionaly, by using Eq.(\ref{anyon-anyon1}) and the non-vanishing components of stress-energy tensor, one will get 
\begin{equation}
 {\cal U}= \frac{\kappa m^2_g}{16 \pi }\bigg \{ \Big ( J^{tot}_1 J^{tot}_2-\frac{m_1 m_2}{m^2_g} \Big )K_2(m_g r)+ \Big ( J^{tot}_1 J^{tot}_2+\frac{m_1 m_2}{m^2_g} \Big )K_0(m_g r ) \bigg \},
\label{tmganyan}
 \end{equation}
where $m^2_g= \mu^2/\kappa^2 $. Thus, in addition to spin-spin and mass-mass interactions, here there also occur spin-mass interactions such that topological mass induces extra spins $ J^{ind}_a= \kappa m_a/\mu $ as in the general case. Observe that depending on the choice of the non-vanishing parameters $ (J_a, m_a) $, Eq.(\ref{tmganyan}) can be either negative or positive but can not be zero. 

Let us now consider its small and large distance behaviours of potential energy: First of all, in the neighbourhood of the sources, one obtains  
\begin{equation}
\begin{aligned}
 {\cal U}  \sim \frac{\frac{\kappa}{8 \pi}  \Big ( J^{tot}_1  J^{tot}_2-\frac{m_1 m_2}{m^2_g} \Big )}{ r^2}  
-\frac{\kappa m^2_g}{16 \pi}  \Big ( J^{tot}_1 J^{tot}_2+\frac{m_1 m_2}{m^2_g} \Big ) \Big (\ln(m_g r)+\gamma_{E} \Big ).
  \end{aligned}
\end{equation}
Finally, for large distances, Eq.(\ref{tmganyan}) asymptotically behaves 
 \begin{equation}
 {\cal U} \sim \frac{\kappa m^2_g J^{tot}_1 J^{tot}_2}{8  \pi} \sqrt{\frac{\pi}{ 2 m_g r}} \, e^{-m_g r},
\end{equation}
and it becomes attractive if $ J^{tot}_1 . J^{tot}_2 <0 $.
  
\subsubsection{ Flat-Space Chiral Gravity}

AdS/CFT correspondence is one of the forthcoming approach to construct well-defined quantum gravity theory: remember that for each $d$-dimensional bulk AdS gravity, there is a corresponding dual  $d-1$-dimensional CFT on the boundary. Therefore, naturally one can ask whether there is such a gauge/gravity duality in flat space or not? Recently, it was proposed that a pure gravitational Chern-Simons term
\begin{equation}
 S=\frac{k}{4 \pi} \int d^3 x \sqrt{-g} \, \eta^{\mu \nu \alpha} \Gamma^\beta{_{\mu \sigma}} \Big (\partial_\nu \Gamma^\sigma{_{\alpha \beta}}+\frac{2}{3} \Gamma^\sigma{_{\nu \lambda}}  \Gamma^\lambda{_{\alpha \beta}} \Big ),
\end{equation}
gives rise to a duality between an asymptotically flat limit of TMG, called flat-space chiral gravity, and a specific 1+1-dimensional Galilean Conformal field (GCF) theory with a central charge of 24 \cite{Bagchi:2012yk}. Therefore, it is also very crucial to find the corresponding potential energy of the bulk flat-space chiral gravity and then analyse whether its short and large distance behaviours are consistent with the Newton's theory. For that purpose, we will evaluate the Newtonian potential energy from one graviton exchange between two locally conserved spinning sources as we have done so far.  To do so, let us observe that by performing the following limits
\begin{equation}
 \kappa \to \infty , \quad \mu \to \frac{2 \pi}{k},
\end{equation}
in Eq.(\ref{mainressct}), one gets the corresponding scattering amplitude as follows
 \begin{equation}
 4{\cal A}=-\frac{4 \pi}{k} T^{'}_{\rho\nu} \frac{1}{\partial^4} \eta^{\rho\mu\sigma} \partial_\mu T_\sigma{^\nu}, 
\label{flatspaelim1}
 \end{equation}
where we set $T=0$ and $m^2=0 $. Note that due to Chern-simons term which can only couple to a covariantly conserved traceless energy momentum tensor, one needs to set a null source. For this purpose, let us consider flat space in null coordinates
\begin{equation}
 ds^2=-dudv+dz^2, 
\end{equation}
where $ u=t-x $ and $ v=t+x$. Then, the vector can be written in the form as  $ \ell_\mu \equiv \partial_\mu u $ with the  $ \ell_\mu$ is a vector field that satisfies
 \begin{equation}
 \ell^\mu =-\delta^\mu_v \qquad  \ell_\mu l^\mu=0.
 \end{equation}
In meantime, one can write the corresponding covariantly conserved null source  as $ T^{\mu\nu}=E \delta(u) \delta(y) \ell^\mu \ell^\nu $ with which Eq.(\ref{flatspaelim1}) gives a trivial  amplitude.

\section[Causality in $3D$ massive gravity theories]{Causality in $3D$ massive gravity theories \footnote{The results of this chapter are published in \cite{kilicarslan3} .}}

Shapiro's time-delay \cite{Shapiro:1964uw} is the fourth test of general relativity (GR) in the solar system. Essentially  Shapiro showed that light wave takes longer time in the round trip due to presence of a massive body. Time delay in GR had been shown 
in many ways \cite{Gao:2000ga}. But some modified gravity theories that have quadratic or cubic curvature terms give a Shapiro time advance instead of time delay since the adding higher order curvature terms lead to causality violation \cite{Camanho:2014apa}. Interestingly, causality violation in these theories can be recovered by adding of an infinite tower of massive higher-spin particles \cite{Camanho:2014apa}.

Here we consider the causality issue in $2+1$ dimensional gravity. At this point, one might think that the Shapiro time-delay does not make any sense in $2+1$ dimensions such that it has a locally trivial structure. However, in $2+1$ dimensions, there are many locally nontrivial massive gravity theories that have been devoted a lot of work in the literature, and we want to study on them from the point of causality. Especially, we want to understand whether the  {\it unitarity} and {\it causality} conditions are compatible or not in these theories.  By unitarity, we mean ghosts and tachyons freedom of the theory, and by causality we mean that time delay is positive definite ---as opposed to a time-advance---, from the point of Shapiro's argument. 

We will only be interested in the local causality issue and avoid getting involved in the global cases which are rather complicated. Recall that since the bare Einstein's gravity does not admit any DOF in 2+1 dimensions, the local causality turns out to be trivial and thus one can only deal with the global causality issues \cite{thooft}. In other words, any vacuum solutions of Einstein's gravity in 3 spacetime dimensions are locally equivalent to flat or constant curvature backgrounds. The Riemann curvature can be written in terms of curvature tensor; the theory has not any local propagating degrees of freedom, and there is no local causality issue. Whereas  3-dimensional Einstein's gravity, there are several dynamical massive gravity theories that have the local massive propagating degree of freedom. Especially, we will consider the topologically massive gravity (TMG) \cite{DJT} and new massive gravity (NMG) \cite{BHT} that have taken more attention in the literature. These theories satisfies unitary conditions with the correct sign choices of Einstein-Hilbert term. Unlike the Einstein-Gauss-Bonnet or cubic theories in higher dimensions, we will show that causality and unitarity are compatible in TMG, NMG and their modifications once the sign of Newton's constant is chosen negative as opposed the $3+1$ dimensional case. As compared the $2+1$-dimensional gravity with the higher dimensional ones, the main reason of why this is exclusive to lower dimension still remains unclear.

To study local causality issue in three dimensional gravity theories, we will consider spaces that asymptote to Minkowski, de sitter and Anti-de sitter space. First of all, we consider the issue in asymptotically Minkowski space; this will be the first part of the study. Secondly, we will consider the asymptotically Anti-de Sitter (AdS) solutions: this will be done in the second part of the study. Shapiro's time-delay can be computed at least in two ways: The most known method is the one that evaluating the time-delay of light moving in a closed curve around the black hole and later comparing it with the ordinary undisturbed one. Another possible way is to compute the time-delay of a massless particle that moves in the shock wave background \cite{shock, Dray} generated by a ultra-relativistic massless particle. The second method is well suitable for our purpose since TMG, NMG and other massive gravity (except purely quadratic gravity theories) theories have not asymptotically flat black hole solutions. In the second part of study, we will consider the case of negative cosmological constant. Unlike the flat space analysis, as we will see,  AdS$_3$ introduces a new scale, time delay will depend the two scales effective mass $m_g $ and cosmological constant $\sqrt{|\Lambda |}$.

\subsection{General Relativity Warm-up in 2+1 Dimensions}
\label{sec2}

It is a well known fact that 2+1 dimensional Einstein's gravity has a locally trivial structure that is, it has not any gravitational degree of freedom. Therefore, there is no local causality issue.  Nonetheless, we will study the causality issue in Einstein's gravity as a warm up exercise to set structures for massive gravity theories. 

Consider  a shock wave metric  generated by massless particle 
\begin{equation}
  ds^2=-dudv+H(u,y)du^2+dy^2,
  \label{metric1}
\end{equation}
where $u=t-x$ and $v=t+x$ null coordinates. We assume that, there is a massless point particle moving in the $+x$ direction with $p^\mu=\lvert p\rvert(\delta^\mu_0+\delta^\mu_x) $, hence the only non-vanishing component of the energy-momentum tensor is
\begin{equation}
  T_{uu}=\lvert p\rvert \delta(y)\delta(u),
\end{equation}
here $ p$ is the momentum of a particle. For the shock wave ansatz, Einstein's field equations  reduces to (we set $|8\pi G| = 1$)
\begin{equation}
\begin{aligned}
\partial_y^2H(u,y)=-2\sigma\lvert p\rvert \delta(y)\delta(u) ~, 
\end{aligned}
\end{equation}
whose the most general solution  is
\begin{equation}
H(u,y)=-2\sigma\lvert p\rvert\delta(u)\theta(y)y\,+\,c_1(u)y+c_2(u) ~,
\end{equation}
where $c_1(u)$ and $c_2(u)$ are functions of $u$ coordinates. These functions can be determined by coordinate transformations. Observe that when we set $c_1(u)=0$ and $c_2(u)=0$ one obtains a non-vanishing profile for $y>0$ but a vanishing one for $y<0$. On the other hand if one sets $c_1(u)=2\lvert p\rvert\delta(u)$ and $c_2(u)=0$ one obtains the following profile 
\begin{equation}
H(u,y)=2\sigma\lvert p\rvert\delta(u)\theta(-y)y ~,\label{profileGR}
\end{equation}
which has a non-trivial profile for $y<0$. Notice that, although space is flat outside the source, metric can  not be written in Cartesian coordinates for both sides of profile with a single chart due to source.  

We will calculate Shapiro time delay in three dimensional massive gravity theories when a particle crossing the shock wave geometry at an  impact parameter $y=b>0$. Here, the delay is determined by an asymptotic observer. Namely, one needs to bring the shock wave metric to the asymptotically flat and Cartesian form with appropriate coordinate transformations. For that reason, in Einstein's gravity case  Eq.(\ref{profileGR}), we set $c_1$ and $c_2$ in such a way that spacetime to be asymptotically flat. This implies that the shock wave profile is trivial for $y>0$. Consequently, a massless particle in this geometry does not experience any time delay. This is expected result in GR since there is not any propagating degree of freedom. Note that this does not imply that moving particles do not interact with each other. In fact there is a instantaneous interaction between particles and causality problem but we will not consider this issue here since this is related to global causality issues.      

On the other hand, we can compute time delay  with using the Eikonal scattering amplitudes and compare the results with geodesic analysis  as was obtained above \cite{thooft2}. For this purpose, let us consider the tree-level scattering amplitude between gravitating massless particles  in the deflectionless limit defined as $\tfrac{t}{s}\rightarrow 0$. In the Einstein's gravity case, it is given by
\begin{equation}
\mathcal{A}_{tree}(s,t)=-\sigma \frac{s^2}{t} ~,
\end{equation}
where $s$ is  square of the center-of-mass energy and $\sqrt{-t}$ is the momentum transfer. Here, the infrared scattering amplitude is governed by the variables within the impact parameter in which the eikonal phase shift corresponds to the Shapiro time-delay and is obtained via the Fourier transform of the amplitude as follows \cite{kabat,giddings}\\
\begin{equation}
\delta(b,s)=\frac{1}{2s}\int\frac{dq}{2\pi}\,e^{i q b}\mathcal{A}_{tree}(s,-q^2)=
\frac{\sigma}{4\pi}\,s \int dq\, \frac{e^{iqb}}{q^2} ~.\label{phaseGR}
\end{equation}
Notice that although the time delay is expected to be vanished in Einstein's gravity, Eq.(\ref{phaseGR}) comes up with an unacceptable physical result. That is, note that Eq.(\ref{phaseGR}) diverges as one goes to the zero momentum limit of the amplitude. This is presumably an artifact effect emerging during the eikonal approximation in which the zero angular momenta generically brings up infinities. As in the ordinary perturbative quantum field theory, one needs to truncate these divergences such that the integral will yield zero in the eikonal limit. We will also use this procedure for other theories to gauge the shock wave profiles by GR results.

\subsection{ Causality in TMG }
\label{CausTMG}

Let us consider the issue of ''causality'' in TMG. For this purpose, we need to calculate shapiro time-delay.
The source-coupled field equations of the TMG are given \cite{DJT}
\begin{equation}
  \sigma G_{\mu\nu}+\frac{1}{\mu} C_{\mu\nu}=T_{\mu\nu},
  \label{tmg}
\end{equation}
where $\sigma$ is dimensionless constant, $ T_{\mu\nu} $ is energy momentum tensor and  $ C_{\mu\nu} $ is Cotton tensor.
This parity non-invariant theory has the single massive spin-2 excitation with a mass
\begin{equation}
 m_{g}=-\sigma\lvert \mu\rvert,
\end{equation}
hence we set $\sigma<0$ for unitarity. Notice that $\mu \to -\mu$ changes the parity of the spin-2 particle and mass remains intact under the parity change. So we fix it by assuming $\mu>0$.
For our shock-wave ansatz in three dimensions, the only non-vanishing components of the Ricci and Cotton tensors can be computed as
\begin{equation}
 R_{uu}=G_{uu}=-\frac{1}{2}\frac{\partial^2}{\partial y^2}H(u,y),\hskip 1 cm C_{uu}=\frac{1}{2}\frac{\partial^3}{\partial y^3}H(u,y).
\end{equation}
Then (\ref{tmg}) reduces to
\begin{equation}
\begin{aligned}
 -\frac{\sigma}{2}\partial_y{^2}H(u,y)+ \frac{1}{2\mu}\partial_y{^3}H(u,y)=\lvert p\rvert \delta(y)\delta(u),
\label{eqn}
 \end{aligned}
\end{equation}
 whose the most general solution can be given as  
\begin{equation}
H(u,y)=-\frac{2 \lvert p \rvert}{m_g\sigma}  \delta(u)  \theta (y)\bigg(e^{-m_g y}+m_g y-1\bigg) +c_1\frac{e^{-m_g y}}{m_g^2}+c_2
   y+c_3,
\end{equation}
where $c_1$, $c_2$ and $c_3$  are  null coordinate dependent functions. We will choose the constants in such a way that the space-time is asymptotically flat. On the other hand, one can transform the metric to Cartesian form in the asymptotic limit with coordinate transformations, but this is not possible with a single chart. Then assuming $\mu>0$ and by a gauge fixing, shock wave of TMG recast as  
\begin{equation}
\begin{aligned}
H(u,y) =-\frac{2\sigma}{m_g} \lvert p \rvert \delta(u)\theta(y) e^{-m_g y}+
\frac{2\sigma}{m_g} \lvert p \rvert \delta(u)\theta(-y)(m_g y-1) ~.
\label{metfluc1}
\end{aligned}
\end{equation}

Consider a massless particle crossing shock wave geometry, with an impact parameter $y=b>0$,
the shock wave of TMG reads  
\begin{equation}
\begin{aligned}
 H(u,y) =-\frac{2}{m_g\sigma} \lvert p \rvert \delta(u)\theta(y) e^{-m_g y}.
 \label{metfluc1}
 \end{aligned}
\end{equation}
When Eq.(\ref{metfluc1}) is inserted in Eq.(\ref{metric1}), the discontinuity in the $u$ coordinate can be cancelled out by re-defining the other null coordinate as 
\begin{equation}
\begin{aligned}
 v \equiv v_{new}-\frac{2}{m_g\sigma} \lvert p \rvert \theta(u) e^{-m_g b}.
 \label{metfluc}
 \end{aligned}
\end{equation}
which gives the time delay 
\begin{equation}
\begin{aligned}
 \Delta v =-\frac{2 \lvert p \rvert }{m_g\sigma} e^{-m_g b}.
 \label{metfluc}
 \end{aligned}
\end{equation}
 Since the shifting in the $v$ coordinate is positive for any $b$, the time delay is positive if $\sigma<0$.
Therefore unitarity and causality are compatible in TMG for these null geodesics. 
\subsubsection{Eikonal Scattering in TMG}

In this part, we will calculate the time delay from the point of view of eikonal approximation. By using the same assumptions just like in the section \ref{sec2}, we compute $2 \rightarrow 2$ tree level scattering amplitude of two massless gravitating scalar particles in TMG. The scalar field coupling can be given as
\begin{equation}
\begin{aligned}
\mathcal{S}_{TMG} = \int d^{3}x \bigg(& \sqrt{-g} \sigma R + \frac{1}{2\mu}\epsilon^{\lambda\mu\nu} \Gamma^{\rho}_{\lambda\sigma}(\partial_{\mu} \Gamma^{\sigma}_{\rho\nu} + \dfrac{2}{3}\Gamma^{\sigma}_{\mu\alpha}\Gamma^{\alpha}_{\nu\rho}) - \frac{1}{2\alpha} k^\nu k_\nu \\&+ \frac{1}{2} g^{\mu\nu} \partial_{\mu}\phi \partial_{\nu}\phi\bigg) ~,
\end{aligned}
\end{equation}
where $k^{\nu}=\partial_{\mu}(\sqrt{-g}g^{\mu\nu})$. Let the incoming momenta for the particles are 
\begin{equation}
p_{1\mu} = (p_{u},\dfrac{q^2}{16p_{u}},\dfrac{q}{2}) ~, \qquad p_{2\mu} = (\dfrac{q^2}{16p_{v}},p_{v},-\dfrac{q}{2}) ~,
\end{equation}\label{delayphoton1}
or, in terms of Mandelstam variables, $s = -2 p_1 \cdot p_2 \simeq p_u p_v$ and $t \simeq - q^2$, $t/s \ll 1$. If one use the graviton propagator in the Feynman gauge ($\alpha = 1$) Eq.(\ref{propTMG}), tree-level amplitude can be obtained as 
\begin{equation}
\begin{aligned}
\mathcal{A}_{tree}=-\sigma \frac{s^2}{t}\dfrac{1}{\left(1-\dfrac{i\sigma}{\mu}\sqrt{-t}\right)} ~.
\end{aligned}
\end{equation}
And if one computes the phase-shift with the same procedure as used in section \ref{sec2}, one gets the same time delay to the one obtained by the geodesic analysis.

%


\subsubsection{Scalar Particle in a Shock Wave}
The Klein-Gordon equation for  massless scalar field is given
\begin{equation}
\square\phi=0 ~.
\label{KG}
\end{equation}
In the shock wave background, Eq.(\ref{KG}) takes the form 
\begin{equation}
\partial_u\partial_v\phi+H(u,y)\partial^2_v\phi-\frac{1}{4}\partial^2_y\phi=0 ~,
\end{equation}
which is not solvable exactly but this is not required for our case. If we try to solve this differential equation near the wave, we can ignore the last term since it is small compared to the middle term ending up at
\begin{equation}
\partial_u\partial_v\phi+H(u,y)\partial^2_v\phi=0 ~.
\label{scalarp}
\end{equation}
By taking an integration in the $v$-coordinate and dropping the integral constant to have zero field in the $v\rightarrow \pm \infty$ limits,  then one gets
\begin{equation}
\partial_u\phi+H(u,y)\partial_v\phi=0 ~.
\label{scalarp1}
\end{equation}
We can solve this equation by the separation of variables technique. For this purpose, suppose that the  solution of the differential equation is in  the form $\phi(u,v,y)=U(u)V(v)Y(y)$. If we plug this into  Eq.(\ref{scalarp1}), we obtain
\begin{equation}
\frac{1}{H(u,y)}\frac{U'(u)}{U(u)}+\frac{V'(v)}{V(v)}=0 ~.
\end{equation}
Let $p_v=-i\partial_v$ be the momentum of a particle in the $v$ direction, so we have
\begin{equation}
\begin{aligned}
\frac{V'(v)}{V(v)}=-\frac{1}{H(u,y)}\frac{U'(u)}{U(u)}=ip_v ~.
\end{aligned}
\end{equation}
Finally, the solution can be written as
\begin{equation}
\phi(u,v,y)=Y(y)U(u_0)V(0)e^{ip_v \big (v-\int^u H(u',y)du'\big )} ~.
\end{equation}
Here we suppose that momentum of the test particle is known, and when scalar particle crosses the shock wave with an impact parameter $b$, it picks up a phase as 
\begin{equation}
\phi(0^{+},v,b)=e^{-ip_{v}\int_{0^{-}}^{0^{+}} du H(u,b)} \phi(0^{-},v,b)=e^{-ip_{v}\Delta v}  \phi(0^{-},v,b) ~,
\end{equation}
here the shift in the $v$ coordinate $\Delta v$ is the same obtained in Eq.(\ref{metfluc}).  Hence, the scalar particle when crossing the shock wave picks up a phase-shift analogy with the Aharanov-Bohm phase. Note that the same result is reproduced as Eq.(\ref{metfluc}) which was obtained by the geodesic analysis.

\subsubsection{Photon in a TMG Shock Wave}

Maxwell theory coupled to the shock wave produces the same result as  the null-geodesic and the scalar field case as was discussed above. Let us now  consider a 2+1 dimensional Maxwell theory coupled to gravity, it is called non-minimally coupling, is given by the action
\begin{equation}
S=-\frac{1}{4}\int d^3 x\sqrt{-g}\bigg(F_{\mu\nu} F^{\mu\nu}+\gamma R^{\mu\nu}\,_{\rho\sigma}F_{\mu\nu}F^{\rho\sigma}\bigg) ~,
\label{non_minimal_photon}
\end{equation}
where $F_{\mu\nu}$ is the field-strength tensor and $\gamma$ is a coupling constant with mass dimension $-2$. For the shock wave ansatz Eq.(\ref{metric1}), the field equations take the form
\begin{equation}
\nabla^\sigma F_{\rho\sigma}-\gamma R_\rho\,^{\sigma\mu\nu}\nabla_\sigma F_{\mu\nu}=0 ~, \hskip 1 cm  R_{uyuy}=-\frac{1}{2}\partial_y^2 H(u,y) ~.
\end{equation}
Explicitly, one can show that 
\begin{equation}
\partial_u F_{v y}+\bigg(H(u,y)+\gamma \partial_y^2 H(u,y)\bigg)\partial_v F_{v y} + \frac{1}{2} \partial_y F_{u v}=0 ~.
\label{waveqn1}
\end{equation}
Now, assume that $\epsilon_y$  is the linear transverse polarization vector of the wave.  The corresponding vector potential can be given as  $A_y=g(u,v)\epsilon_y$ and consequently the field-strength tensor is $F_{vy}=\partial_vg(u,v)\epsilon_y$. By using of these relations, the Eq.(\ref{waveqn1}) reduces the same form as Eq.(\ref{scalarp})
\begin{equation}
\partial_u\partial_v g(u,v)+\bigg(H(u,y)+\gamma \partial_y^2 H(u,y)\bigg)\partial^2_v g(u,v)=0 ~.
\end{equation}
Hence, $\Delta v$ can be calculated by the same way and then one obtains
\begin{equation}
\Delta v =-\frac{2\sigma \lvert p\rvert }{m_g}\big(1+\gamma  m_g^2 \big )e^{-m_g b} ~.
\label{delayphoton11}
\end{equation}
The first term in Eq.(\ref{delayphoton11}) is the same as obtained for the null geodesics and scalar particle, but we have extra second term. whereas the null geodesic or the scalar particle analysis,  choosing $\sigma$ is negative, does not guarantee time delay due to $\gamma m_g^2$ term. With the choice of $\gamma m_g^2$ term, theory may become  acausal  instead of causal. In fact this is an expected result since we have considered a theory non-minimally coupling of photon to gravity. Furthermore, It was shown that non-minimal coupling violates the strong equivalence principle and gives rise to superluminal propagation and  causality violations \cite{Drummond, Shore}. On the other hand, if coupling constant $\gamma$ is positive, theory gives a time-delay and for $ \gamma <0$, $\gamma m_g^2 > -1$ condition leads to a time-delay.

\subsubsection{ Graviton in a TMG Shock Wave} 
In this section, we consider the gravitons in the background of the shock-wave. For this purpose, we need to linearize the source-free field equations of TMG Eq.(\ref{tmg}) then one obtains
\begin{equation}
\sigma \delta G_{\mu\nu}+\frac{1}{\mu} \delta C_{\mu\nu}=0
\label{linearizedtmg},
\end{equation}
Consider the linearized field equations of TMG about the shock-wave background in the axial-like gauge ( which seems to be probably the best choice that we found after some trial and error) defined as
 \begin{equation}
 h_{v\mu}=0.
 \end{equation}
Then, the perturbation metric reads as  
\[ h_{\mu \nu}(u,v,y)= \left( \begin{array}{ccc}
g & 0 &  f \\
0 & 0 & 0 \\
 f & 0 & h\end{array} \right),
\]
there are six independent equations and to figure out these equations (see for details to Appendix \ref{chp:appendixc}), let us define the following functions
\begin{equation}
\begin{aligned}
&f(u,v,y) \equiv e^{-m_g y} \int^vdv' \int^{v'} s(u,v'') \, dv'' ,\\ \, 
&h(u,v,y) \equiv e^{-m_g y} \int^vdv' \int^{v'} r(u,v'') \, dv'' ,\\ \, 
&g(u,v,y) \equiv e^{-m_g y}  \int^vdv' \int^{v'} p(u,v'') \, dv'' ,\\ \, 
\end{aligned}
\end{equation}
where $s(u,v) $, $r(u,v) $ and $p(u,v) $ are arbitrary functions. By substituting these into TMG field equations, one gets
\begin{equation}
\begin{aligned}
&s(u,v)=e^{ip_v \Big (v-\frac{3}{2}\int^u  H(u',y) du' \Big )},\\
&r(u,v) =-\frac{1}{m_g}\partial_vs(u,v),\\
&p(u,v)=-\Big (\frac{ip_v H(u,y)}{2m_g}-\frac{i}{p_v}m_g \Big )s(u,v).
\end{aligned}
\end{equation}
 Observe that $g(u,v,y) $ and $h(u,v,y) $ can be defined in terms of  $f(u,v,y) $.  With the help of these equations, the solution can be written as
\begin{equation}
 \begin{aligned}
 f(0^{+},v,b)&=e^{ \frac{-3ip_v}{2}\int_{0^{-}}^{0^{+}}  H(u',y) du' } f(0^{-},v,b) 
 =e^{-ip_{v}\Delta v}  f(0^{-},v,b).
  \end{aligned}
\end{equation}
Then, time delay reads
\begin{equation}
  \Delta v =-\frac{3\lvert p\rvert }{m_g\sigma}e^{-m_g b},
\end{equation}
 which is positive definite for negative $\sigma $.  We see that this analysis does not bring any extra condition.
 
\subsection{Causality in NMG}
The action of NMG is \cite{BHT}
\begin{equation}
 I=\frac{1}{\kappa^2}\int d^3x\sqrt{-g}\bigg(\sigma R+\frac{1}{m^2}(R^2_{\mu\nu}-\frac{3}{8} R^2)\bigg),\,
 \label{QGaction}
\end{equation}
whose field equations are
\begin{equation}
\begin{aligned}
&\sigma(R_{\mu\nu}-\frac{1}{2}g_{\mu\nu}R)+\frac{1}{2m^2}\bigg(2\square R_{\mu\nu}-\frac{1}{2}\nabla_\mu\nabla_\nu R-\frac{1}{2}g_{\mu\nu}\square R+4R_{\mu\rho\nu\sigma}R^{\rho\sigma}\\&-g_{\mu\nu}R_{\rho\sigma}R^{\rho\sigma}
-\frac{3}{2}RR_{\mu\nu}+\frac{3}{8}g_{\mu\nu}R^2\bigg)=0.
\end{aligned}
 \label{nmgf}
\end{equation}
This theory has a massive spin-$2$ excitation with two degrees of freedom in both flat and (A)dS backgrounds.
For the shock wave metric, Eq.(\ref{nmgf}) transforms into the following form
\begin{equation}
-\sigma\partial_y{^2}H(u,y)-\frac{1}{m^2} \partial_y{^4}H(u,y)=2\lvert p \rvert \delta(y)\delta(u).
\label{eqnnmg}
\end{equation}
whose  the general solution is
\begin{equation}
\begin{aligned}
 H(u,y) =-\frac{ \lvert p \rvert\delta(u)}{ m\sigma}\bigg((e^ {-my}+my)\theta(y)+(e^ {my}-my)\theta(-y)+c_1y+c_2\bigg).
 \end{aligned}
\end{equation}
Observe that both the left and the right parts of the source are curved since NMG is a parity invariant theory, hence general solution has two step functions unlike the case of TMG. After gauge-fixing procedure as was conducted in TMG part, time delay can be obtained as
\begin{equation}
\begin{aligned}
 \triangle v =-\frac{ \lvert p \rvert}{ m\sigma}e^ {-m_g\lvert b\rvert},
 \label{NMGtimedelay}
 \end{aligned}
\end{equation}
which is positive for $\sigma<0$. Hence, causality and unitarity in NMG are not in conflict.
\subsubsection{Eikonal Scattering in NMG}
We can also calculate the time delay in NMG by using the Eikonal approximation. For this purpose, let us consider the following action
 
 \begin{equation}
\mathcal{S}_{NMG} = \int d^{3}x \left( \sqrt{-g} \sigma R + +\frac{1}{m^2}(R^2_{\mu\nu}-\frac{3}{8} R^2) - \frac{1}{2\alpha} k^\nu k_\nu + \frac{1}{2} g^{\mu\nu} \partial_{\mu}\phi \partial_{\nu}\phi\right) ~,
\end{equation}
By using  corresponding graviton propagator Eq.(\ref{propNMG}), scattering amplitude can be given as
\begin{equation}
\begin{aligned}
\mathcal{A}_{tree}=-\sigma \dfrac{s^{2}}{t}\,\frac{1}{ \Big (1 - \sigma \frac{q^{2}}{m_g^2}\Big) }.
\end{aligned}
\end{equation}
This procedure gives the same result obtained by the geodesic analysis.

\subsubsection{Photon in an NMG Shock Wave}
Now, let us consider the non-minimally coupled photon, described by the action Eq.(\ref{non_minimal_photon}), coupled to the NMG shock-wave, the same calculations as in the TMG case give a shift  $\Delta v$ as
\begin{equation}
\begin{aligned}
\Delta v &=-\frac{\sigma \lvert p\rvert }{m_g}(1+  \gamma m_g^2) e^ {-m_g\lvert b\rvert} ~.\\
\end{aligned}
\end{equation}
which may give time advance instead of time delay. If  $\gamma m_g^2  > -1$ and $\sigma<0$, the theory gives a time-delay for any $b \ne 0$ impact parameter.

\subsubsection{Graviton in an NMG Shock Wave}
For this calculation, we shall to a further simplification within the light-cone gauge and suppose that the perturbation is also traceless, in other words $ h=0$, otherwise the linearized equations are too complicated. Then the perturbation is simply given by 
\[ h_{\mu \nu}(u,v,y)= \left( \begin{array}{ccc}
g& 0 &  f \\
0 & 0 & 0 \\
 f & 0 & 0\end{array} \right).
\]
To solve resulting equations depicted in the Appendix \ref{Appdx31}, we define 
\begin{equation}
\begin{aligned}
&f(u,v,y) \equiv e^{-m y} \partial_vs(v,u), \\  
&g(u,v,y) \equiv -m e^{-m y} s(u,v),  
\end{aligned}
\end{equation}
where $s(u,v) $ is the arbitrary function. After plugging these into the NMG field equations, the solution follows as
\begin{equation}
\begin{aligned}
s(u,v)=e^{ip_v \Big (v-2\int^u  H(u',y) du' \Big )}.
\end{aligned}
\end{equation}
 We see that $g(u,v,y) $ can be written in terms of  $f(u,v,y) $. Then one gets
\begin{equation}
 \begin{aligned}
 f(0^{+},v,b)&=e^{ -2ip_v\int_{0^{-}}^{0^{+}}  H(u',y) du' } f(0^{-},v,b) =e^{-ip_{v}\Delta v}  f(0^{-},v,b).
  \end{aligned}
\end{equation}
Then finally, one obtains
\begin{equation}
  \Delta v =-\frac{2\lvert p\rvert }{m \sigma}e^{-m b},
\end{equation}
 which gives time delay for negative $\sigma$. Then causality and unitarity in NMG are not in conflict.

\subsubsection{Born-Infeld Gravity}
The action of Born-Infeld Gravity \cite{gullu} ( aclled BINMG) that has a massive spin-2 particle is
\begin{equation}
 I_{BI-NMG}=-4 m^2\int d^3x\sqrt{-g}F(R,K,S)+{\cal{L}}_{\mbox{matter}},
 \label{QGaction}
\end{equation}
where 
\begin{equation}
\begin{aligned}
 & F(R,K,S)=\sqrt{1-\frac{\sigma}{2m^2}(R+\frac{\sigma}{m^2}K-\frac{1}{12m^4}S)}-(1-\frac{\lambda}{2}),\\
 & K=R^2_{\alpha\beta}-\frac{R^2}{2}, \hskip 0.5cm S=8 R^{\alpha\beta}R_{\alpha\sigma}R^\sigma\,_{\beta}-6RR^2_{\alpha\beta}+R^3.
 \end{aligned}
\end{equation}
The field equations are long and hence we do not depict them here but note that for the shock wave metric,  curvature scalars are constant :
\begin{equation}
 R^2_{\alpha\beta}=0,\hskip .6 cm R_\beta\,^{\sigma}R_{\beta\sigma}=0, \hskip .6 cm R=0,
\end{equation}
reducing the field equations to an easy form of NMG 
\begin{equation}
 \frac{1}{2}T_{\mu\nu}=\sigma R_{\mu\nu}+\frac{\sigma^2}{m^2}\square R_{\mu\nu}.
 \label{BINMGshockeqn}
\end{equation}
Therefore, just like in the NMG, if $\sigma <0$, three dimensional Born-Infeld gravity is both unitary and causal. 

\subsection{Anti-de Sitter Space}

After studying the issue of causality in flat background, in this section, we will study this issue in AdS space with the similar method as in the case of flat space analysis. For this purpose, let us consider the AdS$_{3}$ metric in terms of Poincar\'e coordinates
 \begin{equation}
  ds^2_{\text{AdS}_{3}}=\dfrac{\ell^2}{y^2}(-2dudv+dy^2),
 \end{equation}
in which $y\in \mathbb{R}_{\geq 0}$, $u\in \mathbb{R}$, $v\in \mathbb{R}$. To adapt this metric to our case, let us consider the Kerr-Schild ansatz 
\begin{equation} \label{ansatz1}
 ds^{2}=\dfrac{\ell^2}{y^2}\Big(-2dudv-F(u,y)du^2+dy^2\Big),
\end{equation}
which describes a massless high energetic particle moving in the $+x$ direction. For this metric, corresponding energy momentum tensor for a massless particle can be given as 
\begin{equation}
T_{uu}=|p|\dfrac{\ell}{y_{0}}\delta(u)\delta(y-y_0). \label{ST}
\end{equation}

\subsection{Topologically Massive Gravity}

The source-free field equations of TMG are 
\begin{equation}\label{lasTMG}
-G_{\mu\nu} + \frac{1}{\ell^2} g_{\mu\nu} + \frac{1}{\mu} C_{\mu\nu} =0 ,
\end{equation}
where AdS$_3$ radius is defined as $\ell = 1/\sqrt{|\Lambda |}>0$. Notice that we have chosen the sign of Einstein term to be opposite, that is we take $\sigma =-1$. This choice is needed for ghost freedom about the AdS$_3$ vacuum. 

The source-coupled field 
equations  Eq.(\ref{lasTMG}), for the ansatz Eq.(\ref{ansatz1}), reduce to a single third order equation   
\begin{equation} \label{tmgeq}
-y\dfrac{\partial_{y}^{3}F}{2\ell \mu}-\dfrac{y^2\partial_{y}^2F-y\partial_{y}F}{2y^2}=|p|\dfrac{\ell}{y_{0}}\delta(u)\delta(y-y_{0}).
\end{equation}
The homogeneous part solution (complementary solution) of this equation can be given as \cite{Bending} 
\begin{equation} \label{homo}
 F_{h}(y)=c_1\left(\dfrac{y}{\ell}\right)^{1-\ell \mu}+c_2\left(\dfrac{y}{\ell}\right)^2+c_3,
\end{equation}
where $c_i$'s are functions of null coordinate $u$. On the other hand, the  $c_2$ and $c_3$ functions can be eliminated by a coordinate transformation \cite{Ayon}. One can find a viable solution to Eq.(\ref{tmgeq}) as follows: let us first assume that the generic structure of the solution is 
\begin{equation}
F_{p}(y)=\theta(y-y_0)g(y),
\end{equation}
such that $g(y)$ is a solution of Eq.(\ref{tmgeq}). Then, with the consistency condition at $y_0$, one can determine $c_i$. Subsequently, by inserting $F_{p}(y)$ in Eq.(\ref{tmgeq}) and accordingly performing the integral for the sector $y_{0}-\varepsilon$ and $y_{0}+\varepsilon$ and later dropping the delta function sections via the limit $\varepsilon \to 0$ as well as using the continuity conditions for $F(y)$ and ${F}'(y)$ at $y=y_0$, one will get
\begin{equation}
 {g}''(y_{0})=-2\mu\left(\dfrac{\ell}{y_0}\right)^2\delta(u)|p|,
\end{equation}
where we chose $g'(y_0)={g}(y_0)=0$ for ensuring the continuity of $F(y)$ and ${F}'(y)$. Consequently, the most general solution $F(y)=F_{h}(y)+F_{p}(y)$ can be given as
\begin{eqnarray}
 &F(y)= \ell^2\mu\dfrac{\delta(u)|p|}{1-(\ell \mu)^2}\left[2\left(\dfrac{y}{y_0}\right)^{1-\ell \mu}-(1-\ell \mu)\left(\dfrac{y}{y_0}\right)^2-(1+\ell \mu)\right]\theta(y-y_0)\nonumber \\[0.5em] 
 & + \ell^2\mu\dfrac{\delta(u)|p|}{1-(\ell \mu)^2}\left[2c_1\left(\dfrac{y}{y_0}\right)^{1-\ell \mu}+(1-\ell \mu)c_2\left(\dfrac{y}{y_0}\right)^2+(1+\ell \mu)c_3\right], \label{A39}
\end{eqnarray}
where undetermined coefficients $c_i$'s can be fixed by introducing the appropriate boundary conditions.

\subsubsection{The Flat Spacetime  Limit}

In this section, we will check the consistency of Eq.(\ref{A39}). To do so, one needs to take the flat space limit ($\Lambda \rightarrow 0$) and the result for the shock-wave profile must be the same as in the case of  flat space analysis. The flat space limit will also help us to impose boundary conditions. Before taking this limit, let us define a new coordinate
\begin{equation} \label{change}
 y=\ell e^{z/\ell},
\end{equation}
which reduces the AdS$_3$ metric to
\begin{equation}
 ds^2=-2 e^{-2z/\ell} dudv+dz^2,
\end{equation}
and then $\ell\rightarrow\infty$ limit can be used to obtain flat space. Furthermore, shock wave profile function defined in Section (\ref{CausTMG}) can be written in terms of the new profile function $F(y)$ as
\begin{equation}
 -\dfrac{\ell^2}{y^2}F(y)=H(y).
\end{equation}
Hence, in the  $\ell \to \infty $ limit, shock wave profile can be obtained as
\begin{eqnarray}
 &H(z)=\dfrac{\delta(u)|p|}{\mu}\left[2e^{-\mu(z-z_0)}-2+2\mu(z-z_0)\right]\theta(z-z_0)  \\[0.5em] \nonumber
 &  + \dfrac{\delta(u)|p|}{\mu}\left[2c_1e^{-\mu(z-z_0)}+(c_2+c_3)-\ell \mu(c_2-c_3)-2\mu c_2(z-z_0)\right], \nonumber
\end{eqnarray}
which is compatible with the result obtained in the flat space. we can use this result to set the undetermined coefficients in Eq.(\ref{A39}). Therefore, one needs to impose $c_1=0$ and $c_2=c_3=1$ to reproduce the same result as obtained in our flat space analysis.

\subsubsection{Asymptotically AdS$_3$ Boundary Conditions}

In the previous section, we found the integral constants in Eq.(\ref{A39}) by imposing flat space gauge fixing conditions. Let us now do the same analysis for asymptotically AdS$_3$ boundary conditions. For this purpose, one can consider the Brown-Henneaux (BH) boundary conditions \cite{BH} which is defined by the following linearized metric perturbations $h_{\mu \nu }=g_{\mu \nu}-g^{\text{AdS}}_{\mu \nu}$ 
\begin{equation}
 h_{uu}\simeq h_{uv}\simeq h_{vv}\simeq h_{yy}\simeq \mathcal{O}(1) , \quad h_{uy}\simeq h_{vy}\simeq \mathcal{O}(y).
\end{equation}
Here, $\mathcal{O}(y^n)$ stands for that any component decays as $y^n$ or faster as one goes toward the $y=0$ (that is, the boundary of AdS$_3$). With the help of these boundary conditions, one needs to impose $F(y)\sim\mathcal{O}(y^2)$. If we suppose that $\mu>{1}/{\ell}$, we need to choose $c_1=0$ to satisfy the BH boundary conditions. On the other side, to have regular AdS$_3$ space,  we need to set $c_2=c_3=1$. Notice that, the same values for integral constants are found by imposing BH boundary conditions. Then, the gauge-fixed solution can be written as
\begin{equation}
\begin{aligned}
 F(y)=& \ell^2\mu\dfrac{\delta(u)|p|}{1-(\ell \mu)^2}\bigg[2(\dfrac{y}{y_0})^{1-\ell \mu}\bigg]\theta(y-y_0) \\
& + \ell^2\mu\dfrac{\delta(u)|p|}{1-(\ell \mu)^2}\left[(1-\ell \mu)\left(\dfrac{y}{y_0}\right)^2+(1+\ell \mu)\right]\theta(-(y-y_0)). 
\end{aligned}
\end{equation}
Observe that $F(y)$ has $y^2$ dependence for $y<y_0$ since AdS$_{3}$ is defined in terms of different coordinates.

\subsubsection{Shapiro Time-delay in AdS}

In the light of above computations, we can calculate Shapiro time delay for massless particle moving in the shock wave geometry. The main idea is to show when particle crosses the shock wave the shifting in the coordinate $\Delta v$ positive or not. If $\Delta v$ is positive,  the particle experiences time delay; otherwise, it leads to causality violation. To compute shapiro time delay, we follow the same procedure  as in the case of the geodesic analysis in Section (\ref{CausTMG}). The Klein-Gordon equation for a massless scalar field in the shock-wave background takes the form
\begin{equation}
\partial_{u}\partial_{v}\phi+F(u,y)\partial_{v}^2\phi=0.
\end{equation}
By using of this equation, the time shift $\Delta v$ can be expressed by the following integral 
\begin{equation}
 \Delta v=\int_{0^{-}}^{0^{+}}du\ F(u,y) .
\end{equation}
For a massless particle traversing the shock-wave at $z>z_0$, one gets
\begin{equation}
 \Delta v=\dfrac{2\mu|p|}{m_g^2}e^{-m_g(z-z_0)}, \label{BoundTMG}
\end{equation}
where $m_g$ is the mass of spin-2 excitation is defined as $m_g^2=\mu^2  - 1/\ell^2$. In fact, shapiro time delay Eq.(\ref{BoundTMG}) is dependent on sign of the Einstein-Hilbert term in the action.  Here, the ghost-freedom about AdS vacuum required the sign of Einstein term to be opposite. Consequently, as we have shown in flat space analysis, $\Delta v$ is positive definite and causality and unitarity are not in conflict. On the other hand, if one takes $\ell \to \infty $ limit, this reproduces the flat space result Eq.(\ref{metfluc}).
  
 \subsubsection{Chiral Gravity}

In TMG, there is a conflict between the unitarity of the gravitons and the positive energy of the black holes. Namely, the positivity of the energy of bulk excitations and Brown-Henneaux boundary conditions cannot be simultaneously satisfied. A particular way out of this controversial issue was given in \cite{Chiral} where has shown that TMG is only consistent at the special limit, so called chiral point $\mu\ell = 1$ where graviton mass vanishes ($m_g=0$). At the chiral gravity limit, theory has only right-moving central charge of the boundary theory. On the other hand, chiral gravity has potentially problematic log mode solutions which do not satisfy BH boundary conditions \cite{GrumillerJohansson, GrumillerJohansson2}. Namely, in addition to BH boundary conditions, the theory has also other type of asymptotically AdS$_3$ boundary conditions. Therefore, in this section, we will impose both of the boundary conditions at the chiral point.

To find the shock wave profile function at the chiral point, firstly we need to take $\mu\ell\rightarrow1$  limit of Eq.(\ref{tmgeq}). For this purpose, let us consider the natural basis $\{ (y^{1-\ell \mu}-1)/(1-\ell \mu), y^2, 1\}$ for $F_h(y)$. At the chiral limit, the modes of $F_h(y)$ transform as $\{\log(y), y^2,1\} $. In this limit,  Eq.(\ref{tmgeq}) can be recast as
\begin{eqnarray}
 & F(y)=\dfrac{\delta(u)|p|}{\mu}\left[\log\left(\dfrac{y}{y_0}\right)-\dfrac{1}{2}\left(\left(\dfrac{y}{y_0}\right)^2-1\right)\right]\theta(y-y_0)\nonumber\\[0.5em]
 & +\dfrac{\delta(u)|p|}{\mu}\left[c_1\log\left(\dfrac{y}{y_0}\right)+c_2\left(\dfrac{y}{y_{0}}\right)^2+c_3\right]. \label{GHG}
\end{eqnarray}

As discussed above, at the chiral point, the theory admits two different boundary conditions. Let us first consider the BH boundary conditions: we can eliminate the $\sim \log (y)$ modes by imposing the BH boundary conditions, that is $c_1=0$. The other coefficients can be fixed by the choice of coordinates. Let gauge-fix them by removing the quadratic and constant terms for $y>y_0$. The gauge-fixed profile of chiral gravity is
 \begin{equation}
  F(y)=\theta(y-y_0)\dfrac{\delta(u)|p|}{\mu}\log\left(\dfrac{y}{y_0}\right)+\theta(y_0-y)\dfrac{\delta(u)|p|}{2\mu}\left[\left(\dfrac{y}{y_0}\right)^2-1\right].
 \end{equation}
Now we impose the other boundary conditions  given in \cite{GrumillerJohansson}, which in our coordinates would permit for $F(y)\sim\mathcal{O}(\log(y))$. If one imposes in Eq.(\ref{GHG}), one obtains 
 \begin{equation}
 \begin{aligned}
  F(y)=&\theta(y-y_0)\dfrac{\delta(u)|p|}{\mu}\log\left(\dfrac{y}{y_0}\right)+\dfrac{\delta(u)|p|}{2\mu}\left[c_1 \log\left(\dfrac{y}{y_0}\right)\right]\\&+\theta(y_0-y)\dfrac{\delta(u)|p|}{2\mu}\left[\left(\dfrac{y}{y_0}\right)^2-1\right],
 \end{aligned}
 \end{equation}
where  $c_1$ could not be gauge-fixed.  This phenomenon is the non-linear analog of the logarithmic modes of \cite{GrumillerJohansson}. 

\subsection{The New Massive Gravity}
Now, let us study the causality issue in NMG for asymptotically AdS spaces.  Parity even theory NMG has two massive spin-2 polarizations. The field equations of the NMG Eq.(\ref{nmgf}), which contains four derivative with respect to metric, are given. 
\begin{equation}
-G_{\mu\nu} + |\Lambda |g_{\mu\nu}+\frac{1}{2m^2}K_{\mu\nu} = 0, \label{Ufa}
\end{equation}
where K-tensor is defined as $K_{\mu\nu} = 2\square R_{\mu\nu}-(1/2)\nabla_{\mu}\nabla_{\mu}R-(1/2)g_{\mu\nu}\square R+4R_{\mu\alpha\nu\beta}R^{\alpha\beta}-(3/2)RR_{\mu\nu}-g_{\mu\nu}K$. It is also important to note that, whereas the TMG case, in NMG the AdS$_3$ radius $\ell $ depends not only $\Lambda $ but also  the mass parameter $m$ \cite{BHT}.
For the shock wave metric Eq.(\ref{ansatz1}), the field equations of NMG Eq.(\ref{Ufa}) reduces to
 \begin{equation}\label{nmgfield}
 \left[y^4\partial_{y}^{4}F+2y^3\partial_{y}^{3}F-\dfrac{(1+2\ell^2 m^2)}{2}(y^2\partial_{y}^{2}F-y\partial_{y}F)\right]\dfrac{1}{2\ell^2 m^2 y^2}=|p|\dfrac{\ell}{y_{0}}\delta(u)\delta(y-y_{0}),
 \end{equation}
whose complementary solution is
\begin{equation} \label{nmghomo}
 F_{h}(y)=c_{+}\left(\dfrac{y}{\ell}\right)^{1+\beta}+c_{-}\left(\dfrac{y}{\ell}\right)^{1-\beta}+c_2\left(\dfrac{y}{\ell}\right)^2+c_3,
\end{equation}
where $c_{\pm}$ are functions depending the null coordinates $u$ and $\beta$ is defined as $\beta=\sqrt{1/2+\ell^2 m^2}$. As was done above, the generic solution can similarly be found as follows: considering the generic structure $F_{p}=\theta(y-y_{0})g(y)$ with $g(y)$ is a solution of Eq.(\ref{nmgfield}) and then following above-given track, one will arrive at $g(y_0)= {g}'(y_0)= {g}''(y_0)=0$ with which one will finally obtain
\begin{eqnarray}
{g}'''(y_0)=2m^2|p|\left(\dfrac{\ell}{y_0}\right)^{3}\delta(u).
\end{eqnarray}
The most general solution to Eq.(\ref{nmgfield}) reads
\begin{eqnarray}
 &  F(y)=\theta(y-y_0)m^2|p|\delta(u)\left(\dfrac{\ell^3}{\beta^2-1}\right)\nonumber \\[0.5em]
 &\times \left[1-\left(\dfrac{y}{y_0}\right)^2+\dfrac{1}{\beta}\left(\dfrac{y}{y_0}\right)^{1+\beta}-\dfrac{1}{\beta}\left(\dfrac{y}{y_0}\right)^{1-\beta}\right] \nonumber \\[0.5em]
 & m^2|p|\delta(u)\left(\dfrac{\ell^3}{\beta^2-1}\right)\left[c_1-c_2\left(\dfrac{y}{y_0}\right)^2+\dfrac{c_3}{\beta}\left(\dfrac{y}{y_0}\right)^{1+\beta}-\dfrac{c_4}{\beta}\left(\dfrac{y}{y_0}\right)^{1-\beta}\right],
\end{eqnarray}
where $c_4$ is a  constant coefficient. As in the case of TMG, these functions  can be fixed by imposing appropriate boundary conditions. To determine boundary conditions, we can apply again the flat space limit of shock wave profile. With the help of this, one can obtain 
\begin{eqnarray}
 &H(z)=\theta(z-z_0)|p|\delta(u)\left[2(z-z_0)-\dfrac{1}{m}\left(e^{m(z-z_0)}-e^{-m(z-z_0)}\right)\right] \nonumber \\[0.5em]
 & + |p|\delta(u)\left[-\ell(c_1-c_2)+2c_2(z-z_0)-\dfrac{1}{m}\left(c_3e^{m(z-z_0)}-c_4e^{-m(z-z_0)}\right) \right].
 \label{HProfile}
\end{eqnarray}
Now, we can fix  undetermined constants by demanding the spacetime to be asymptotically flat in the limit $\ell \rightarrow \infty$.  Due to the fact that, one needs to set $c_3=-1$, $c_4=0$, and $c_1=c_2$. In addition to that, one needs to choose $c_1=c_2=-1$ to have Cartesian coordinates at $z>z_0$. Finally, Eq.(\ref{HProfile}) can be recast as
\begin{equation}
H(z)=\dfrac{|p|\delta(u)}{m}e^{-m|z-z_0|}-\theta(-(z-z_0))\dfrac{2|p|\delta(u)}{m}(z-z_0),
\end{equation}
which is the same as obtained in the flat space analysis. Furthermore, as in the case of TMG,  $c_i$ functions can be determined by  demanding BH boundary conditions for a finite $\ell$ instead of  the spacetime to be asymptotically flat in the limit $\ell \rightarrow \infty$. Now let us show that both ways give the same fixing conditions for $c_i$ functions: For the sake of simplicity, suppose that $m^2>1/(2\ell^2)$, leads to $\beta>1$, so we set $c_4=0$. On the other hand, we need to set $c_3=-1$ to have regular AdS$_3$ deep into the bulk (i.e. $y\rightarrow\infty$). Finally, one needs to set $c_1=c_2=-1$ to describe AdS$_3$ in the usual coordinates at $y\rightarrow\infty$. Therefore, the shock wave profile for finite $\ell$ can be written as 
\begin{equation}
\begin{aligned}
 F(y)=\left(\dfrac{m^2|p|\ell^3}{\beta(1-\beta^2)}\right)\bigg[&\theta(y-y_0)\left(\dfrac{y}{y_0}\right)^{1-\beta}
+\theta(y_0-y)\bigg( \left(\dfrac{y}{y_0}\right)^{1+\beta}\\&+\beta \left( \left(\dfrac{y}{y_0}\right)^2-1 \right) \bigg) \bigg].
\end{aligned}
\end{equation}
Shapiro time delay can be calculated with the same procedure follows as before and time shift when particle crosses the shock wave at $z>z_0$ is
\begin{equation}\label{shiftnmg}
 \Delta v=\left(\dfrac{2m\ell^2}{2m^2\ell^2-1}\right)\dfrac{|p|}{\sqrt{1+1/(2m^2\ell^2)}}e^{\left(1/\ell-m\sqrt{1+1/(2m^2\ell^2)}\right)(z-z_0)},
\end{equation}
which can also be written:
\begin{equation}\label{shiftnmg2}
 \Delta v= \left(\dfrac{2m\ell^2}{2m^2\ell^2-1}\right)\dfrac{|p|}{\sqrt{1+1/(2m^2\ell^2)}} e^{-m_{g}(z-z_0)},
\end{equation}
where the effective mass $m_{g}$ is defined as $m_{g}= m\sqrt{1+1/(2m^2\ell^2)}-1/\ell$. Time delay in NMG is positive, and taking the $\ell\to\infty $ limit which yields the flat space result (where $m_{\text{g}}=m$).

\subsubsection{Critical Points of NMG}
As was discussed in the TMG, at the special points, that is  graviton mass $m_g$ vanishes, we can consider the critical points of NMG. This can be thought as the analogy of chiral point of TMG.  Unlike the TMG case, NMG has two critical points $m^2\ell^2=1/2$ ($\beta = 1$) and $m^2\ell^2= -1/2$ ($\beta =0 $). Furthermore, the boundary theory has left and right central charges at these special points. Let us consider both critical points, respectively.

First of all, for the first critical point ($\beta =1$), the natural basis of $F_h(y)$ can be taken as $\{y^2\log(y),\log(y), y^2,1\} $. 
In NMG, in addition, there exist another critical point, which corresponds to $m^2\ell^2= -1/2$ (that is, $\beta =0 $). Taking the $\beta \to 1$ limit of Eq.(\ref{nmgfield}) yields
\begin{eqnarray}
  & F(y)=\theta(y-y_0)|p|\delta(u)\dfrac{\ell}{4}\left(\log\left(\dfrac{y}{y_0}\right)\left[\left(\dfrac{y}{y_0}\right)^2+1\right]+\left[1-\left(\dfrac{y}{y_0}\right)^2\right]\right) \\[0.5em] \nonumber
  & +|p|\delta(u)\dfrac{\ell}{4}\left(\log\left(\dfrac{y}{y_0}\right)\left[c_1\left(\dfrac{y}{y_0}\right)^2+c_2\right]+c_3+c_4\left(\dfrac{y}{y_0}\right)^2\right) .
\end{eqnarray}
Just like in TMG, at the critical point, the theory admits many possible boundary conditions that can be imposed. Imposing the BH boundary conditions yields
 \begin{eqnarray}
  & F(y)=\theta(y-y_0)|p|\delta(u)\dfrac{\ell}{4}\log\left(\dfrac{y}{y_0}\right)\left[\left(\dfrac{y}{y_0}\right)^2+1\right] \\[0.5em] \nonumber
  & +|p|\delta(u)\dfrac{\ell}{4}\left(c_1\log\left(\dfrac{y}{y_0}\right)\left(\dfrac{y}{y_0}\right)^2\right)+\theta(y_0-y)|p|\delta(u)\dfrac{\ell}{4}\left(\left(\dfrac{y}{y_0}\right)^2-1\right),
 \end{eqnarray}
where the function $c_1$ could not be determined by gauge fixing procedure. On the other hand, if one imposes the weakened boundary conditions which is given by \cite{GrumillerJohansson}, one obtains
 \begin{eqnarray}
  & F(y)=\theta(y-y_0)|p|\delta(u)\dfrac{\ell}{4}\log\left(\dfrac{y}{y_0}\right)\left[\left(\dfrac{y}{y_0}\right)^2+1\right] \\[0.5em] \nonumber
  & +|p|\delta(u)\dfrac{\ell}{4}\log\left(\dfrac{y}{y_0}\right)\left[c_1\left(\dfrac{y}{y_0}\right)^2+c_2\right]+\theta(y_0-y)|p|\delta(u)\dfrac{\ell}{4}\left(\left(\dfrac{y}{y_0}\right)^2-1\right),
 \end{eqnarray}
which contains additional logarithmic modes with coefficient $c_2$.

On the other side, for other critical point ($\beta = 0$), for the homogeneous solution $F_h(y)$  one can set the natural basis as $\{y\log(y), y,  y^2, 1\}$. Notice that, this critical point forces the $m^2$ to be negative which corresponds to partially massless point for  de Sitter solutions. Therefore, in the $\beta \to 0$ limit,    Eq.(\ref{nmgfield}) reduces to
\begin{eqnarray}
   & F(y)=\theta(y-y_0)|p|\delta(u)\dfrac{\ell}{2}\left(2\dfrac{y}{y_0}\log\left(\dfrac{y}{y_0}\right)+\left[1-\left(\dfrac{y}{y_0}\right)^2\right]\right) \nonumber \\[0.5em]
   & +|p|\delta(u)\dfrac{\ell}{2}\left(2c_1\left(\dfrac{y}{y_0}\right)\log\left(\dfrac{y}{y_0}\right)+c_2\left(\dfrac{y}{y_0}\right)+c_3+c_4\left(\dfrac{y}{y_0}\right)^2\right),
\end{eqnarray}
which also contains $\sim y$ type logarithmic term which is characteristic of conformal gravity \cite{conformal}. 

\section{Conclusions}
In this thesis, we have firstly constructed Weyl-invariant version of TMG and shown that weyl invariant version of gravitational Chern-Simons term produces Abelian Chern-Simons term. This result, integrated with Weyl-gauged version of Maxwell and Einstein theories, leads to Weyl-invariant TMG coupled to TME-Proca theory. We have studied the perturbative spectrum of the Weyl-invariant TMG and its particle spectrum in detail. Both spin-$2$ and spin-$1$ particles acquire masses via symmetry breaking of the Weyl's symmetry either spontaneously in $AdS$ vacua as in the Higgs mechanism or radiatively in flat vacuum in an analogy with the Coleman-Weinberg mechanism. One can ask that whether chiral gravity appears in broken phase or not. We have considered this issue and see that chiral gravity does not appears in AdS backgrounds as a critical point of Weyl-invariant TMG such that it violates the positivity of the central charges.      

Secondly, we have studied cosmological TMG integrated with Fierz-Pauli mass term in maximally symmetric backgrounds. We have found the particle spectrum of the theory and shown that there is no unitarily Fierz-Pauli extension of chiral gravity. We also calculated tree-level scattering amplitude and associated non-relativistic potential energy between two covariantly conserved point-like spinning sources for various theories in flat spacetimes. We attained the gravitational anyon behaviours of particles due to presence of gravitational Chern-Simons term.  We also studied the flat space chiral limit of the scattering amplitude, yields trivial amplitude. 

Finally, we have discussed the issue of local causality in $2+1$ dimensional gravity theories  both in asymptotically Minkowski and asymptotically $AdS_3$ spacetimes.  We have calculated the Shapiro's time delay for massive gravity theories and unitarity and causality requirements bring the constraint on Newton's constant. We have shown that Einstein's gravity, topologically massive gravity and the new massive gravity are causal as long as the sign of the Newton's constant is negative. We have also studied the same problem in the Born-infeld gravity, which is extension of NMG, showing that the situation is similar as in the case of NMG and TMG. It is refreshing to see that causality and unitarity are not in conflict
for the massive gravity theories. This is in sharp contrast to the Einstein-Gauss Bonnet and cubic theories in higher dimensions ($n\geq 4$) where causality and unitarity are in conflict. 
 
\section{Acknowledgements} 
First of all, I would like to express my most sincere gratitude to my supervisor Professor Bayram Tekin for his long-lived guidance, advices, encouragements, motivations and tremendous expertise in the field. His unique experience and knowledge have always encouraged me to continue learning and researching. His guidance and wisdom contributed remarkably to the production of this study. One could not wish for a better supervisor.

My special thanks go to my close friend, room-mate and collaborator Dr. Suat Dengiz for his all support during my PhD and also his critical reading and comments on the thesis. 

I would like to thank my collaborator Gökhan Alka\c{c} for his help, support, hospitality and allowing me stay his home when I came to the Groningen for a one year as a guest researcher. I am very thankful to Groningen University for hospitality during my visit. I would also like to thank to the following people who I have interacted throughout my PhD: Professor Atalay Karasu, Professor Namık Kemal Pak, Professor A. Nihat Berker, Professor Eric Bergshoeff,  Professor Yıldıray Ozan,  Assoc. Professor Tahsin \c{C}. {\c{S}}i{\c{s}}man, Assoc. Professor Ibrahim G\"{u}ll\"{u}, Assoc. Professor Çetin Şentürk, Dr. Shankhadeep Chakrabortty, Dr. Cesim Dumlu, M. Gökhan Şensoy, M. Mira\c{c} Serim, M. Cem Lider, Mustafa Tek, Emel Alta{\c{s}}, Kezban Ta{\c{s}}seten Ata, Deniz \"Ozen, Şahin Kürek\c{c}i.

A special paragraph must be devoted my beloved family for their continuous and unflagging love, encouragements and supports through my studies. I am grateful to my wife Ebru and son Yi\u{g}it Feza who have been light of my life. Their unconditional love, limitless patience and company during this journey have helped me towards the completion of this study. To my parents, thank you for your help, support and inspiring me to follow my dreams.

During my Ph.D. education, I have been supported by The Scientific and Technological Research Council of Turkey (T\"{U}B\.{I}TAK) with the scholarships in two $2211$ and $2214-A$  International
Doctoral Research Fellowship Programme No:1059B141401079.

\appendix

\section{Field Equations of Topologically Massive Gravity}
\label{chp:appendixa}
The Lagrangian density of Topologically massive gravity is given by 
 \begin{equation}
 \begin{aligned}
 {\cal{L}}_{TMG}=\sqrt{-g} \Bigg \{\,  R +\frac{1}{2 \mu} \, \eta^{\mu \nu \alpha} \Gamma^\beta{_{\mu \sigma}} \Big (\partial_\nu \Gamma^\sigma{_{\alpha \beta}}+\frac{2}{3} \Gamma^\sigma{_{\nu \lambda}}  \Gamma^\lambda{_{\alpha \beta}} \Big ) \Bigg \}.
\label{TMGFP1}
\end{aligned}
\end{equation}
In order to obtain the field equations, by varying the Lagrangian density with respect to metric, one will get 
\begin{equation}
\delta{\cal{L}}_{TMG}=\delta{\cal{L}}_{EH}+\delta{\cal{L}}_{CS},
\end{equation}
where 
\begin{equation}
\begin{aligned}
&{\cal{L}}_{EH}=\sqrt{-g}R,\\&
{\cal{L}}_{CS}=\frac{1}{2 \mu} \, \epsilon^{\mu \nu \alpha} \Gamma^\beta{_{\mu \sigma}} \Big (\partial_\nu \Gamma^\sigma{_{\alpha \beta}}+\frac{2}{3} \Gamma^\sigma{_{\nu \lambda}}  \Gamma^\lambda{_{\alpha \beta}} \Big ).
\end{aligned}
\end{equation}
Let us study these terms separately.
\subsection*{Variation of Einstein Hilbert term:}
The variation of the Einstein-Hilbert term with respect to $g^{\mu\nu}$ becomes 
\begin{equation}
\begin{aligned}
 \delta{\cal{L}}_{EH}&=R\, \delta \sqrt{-g} +\sqrt{-g}\, \delta R\\
 &=R\,\delta \sqrt{-g} +\sqrt{-g}R_{\mu\nu}\,\delta g^{\mu\nu}+\sqrt{-g}g^{\mu\nu}\, \delta R_{\mu\nu}.
\label{varofI1}
\end{aligned}
\end{equation}
First of all, the variation of the first part is $(\delta \sqrt{-g})=-\frac{1}{2} \sqrt{-g} \,\, g_{\mu\nu} \delta g^{\mu\nu}$. 
On the other side, the second term on the right hand side is already in the desired form, but the third term is not. To vary the third term, one needs to use the Palatini identity 
\begin{equation}
 \delta R_{\mu\nu}=\nabla_\alpha \delta \Gamma^\alpha_{\mu\nu}-\nabla_\mu \delta \Gamma^\alpha_{\alpha\nu},
 \label{palantr}
\end{equation}
with which the third term can be evaluated as
 \begin{equation}
 \sqrt{-g}g^{\mu\nu}\, \delta R_{\mu\nu}=\partial_\alpha\bigg(\sqrt{-g}  g^{\mu\nu}\delta \Gamma^\alpha_{\mu\nu}-\sqrt{-g}g^{\mu\alpha} \delta \Gamma^\alpha_{\alpha\nu}\bigg).
 \label{Pal1}
 \end{equation}
Here we have used $\sqrt{-g}\nabla_\mu V^\mu=\partial_\mu(\sqrt{-g}V^\mu)$ where $V^\mu$ is a vector. Thus, by using Eq.(\ref{Pal1}), one can obtain the variation of Einstein-Hilbert term as follows  
\begin{equation}
 \delta{\cal{L}}_{EH}= \sqrt{-g}\,\delta g^{\mu\nu}\bigg(R_{\mu\nu}-\frac{1}{2}\mbox{g}^{\mu\nu}R\bigg) +\partial_\alpha J_{EH}^\alpha,
\end{equation}
where $J_{EH}^\alpha=\sqrt{-g}\, g^{\mu\nu}\delta \Gamma^\alpha_{\mu\nu}-\sqrt{-g}\,g^{\mu\alpha} \delta \Gamma^\alpha_{\alpha\nu}$ is the corresponding boundary term. 
\subsection*{Variation of Chern-Simons term:}
The variation of the Chern-Simons term with respect to $\mbox{g}^{\mu\nu}$ reads 
\begin{equation}
\begin{aligned}
\delta L_{CS}=& \epsilon^{\lambda\mu \nu} \delta\Gamma^\rho{_{\lambda \sigma}} \Big (\partial_\mu \Gamma^\sigma{_{\rho \nu}}+\frac{2}{3} \Gamma^\sigma{_{\mu \beta}}  \Gamma^\beta{_{\nu \rho}} \Big )\\&+\epsilon^{\lambda\mu \nu } \Gamma^\rho{_{\lambda \sigma}} \Big (\partial_\mu \delta\Gamma^\sigma{_{\rho \nu}}+\frac{2}{3}\delta \Gamma^\sigma{_{\mu \beta}}  \Gamma^\beta{_{\nu \rho}}+\frac{2}{3} \Gamma^\sigma{_{\mu \beta}}  \delta\Gamma^\beta{_{\nu \rho}}  \Big ).
\label{CSeq1}
\end{aligned}
\end{equation}
Here, the multiplicative term $\frac{1}{2\mu}$ is suppressed. Note that the second and fourth terms in Eq.(\ref{CSeq1}) can respectively be recast as follows  
\begin{equation}
\begin{aligned}
&\frac{2}{3}\epsilon^{\lambda\mu \nu } \Gamma^\rho{_{\lambda \sigma}} \delta \Gamma^\sigma{_{\mu \beta}}  \Gamma^\beta{_{\nu \rho}}=
\frac{2}{3}\epsilon^{\lambda\mu \nu} \Gamma^\sigma{_{\mu \beta}}  \Gamma^\rho{_{\lambda \sigma}}  \delta\Gamma^\beta{_{\nu \rho}},\\&
\frac{2}{3}\epsilon^{\lambda\mu \nu} \delta\Gamma^\rho_{\lambda \sigma}  \Gamma^\sigma_{\mu \beta}  \Gamma^\beta_{\nu \rho}=\frac{2}{3}\epsilon^{\lambda\mu \nu} \Gamma^\rho_{\lambda \sigma}\Gamma^\sigma_{\mu \beta}  \delta\Gamma^\beta_{\nu \rho},
\end{aligned}
\end{equation}
with which Eq.(\ref{CSeq1}) takes the form
\begin{equation}
\delta L_{CS}= \epsilon^{\lambda\mu \nu}\bigg( \delta\Gamma^\rho_{\lambda \sigma}\partial_\mu \Gamma^\sigma_{\rho \nu}+\Gamma^\rho_{\lambda \sigma} \partial_\mu \delta\Gamma^\sigma_{\rho \nu}+2\Gamma^\rho_{\lambda \sigma} \Gamma^\sigma{_{\mu \beta}}  \delta\Gamma^\beta{_{\nu \rho}}\bigg).  
\label{CSeq3}
\end{equation}
To simplify further, notice that the second term in Eq.(\ref{CSeq3}) can also be written as
\begin{equation}
 \epsilon^{\lambda\mu \nu} \Gamma^\rho_{\lambda \sigma} \partial_\mu \delta\Gamma^\sigma_{\rho \nu}=\partial_\mu(\epsilon^{\lambda\mu \nu} \Gamma^\rho_{\lambda \sigma}\delta\Gamma^\sigma_{\rho \nu})+\epsilon^{\lambda\mu \nu}\delta\Gamma^\rho_{\lambda \sigma} \partial_\mu \Gamma^\sigma_{\rho \nu}.
\end{equation}
Thus, by substituting this back into Eq.(\ref{CSeq3}), one gets
\begin{equation}
\delta L_{CS}= \partial_\mu(\epsilon^{\lambda\mu \nu} \Gamma^\rho_{\lambda \sigma}\delta\Gamma^\sigma_{\rho \nu})+2\epsilon^{\lambda\mu \nu}\delta\Gamma^\rho_{\lambda \sigma}\bigg( \partial_\mu \Gamma^\sigma_{\rho \nu}+\Gamma^\sigma{_{\mu \beta}}\Gamma^\beta{_{\nu \rho}}\bigg).  
\label{CSeq5}
\end{equation}
Recall that Riemann tensor is defined as follows
\begin{equation}
R^\rho{_{\sigma\mu\nu}}=\partial_\mu\Gamma^\rho_{\nu \sigma}+\Gamma^\rho_{\mu \lambda}\Gamma^\lambda_{\nu \sigma}+\mu \leftrightarrow \nu.
\end{equation}
Thereupon, by virtue of this definition, Eq.(\ref{CSeq5}) reduces to
\begin{equation}
\delta L_{CS}= \partial_\mu(\epsilon^{\lambda\mu \nu} \Gamma^\rho_{\lambda \sigma}\delta\Gamma^\sigma_{\rho \nu})+\epsilon^{\lambda\mu \nu}\delta\Gamma^\rho_{\lambda \sigma}R^\sigma{_{\rho\mu\nu}}.
\label{CSeq66}
\end{equation}
On the other hand, in $2+1$ dimensions, Riemann tensor can be expressed in terms of Ricci tensor as
\begin{equation}
R^\sigma{_{\rho\mu\nu}}=\delta^\sigma_\mu \tilde{R}_{\rho\nu}+g_{\rho\nu}\tilde{R}^\sigma{_{\mu}}
-\delta^\sigma_\nu \tilde{R}_{\rho\nu}-g_{\rho\nu}\tilde{R}^\sigma{_{\mu}},
\end{equation}
where $\tilde{R}_{\mu\nu}=R_{\mu\nu}-\frac{1}{4}g_{\mu\nu}R$. With this unique property, the second term in the Eq.(\ref{CSeq66}) can be rewritten as
\begin{equation}
\epsilon^{\lambda\mu \nu}\delta\Gamma^\rho_{\lambda \sigma}R^\sigma{_{\rho\mu\nu}}=2\epsilon^{\lambda\mu \nu}\tilde{R}^\sigma{_{\mu}}\nabla_\lambda\delta g_{\nu\sigma}.
\end{equation}
Accordingly, with the help of this, Eq.(\ref{CSeq66}) reduces to 
\begin{equation}
\delta L_{CS}= \partial_\mu(\epsilon^{\lambda\mu \nu} \Gamma^\rho_{\lambda \sigma}\delta\Gamma^\sigma_{\rho \nu})+2\epsilon^{\lambda\mu \nu}\tilde{R}^\sigma{_{\mu}}\nabla_\lambda\delta g_{\nu\sigma}.
\label{CSeq6}
\end{equation}
By using  the expansion of $\nabla_\lambda\delta g_{\nu\sigma}$ and reorganizing the dummy indices, Eq.(\ref{CSeq6}) can be rewritten as
\begin{equation}
\delta L_{CS}= \partial_\mu\bigg(\epsilon^{\lambda\mu \nu} \Gamma^\rho_{\lambda \sigma}\delta\Gamma^\sigma_{\rho \nu}-2\epsilon^{\lambda\mu \nu}\tilde{R}^\sigma{_{\lambda}}\delta g_{\nu\sigma}\bigg)-2\delta g_{\nu\sigma}\epsilon^{\lambda\mu \nu}\nabla_\lambda\tilde{R}^\sigma{_{\mu}},
\label{CSeq7}
\end{equation}
By keeping in mind the cotton tensor 
\begin{equation}
\begin{aligned}
C^{\mu\nu}&=\frac{\epsilon^{\lambda\mu \nu}}{\sqrt{-g}}\nabla_\lambda\tilde{R}^\sigma{_{\mu}},\\&=\eta^{\lambda\mu \nu}\nabla_\lambda\tilde{R}^\sigma{_{\mu}}.
\end{aligned}
\end{equation}
Finally, one will eventually get
\begin{equation}
\delta L_{CS}= -\frac{1}{\mu}\sqrt{-g}\delta g_{\nu\sigma}C^{\nu\sigma}+\partial_\mu j_{CS}^\mu,
\label{CSeq8}
\end{equation}
where $j_{CS}^\mu$ is a boundary term for Chern-Simons term and given as $j_{CS}^\mu=\frac{1}{2\mu}(\epsilon^{\lambda\mu \nu} \Gamma^\rho_{\lambda \sigma}$ $\delta\Gamma^\sigma_{\rho \nu}-2\epsilon^{\lambda\mu \nu}\tilde{R}^\sigma{_{\lambda}}\delta g_{\nu\sigma})$. Note that we re-added the term $\frac{1}{2\mu}$. 
Together with the Einstein-Hilbert term, the variation takes the following form
\begin{equation} 
\delta L_{TMG}=  \sqrt{-g}\,\delta g^{\mu\nu}\bigg(R_{\mu\nu}-\frac{1}{2}g^{\mu\nu}R+\frac{1}{\mu}C_{\mu\nu}\bigg)+\partial_\mu\bigg(j^\mu_{EH}+j^\mu_{CS}\bigg).
\label{CSeq9}
\end{equation}
Consequently, up to a boundary term, the vacuum field equations of TMG can be obtained
\begin{equation}
R_{\mu\nu}-\frac{1}{2}g^{\mu\nu}R+\frac{1}{\mu}C_{\mu\nu}=0.
\end{equation}

\section{Topologically Massive Gravity with Fierz-Pauli mass Term in (A)dS Background}
\label{chp:appendixb}
The action of Topologically massive gravity with a Fierz-Pauli mass term is given by 
 \begin{equation}
 \begin{aligned}
 {I}=\int d^3x \sqrt{-g} \, \Big \{& \frac{1}{ \kappa} ( R -2 \Lambda)-\frac{m^2}{4 \kappa}(h^2_{\mu\nu}-h^2)\\&+\frac{1}{2 \mu} \, \eta^{\mu \nu \alpha} \Gamma^\beta{_{\mu \sigma}} \Big (\partial_\nu \Gamma^\sigma{_{\alpha \beta}}+\frac{2}{3} \Gamma^\sigma{_{\nu \lambda}}  \Gamma^\lambda{_{\alpha \beta}} \Big )+ {\cal L}_{matter} \Bigg \},
\label{TMGFP1}
\end{aligned}
\end{equation}
whose field equations read 
\begin{equation}
 \frac{1}{\kappa} (R_{\mu \nu}-\frac{1}{2} g_{\mu \nu} R+ \Lambda g_{\mu \nu})+\frac{1}{\mu} C_{\mu \nu}+\frac{m^2}{2 \kappa}(h_{\mu\nu}-g_{\mu\nu}h) =\tau_{\mu \nu}.
\label{tmgfpfeq}
 \end{equation}
The Cotton tensor $ C_{\mu \nu} $ can be rewritten in the following explicitly symmetric version 
\begin{equation}
 C^{\mu \nu}=\frac{1}{2} \eta^{\mu \alpha \beta} \nabla_\alpha G^\nu{_\beta}+\frac{1}{2} \eta^{\nu \alpha \beta} \nabla_\alpha G^\mu{_\beta},
 \label{linearcottmgfp}
\end{equation}
with which the linearization of the TMG field equation in Eq.(\ref{TMGFP1}) about a generic background $ g_{\mu\nu}=\bar{g}_{\mu\nu}+h_{\mu\nu} $ reads
\begin{equation}
 \frac{1}{\kappa} {\cal G}^L_{\mu \nu}+\frac{1}{2 \mu} \eta_{\mu \alpha \beta} \bar{\nabla}^\alpha {\cal G}^L_{\nu}{^\beta}+\frac{1}{2 \mu} \eta_{\nu \alpha \beta} \bar{\nabla}^\alpha {\cal G}^L_{\mu}{^\beta}+\frac{m^2}{2 \kappa}(h_{\mu\nu}-\bar{g}_{\mu\nu}h)=T_{\mu \nu}.
\label{linertmgfpfeq}
\end{equation}
To write all the curvature tensors in terms of the source terms, one needs to take the divergence of Eq.(\ref{linertmgfpfeq}) with which one gets 
\begin{equation}
\bar{\nabla}^\mu h_{\mu\nu}-\bar{\nabla}_\nu h=0 ,
\end{equation}
that gives $ R^L=-2 \Lambda h$. With the help of these as well as trace of the linearized field equations, one can easily obtain
\begin{equation}
 h=\frac{\kappa}{\Lambda-m^2} T, \qquad {\cal G}^L=\frac{\Lambda \kappa}{\Lambda-m^2} T.
\end{equation}
Let us now multiply Eq.(\ref{linertmgfpfeq}) by $ \eta^{\mu \sigma \rho}\bar{\nabla}_\sigma $, gives 
\begin{equation}
\begin{aligned}
 & \frac{1}{\kappa} \eta^{\mu \sigma \rho} \bar{\nabla}_\sigma {\cal G}^{L}_{\mu \nu}+\frac{1}{2 \mu} \eta^{\mu \sigma \rho} \eta_{\mu \alpha \beta} \bar{\nabla}_\sigma  \bar{\nabla}^\alpha {\cal G}^{L}_{\nu}{^\beta}
 +\frac{1}{2 \mu} \eta^{\mu \sigma \rho} \eta_{\nu \alpha \beta} \bar{\nabla}_\sigma  \bar{\nabla}^\alpha {\cal G}^{L}_{\mu}{^\beta} \\
 &+\frac{m^2}{2 \kappa} \eta^{\mu \sigma \rho} \bar{\nabla}_\sigma (h_{\mu\nu}-\bar{g}_{\mu\nu}h)= \eta^{\mu \sigma \rho} \bar{\nabla}_\sigma T_{\mu \nu}.
\label{lineartmgfp2}
\end{aligned}
\end{equation}
The second term in Eq.(\ref{lineartmgfp2}) can be recast as
\begin{equation} 
\begin{aligned}
\eta^{\mu \sigma \rho} \eta_{\mu \alpha \beta} \bar{\nabla}_\sigma  \bar{\nabla}^\alpha {\cal G}^{L}_{\nu}{^\beta}&=-\bar{\square} {\cal G}^{L \rho}{_\nu}+\bar{\nabla}_\sigma \bar{\nabla}^\rho {\cal G}^{L}_\nu{^\sigma} \\
&=-\bar{\square} {\cal G}^{L \rho}{_\nu}+\bar{R}_\sigma{^\rho}{\nu}{^\lambda} {\cal G}^{L}_\lambda{^\sigma}+\bar{R}^\rho{_\lambda}{\cal G}^{L}_\nu{^\lambda}.
\label{secondtermtmgfp}
\end{aligned}
\end{equation}
In the maximally symmetric $2+1$ dimensional curved backgrounds, Riemann and Ricci tensors are defined as
\begin{equation}
 \bar{R}_{\mu\nu\alpha\beta}=\Lambda (\bar{g}_{\mu \alpha} \bar{g}_{\nu \beta}-\bar{g}_{\mu \beta} \bar{g}_{\nu \alpha}), \qquad \bar{R}_{\mu\nu}=2 \Lambda \bar{g}_{\mu\nu},
\end{equation}
with which Eq.(\ref{secondtermtmgfp}) turns into
\begin{equation}
 \eta^{\mu \sigma \rho} \eta_{\mu \alpha \beta} \bar{\nabla}_\sigma  \bar{\nabla}^\alpha {\cal G}^{L}_{\nu}{^\beta}=-\bar{\square} {\cal G}^{L \rho}{_\nu}
+ 3 \Lambda {\cal G}^{L}_\nu{^\rho}-\Lambda \delta^\rho{_\nu} {\cal G}^{L}.
\label{secondtermtmgfp2}
\end{equation}
On the other hand, one can also recast the third term in  Eq.(\ref{lineartmgfp2}) as follows
\begin{equation}
\begin{aligned}
 \eta^{\mu \sigma \rho} \eta_{\nu \alpha \beta} \bar{\nabla}_\sigma \bar{\nabla}^\alpha {\cal G}^{L}_\mu{^\beta}&= -\bar{\square} {\cal G}^{L}_\nu{^\rho}
 +\bar{\nabla}_\sigma \bar{\nabla}^\rho {\cal G}^{L}_\nu{^\sigma}-\bar{\nabla}_\nu \bar{\nabla}^\rho {\cal G}^{L}+\delta^\rho{_\nu} {\cal G}^{L}\\
 &=-\bar{\square} {\cal G}^{L}_\nu{^\rho}
 + \bar{R}_\sigma{^\rho}{\nu}{^\lambda} {\cal G}^{L}_\lambda{^\sigma}+\bar{R}^\rho{_\lambda}{\cal G}^{L}_\nu{^\lambda} -\bar{\nabla}_\nu \bar{\nabla}^\rho {\cal G}^{L}+\delta^\rho{_\nu} \bar{\square}  {\cal G}^{L}\\
 &=-\bar{\square} {\cal G}^{L}_\nu{^\rho}
 +3 \Lambda {\cal G}^{L}_\nu{^\rho}-\Lambda \delta^\rho{_\nu} {\cal G}^{L}-\bar{\nabla}_\nu \bar{\nabla}^\rho {\cal G}^{L}+\delta^\rho{_\nu} \bar{\square} {\cal G}^{L},
 \label{thirdtermtmgfp}
 \end{aligned}
\end{equation}
where we have availed the identity Eq.(\ref{identity}) in the first line and also the covariantly conserved condition of $ \bar{\nabla}_\mu {\cal G}^{L\mu \nu}=0 $. 
Finally, by substituting Eq.(\ref{secondtermtmgfp2}) and Eq.(\ref{thirdtermtmgfp}) into Eq.(\ref{lineartmgfp2}), one will finally get
\begin{equation}
\begin{aligned}
 & \frac{1}{\kappa} \eta^{\mu \sigma \rho} \bar{\nabla}_\sigma {\cal G}^{L}_{\mu\nu}-\frac{1}{\mu} \bar{\square}{\cal G}^{L}_\nu{^\rho}+\frac{3 \Lambda}{\mu}{\cal G}^{L}_\nu{^\rho}
  -\frac{\Lambda}{\mu} \delta^\rho{_\nu} {\cal G}^{L} \\
  &-\frac{1}{2 \mu} \bar{\nabla}_\nu \bar{\nabla}^\rho {\cal G}^{L}+\frac{1}{2 \mu} \delta^\rho{_\nu} \bar{\square} {\cal G}^{L}+\frac{m^2}{2 \kappa} \eta^{\mu \sigma \rho} \bar{\nabla}_\sigma (h_{\mu\nu}-\bar{g}_{\mu\nu}h)= \eta^{\mu \sigma \rho} \bar{\nabla}_\sigma T_{\mu\nu},
\label{ghdghfp1}
  \end{aligned}
  \end{equation}
which is manifestly traceless. Now let us try to rewrite the first term in Eq.(\ref{ghdghfp1}) in compact form. For this purpose, with the help of Eq.(\ref{linearcottmgfp}) and Eq.(\ref{linertmgfpfeq}), one gets
\begin{equation}
 \eta_{\rho\sigma\mu} \bar{\nabla}^\sigma {\cal G}^{L}_\nu{^\mu}=\mu T_{\rho\nu}-\eta_{\rho\sigma\nu} \bar{\nabla}^\sigma R^{L}-\frac{\mu}{\kappa}{\cal G}^{L}_{\rho \nu} -\frac{\mu m^2}{2 \kappa} (h_{\rho\nu}-\bar{g}_{\rho\nu}h).
\label{newrel1}
 \end{equation}
Thus, by plugging this back into the Eq.(\ref{ghdghfp1}), one arrives at
\begin{equation}
 \begin{aligned}
 &\Big (\bar{\square}-3 \Lambda-\frac{\mu^2}{\kappa^2} \Big ) {\cal G}^{L}{_{\rho \nu}}-\frac{\mu^2 m^2}{2 \kappa^2} (h_{\rho\nu}-\bar{g}_{\rho\nu}h)-\frac{\mu m^2}{2 \kappa} \eta^{\mu \sigma}{_\rho} \bar{\nabla}_\sigma (h_{\mu\nu}-\bar{g}_{\mu\nu}h)\\
 &=\frac{\mu}{2} \, \eta_\rho{^{\mu\sigma}} \bar{\nabla}_\mu T_{\sigma\nu}+ \frac{\mu}{2} \, \eta_\nu{^{\mu\sigma}} \bar{\nabla}_\mu T_{\sigma\rho}-\frac{\mu^2}{\kappa} T_{\rho \nu} -\Lambda \bar{g}_{\rho \nu} {\cal G}^{L}-\frac{1}{2} \bar{\nabla}_\nu \bar{\nabla}_\rho {\cal G}^{L}+\frac{1}{2} \bar{g}_{\rho\nu} \bar{\square} {\cal G}^{L}.
 \label{klmapt2fp2}
 \end{aligned}
\end{equation}
Observe that the symmetry allows us to drop $ \eta_{\rho\sigma\nu} \bar{\nabla}^\sigma R^{L}$ term. To convert this equation into a wave-type equation, one needs to rewrite the Fierz-Pauli term in terms of $ {\cal G}^L_{\mu \nu} $ and its contractions. To do this, let us define
\begin{equation}
 {\cal B}_{\mu\nu}=\eta_{\mu\alpha\beta} \bar{\nabla}^\alpha {\cal G}^L_{\nu}{^\beta},
\end{equation}
which verifies $\bar{g}^{\mu\nu}{\cal B}_{\mu\nu}={\cal B}=0$ and $\bar{\nabla}^\mu {\cal B}_{\mu\nu}=0$. Then, Eq.(\ref{klmapt2fp2}) boils  down to 
\begin{equation}
 -\frac{1}{\kappa} {\cal B}^\rho{_\nu}+\frac{1}{2 \mu} \eta^{\mu \sigma \rho}\bar{\nabla}_\sigma {\cal B}_{\mu \nu}+\frac{1}{2 \mu} \eta^{\mu \sigma \rho}\bar{\nabla}_\sigma {\cal B}_{\nu \mu}
 +\frac{m^2}{2 \kappa} \eta^{\mu \sigma \rho} \bar{\nabla}_\sigma (h_{\mu\nu}-\bar{g}_{\mu\nu}h)= \eta^{\mu \sigma \rho} \bar{\nabla}_\sigma T_{\mu \nu}.
\label{newdef1}
 \end{equation}
Applying $ \eta_{\rho \alpha \beta}\bar{\nabla}^\alpha $ to this equation, one obtains
\begin{equation}
\begin{aligned}
 &-\frac{1}{\kappa} \eta_{\rho \alpha \beta}\bar{\nabla}^\alpha {\cal B}^\rho{_\nu}+\frac{1}{2 \mu} \eta_{\rho \alpha \beta} \eta^{\mu \sigma \rho}\bar{\nabla}^\alpha\bar{\nabla}_\sigma {\cal B}_{\mu \nu}+\frac{1}{2 \mu} \eta_{\rho \alpha \beta} \eta^{\mu \sigma \rho}\bar{\nabla}^\alpha \bar{\nabla}_\sigma {\cal B}_{\nu \mu} \\
 &+\frac{m^2}{2 \kappa}\eta_{\rho \alpha \beta} \eta^{\mu \sigma \rho} \bar{\nabla}^\alpha \bar{\nabla}_\sigma (h_{\mu\nu}-\bar{g}_{\mu\nu}h)= \eta_{\rho \alpha \beta}\eta^{\mu \sigma \rho} \bar{\nabla}^\alpha  \bar{\nabla}_\sigma T_{\mu \nu}.
\label{newdef2fp2}
 \end{aligned}
  \end{equation}
Accordingly, by using the above developed tools, Eq.(\ref{newdef2fp2}) can be recast as
\begin{equation}
\begin{aligned}
& -\frac{1}{\kappa} \eta_{\rho \alpha \beta}\bar{\nabla}^\alpha {\cal B}^\rho{_\nu}-\frac{1}{2 \mu}\Big (\bar{R}^\mu{_{\beta \mu}}{^\lambda} {\cal B}_{\lambda \nu}+\bar{R}^\mu{_{\beta \nu}}{^\lambda} {\cal B}_{\mu \lambda}-\bar{\square}{\cal B}_{\beta \nu}  \Big )-\frac{1}{2 \mu}\Big (\bar{R}^\mu{_{\beta \nu}}{^\lambda} {\cal B}_{\lambda \mu}\\&+\bar{R}^\mu{_{\beta \mu}}{^\lambda} {\cal B}_{\nu \lambda}-\bar{\square}{\cal B}_{\nu \beta} \Big ) -\frac{m^2}{2 \kappa}\Big(\bar{R}^\mu{_{\beta \mu}}{^\lambda} h_{\lambda \nu}+\bar{R}^\mu{_{\beta \nu}}{^\lambda}h_{\mu \lambda}-\bar{\square}h_{\beta \nu}+ \bar{g}_{\nu \beta} \bar{\square} h \Big)\\&= -\bar{R}^\mu{_{\beta \mu}}{^\lambda} T_{\lambda \nu}-\bar{R}^\mu{_{\beta \nu}}{^\lambda}T_{\mu \lambda}+\bar{\square}T_{\beta \nu},
\label{newdef5}
 \end{aligned}
  \end{equation}
or equivalently
\begin{equation}
\begin{aligned}
& -\frac{1}{\kappa} \eta_{\rho \alpha \beta}\bar{\nabla}^\alpha {\cal B}^\rho{_\nu}+\frac{1}{2 \mu}(\bar{\square}-3 \Lambda) \Big ({\cal B}_{\beta \nu}+{\cal B}_{\nu \beta} \Big ) \\
 &-\frac{m^2}{2 \kappa}\Big(3 \Lambda h_{\beta \nu}-\Lambda \bar{g}_{\beta\nu} h-\bar{\square}h_{\beta \nu}+ \bar{g}_{\beta\nu } \bar{\square} h \Big)= -3 \Lambda T_{\beta \nu}+\Lambda \bar{g}_{\beta\nu} T+\bar{\square}T_{\beta \nu}.
\label{newdef7fp2}
 \end{aligned}
  \end{equation}
On the other hand, the linearized Einstein tensor can be written as ${\cal G}^L_{\beta \nu}=\Lambda h_{\beta \nu}+\frac{1}{2} \bar{\nabla}_\beta \bar{\nabla}_\nu h-\frac{1}{2} \bar{\square}h_{\beta\nu}$ which yields
\begin{equation}
    -\bar{\square}h_{\beta\nu}=2 {\cal G}^L_{\beta \nu}-2 \Lambda h_{\beta \nu}- \bar{\nabla}_\beta \bar{\nabla}_\nu h.
 \label{relatbfp2}
 \end{equation}
With the help of Eq.(\ref{newdef7fp2}) and Eq.(\ref{relatbfp2}), the Fierz-Pauli term takes the following form
 \begin{equation}
\begin{aligned}
\frac{m^2}{2 \kappa}\Big( h_{\beta \nu}- \bar{g}_{\beta\nu} h \Big)& =-\frac{1}{\kappa \Lambda} \eta_{\rho \alpha \beta}\bar{\nabla}^\alpha {\cal B}^\rho{_\nu}+\frac{1}{2 \mu \Lambda}(\bar{\square}-3 \Lambda) \Big ({\cal B}_{\beta \nu}+{\cal B}_{\nu \beta} \Big )-\bar{g}_{\beta\nu} T \\
&-\frac{m^2}{2 \kappa \Lambda} \Big(2 {\cal G}^L_{\beta \nu}- \bar{\nabla}_\beta \bar{\nabla}_\nu h + \bar{g}_{\beta\nu } \bar{\square} h \Big)-\frac{1}{\Lambda}(\bar{\square}- 3\Lambda) T_{\beta \nu}.
 \label{newdef8fp2}
 \end{aligned}
  \end{equation}
The first and second term in the right hand side of the Eq.(\ref{newdef8fp2}) can be recast as, respectively
  \begin{equation}
  \begin{aligned}
   -\frac{1}{\kappa \Lambda} \eta_{\rho \alpha \beta}\bar{\nabla}^\alpha {\cal B}^\rho{_\nu}=\frac{1}{\kappa \Lambda} ( \bar{\square}-3 \Lambda ){\cal G}^L_{\nu\beta}+\frac{1}{\kappa}\bar{g}_{\beta \nu} {\cal G}^L, 
   \end{aligned}
  \end{equation}
and 
\begin{equation}
\begin{aligned}
\frac{1}{2 \mu \Lambda}(\bar{\square}-3 \Lambda) \Big ({\cal B}_{\beta \nu}+{\cal B}_{\nu \beta} \Big )=&\frac{1}{\Lambda}(\bar{\square}-3 \Lambda)T_{\beta\nu}-\frac{1}{\kappa \Lambda}(\bar{\square}-3 \Lambda){\cal G}^{L}_{\beta \nu}\\&-\frac{m^2}{2 \kappa \Lambda}(\bar{\square}-3 \Lambda) (h_{\beta\nu}-\bar{g}_{\beta\nu}h). 
\end{aligned}
\end{equation}
By using of these equations and after some lengthy computations, the Fierz-Pauli mass term Eq.(\ref{newdef8fp2}) can be rewritten as
\begin{equation}
\begin{aligned}
 \frac{m^2}{2 \kappa} ( h_{\beta \nu}- \bar{g}_{\beta\nu} h)=&-\frac{m^2}{ \kappa}(\bar{\square}-2 \Lambda)^{-1} {\cal G}^L_{\beta \nu}+\frac{\Lambda}{\kappa} (\bar{\square}-2 \Lambda)^{-1} \bar{g}_{\beta \nu} {\cal G}^L \\
 &-\frac{m^2}{2 \kappa}(\bar{\square}-2 \Lambda)^{-1} \Big( \bar{g}_{\beta\nu } \bar{\square}-\bar{\nabla}_\beta \bar{\nabla}_\nu \Big)h-\Lambda (\bar{\square}-2 \Lambda)^{-1} \bar{g}_{\beta\nu} T. \\
\label{explfierzpaul2}
 \end{aligned}
 \end{equation}
Moreover, we need to write the third term in the Eq.(\ref{klmapt2fp2}) in terms of $ {\cal G}^L_{\mu \nu} $ and its contractions. For that reason, applying the operator $ \eta^{\beta\sigma\rho} \bar{\nabla}_\sigma $ to Eq.(\ref{explfierzpaul2}), one ultimately gets
\begin{equation}
 \begin{aligned}
 \frac{m^2}{2 \kappa} \eta^{\beta\sigma\rho} \bar{\nabla}_\sigma ( h_{\beta \nu}- \bar{g}_{\beta\nu} h)=&-\frac{m^2}{ \kappa} (\bar{\square}-2 \Lambda)^{-1} \eta^{\beta\sigma\rho} \bar{\nabla}_\sigma {\cal G}^L_{\beta \nu}\\
 &-\frac{m^2}{2 \kappa} (\bar{\square}-2 \Lambda)^{-1} \eta^{\beta\sigma\rho} \bar{\nabla}_\sigma \Big( \bar{g}_{\beta\nu } \bar{\square} h- \bar{\nabla}_\beta \bar{\nabla}_\nu h  \Big)\\&+\frac{\Lambda}{\kappa} (\bar{\square}-2 \Lambda)^{-1} \eta_\nu{^{\sigma\rho}} \bar{\nabla}_\sigma {\cal G}^L -\Lambda  (\bar{\square}-2 \Lambda)^{-1} \eta_\nu{^{\sigma\rho}} \bar{\nabla}_\sigma T. \\
\label{fierz-peq3fp2}
 \end{aligned}
  \end{equation}
By virtue of Eq.(\ref{newrel1}) and Eq.(\ref{explfierzpaul2}), the first term in the right hand side of Eq.(\ref{fierz-peq3fp2}) reduces to
\begin{equation}
\begin{aligned} 
 -\frac{m^2}{ \kappa} (\bar{\square}-2 \Lambda)^{-1}\eta^{\beta\sigma\rho} \bar{\nabla}_\sigma {\cal G}^L_{\beta \nu}&=\frac{\mu m^2}{ \kappa} (\bar{\square}-2 \Lambda)^{-1} T^\rho{_\nu}-\frac{m^2}{ \kappa} (\bar{\square}-2 \Lambda)^{-1} \eta^\rho{_{\sigma\nu}} \bar{\nabla}^\sigma R^{L}\\&-\frac{\mu m^2}{ \kappa^2} (\bar{\square}-2 \Lambda)^{-1}{\cal G}^{L \rho}_{\nu} +\frac{\mu m^2\Lambda}{ \kappa}  (\bar{\square}-2 \Lambda)^{-2} \delta^\rho_\nu T\\
&+\frac{\mu m^4}{ \kappa^2}(\bar{\square}-2 \Lambda)^{-2} {\cal G}^{L\rho}_\nu-\frac{\mu m^2 \Lambda}{\kappa^2} (\bar{\square}-2 \Lambda)^{-2} \delta^\rho_\nu {\cal G}^L \\
 &+\frac{\mu m^4}{2 \kappa^2}(\bar{\square}-2 \Lambda)^{-2} \Big( \delta^\rho_\nu \bar{\square}-\bar{\nabla}^\rho \bar{\nabla}_\nu \Big)h, \\
\end{aligned}
\end{equation}
with follows from that, one obtains
\begin{equation}
 \begin{aligned}
 \frac{m^2}{2 \kappa} \eta^{\beta\sigma\rho} \bar{\nabla}_\sigma ( h_{\beta \nu}- \bar{g}_{\beta\nu} h)&=\frac{1}{ \kappa} (\bar{\square}-2 \Lambda)^{-1}\Bigg(\mu m^2 T^\rho{_\nu}-m^2 \eta^\rho{_{\sigma\nu}} \bar{\nabla}^\sigma R^{L}-\frac{\mu m^2}{ \kappa} {\cal G}^{L \rho}_{\nu}\\&+\Lambda \eta_\nu{^{\sigma\rho}} \bar{\nabla}_\sigma {\cal G}^L-\frac{m^2}{2 }  \eta^{\beta\sigma\rho} \bar{\nabla}_\sigma \Big( \bar{g}_{\beta\nu } \bar{\square} h- \bar{\nabla}_\beta \bar{\nabla}_\nu h  \Big)\\&-\Lambda \kappa  \eta_\nu{^{\sigma\rho}} \bar{\nabla}_\sigma T \bigg) +\frac{\mu m^2}{ \kappa}(\bar{\square}-2 \Lambda)^{-2}\bigg( \frac{ m^2}{\kappa}{\cal G}^{L\rho}_\nu-\frac{\Lambda}{\kappa} \delta^\rho_\nu {\cal G}^L \\&+\frac{ m^2}{2 \kappa} \Big( \delta^\rho_\nu \bar{\square}-\bar{\nabla}^\rho \bar{\nabla}_\nu \Big)h+ \Lambda  \delta^\rho_\nu T\bigg). \\
\label{fierz-peq4fp2}
 \end{aligned}
  \end{equation}
Accordingly, by inserting Eq.(\ref{explfierzpaul2}) and Eq.(\ref{fierz-peq4fp2}) into Eq.(\ref{klmapt2fp2}), one finally arrives at
   \begin{equation}
 \begin{aligned}
   &\Bigg ((\bar{\square}-3 \Lambda-\frac{\mu^2}{\kappa^2})+\frac{2 \mu^2 m^2}{ \kappa^2}(\bar{\square}-2 \Lambda)^{-1}-\frac{\mu^2 m^4}{ \kappa^2}(\bar{\square}-2 \Lambda)^{-2} \Bigg) {\cal G}^{L}{_{\rho \nu}} \\
&=\frac{\mu}{2} \, \eta_\rho{^{\mu\sigma}} \bar{\nabla}_\mu T_{\sigma\nu}+ \frac{\mu}{2} \, \eta_\nu{^{\mu\sigma}} \bar{\nabla}_\mu T_{\sigma\rho}-\frac{\mu^2}{\kappa} T_{\rho \nu}+\frac{\mu^2 m^2}{ \kappa} (\bar{\square}-2 \Lambda)^{-1} T_{\rho\nu}\\
&-\frac{\mu^2 m^2}{2 \kappa \Lambda(1-\frac{m^2}{\Lambda})}\Bigg \{(\bar{\square}-2 \Lambda)^{-1} \Big (1-m^2 (\bar{\square}-2 \Lambda)^{-1}\Big)-\frac{\kappa^2 \Lambda}{\mu^2 m^2} \Bigg \}\\&\times \Big(\bar{g}_{\rho\nu }(\bar{\square}-2 \Lambda)-\bar{\nabla}_\rho \bar{\nabla}_\nu \Big)T.
\label{res122345}
\end{aligned}
\end{equation}
Here, one has 
\begin{equation}
 {\cal G}^L_{\rho\nu}= -\frac{1}{2} (\bar{\square}-2 \Lambda)h_{\rho\nu}+\frac{1}{2} \bar{\nabla}_\rho \bar{\nabla}_\nu h.
  \end{equation}
  
\section{Topologically Massive Gravity: Anyon-Anyon Scattering}
\label{chp:appendixb1}
In this part, we will work on the scattering amplitude of anyons and associated Newtonian potential energy for the TMG case in detail. For this purpose, taking the $ m^2 \to 0$ limit in Eq.(\ref{mainressct}) as well as flat space limit ($\Lambda \to 0$), one gets the scattering amplitude as
\begin{equation}
 \begin{aligned}
  4{\cal A}&=-2\mu T^{'}_{\rho\nu} \frac{1}{(\partial^2-\frac{\mu^2}{\kappa^2})\partial^2} \eta^{\rho\mu\sigma}\partial_\mu T_\sigma{^\nu}+\frac{2 \mu^2}{\kappa}T^{'}_{\rho\nu} \frac{1}{(\partial^2-\frac{\mu^2}{\kappa^2})\partial^2}   T^{\rho\nu} \\
  &-\frac{\mu^2}{\kappa} T^{'} \frac{1}{(\partial^2-\frac{\mu^2}{\kappa^2})\partial^2} T + \kappa T^{'} \frac{1}{\partial^2} T,
\label{anyon-anyon1tmg1}
\end{aligned}
\end{equation}
where we have used the fact that $ \triangle^{(2)}_L$ reduces to $-\partial^2$ in the flat-space limit. After using the fractional decomposition, Eq.(\ref{anyon-anyon1tmg1}) boils down to  
 \begin{equation}
 \begin{aligned}
  4{\cal A}&=-\frac{2\mu}{m^2_g}  T^{'}_{\rho\nu} \Big( \frac{1}{\partial^2-m^2_g}-\frac{1}{\partial^2} \Big ) \eta^{\rho\mu\sigma}\partial_\mu T_\sigma{^\nu}+\frac{2 \mu^2}{\kappa m^2_g}T^{'}_{\rho\nu} \Big( \frac{1}{\partial^2-m^2_g}-\frac{1}{\partial^2}  \Big )   T^{\rho\nu} \\
  &-\frac{\mu^2}{\kappa m^2_g} T^{'} \Big( \frac{1}{\partial^2-m^2_g}-\frac{1}{\partial^2} \Big ) T + \kappa T^{'} \frac{1}{\partial^2} T.
\label{anyon-anyon4tmg2}
\end{aligned}
\end{equation}
In order to find the Newtonian Potential Energy between two localized spinning point-like sources, let us consider the following energy-momentum tensors 
 \begin{equation}
T_{00}= m_a \delta^{(2)}({\bf x}-{\bf x}_a),\qquad T^{i}{_0}=-1/2 J_a \epsilon^{ij} \partial_j \delta^{(2)}({\bf x}-{\bf x}_a).
\end{equation}
Here $a=1,2$ and $m_a $ and $ J_a $ are the mass and spin of the sources, respectively. With these definitions, the scattering amplitude in Eq.(\ref{anyon-anyon4tmg2}) reduces to
  \begin{equation}
   \begin{aligned}
    4 {\cal A}&=-\frac{2 \mu}{m^2_g} T^{'}_{00} G_{(1)}({\bf x},{\bf x}^{'},t,t^{'}) \eta^{0ij} \partial_j T_{i0}+\frac{2 \mu}{m^2_g} T^{'}_{i0} G_{(1)}({\bf x},{\bf x}^{'},t,t^{'}) \eta^{0ij} \partial_j T_{00} \\
    &+\frac{\mu^2}{\kappa m^2_g}T^{'}_{00}G_{(1)}({\bf x},{\bf x}^{'},t,t^{'})T_{00}-\frac{4 \mu^2}{\kappa m^2_g} T^{'}_{0i} G_{(1)}({\bf x},{\bf x}^{'},t,t^{'}) T_0{^i} \\
&+ \kappa \, T^{'}_{00} G_{(2)}({\bf x},{\bf x}^{'},t,t^{'}) T_{00}, 
   \end{aligned}
  \end{equation}
where $ \hat{G}_{(1)}({\bf x},{\bf x}^{'})$ and $ \hat{G}_{(2)}({\bf x},{\bf x}^{'})$ denotes the corresponding green functions:
\begin{equation}
 G_{(1)}({\bf x},{\bf x}^{'},t,t^{'})= \frac{1}{\partial^2-m^2_g}-\frac{1}{\partial^2}, \qquad G_{(2)}({\bf x},{\bf x}^{'},t,t^{'})=\frac{1}{\partial^2}.
 \end{equation}
Thereupon, by virtue of Eq.(\ref{scatdef}) and taking the time integral, one can obtain
\begin{equation}
 \begin{aligned}
  4 {\cal U}&=\frac{\mu m_2 J_1}{m^2_g} \int d^2 x \int d^2 x^{'} \delta^{(2)}({\bf x}^{'}-{\bf x}_2) \hat{G}_{(1)}({\bf x},{\bf x}^{'}) \partial_i \partial^i \delta^{(2)}({\bf x}-{\bf x}_1) \\
&-\frac{\mu m_1 J_2}{m^2_g} \int d^2 x \int d^2 x^{'} \partial^{'}_i \delta^{(2)}({\bf x}^{'}-{\bf x}_2) \hat{G}_{(1)}({\bf x},{\bf x}^{'}) \partial^i \delta^{(2)}({\bf x}-{\bf x}_1) \\
&+\frac{\mu^2 m_1 m_2}{\kappa m^2_g} \int d^2 x \int d^2 x^{'} \delta^{(2)}({\bf x}^{'}-{\bf x}_2) \hat{G}_{(1)}({\bf x},{\bf x}^{'}) \delta^{(2)}({\bf x}-{\bf x}_1) \\
&-\frac{\mu^2 J_1 J_2}{\kappa m^2_g} \int d^2 x \int d^2 x^{'} \partial^{'}_i \delta^{(2)}({\bf x}^{'}-{\bf x}_2) \hat{G}_{(1)}({\bf x},{\bf x}^{'}) \partial^i \delta^{(2)}({\bf x}-{\bf x}_1) \\
&+ \kappa m_1 m_2 \int d^2 x \int d^2 x^{'} \delta^{(2)}({\bf x}^{'}-{\bf x}_2) \hat{G}_{(2)}({\bf x},{\bf x}^{'}) \delta^{(2)}({\bf x}-{\bf x}_1),
\label{scatnor}
\end{aligned}
\end{equation}
where the time integrated Green's functions are
\begin{equation}
\begin{aligned}
 \hat{G}_{(1)}({\bf x},{\bf x}^{'})&=\int d t^{'} \, G_{(1)}({\bf x},{\bf x}^{'},t,t^{'})=\frac{1}{2 \pi} \Big ( K_0 (m_g \lvert {\bf x}-{\bf x}^{'} \rvert)- \ln (m_g \lvert {\bf x}-{\bf x}^{'}  \rvert) \Big ), \\
 \hat{G}_{(2)}({\bf x},{\bf x}^{'})&=\int d t^{'} \, G_{(2)}({\bf x},{\bf x}^{'},t,t^{'})=\frac{1}{2 \pi} \ln (m_g \lvert {\bf x}-{\bf x}^{'}  \rvert).
 \end{aligned}
 \end{equation}
Thus, by evaluating the integrals, one will eventually get
\begin{equation}
\begin{aligned}
  {\cal U}&= \frac{\kappa m^2_g}{16 \pi} \bigg (\frac{\kappa m_1 J_2}{\mu}+\frac{\kappa m_2 J_1}{\mu}+J_1 J_2 \bigg ) \,\, K_2 (m_g \lvert {\bf x}_1-{\bf x}_2 \rvert) \\ 
&+\frac{\kappa m^2_g}{16 \pi} \bigg \{ \frac{2 m_1 m_2 }{m^2_g}+ \bigg (\frac{\kappa m_1 J_2}{\mu}+\frac{\kappa m_2 J_1}{\mu}+J_1 J_2 \bigg ) \bigg \} K_0 (m_g \lvert {\bf x}_1-{\bf x}_2 \rvert),
 \label{neqt}
 \end{aligned}
\end{equation}
here we have used the identity in Eq.(\ref{recur}). Observe that, in addition to spin-spin and mass-mass interactions, there also occur spin-mass interactions due to topological mass $\mu$ that induces a additional spin defined by 
\begin{equation}
 J^{ind}_a= \frac{\kappa m_a}{\mu},
\end{equation}
which changes the initial spin of the particle as
 \begin{equation}
 J^{tot}_a=J_a+J^{ind}_a,\hskip 2 cm a=1,2.
 \label{totspin1}
 \end{equation}
Finally, the associated Newtonian potential energy reads 
\begin{equation}
 {\cal U}= \frac{\kappa m^2_g}{16 \pi }\bigg \{ \Big ( J^{tot}_1 J^{tot}_2-\frac{m_1 m_2}{m^2_g} \Big )K_2(m_g r)+ \Big ( J^{tot}_1 J^{tot}_2+\frac{m_1 m_2}{m^2_g} \Big )K_0(m_g r ) \bigg \},
 \end{equation}
where $ r = \lvert {\bf x}_1-{\bf x}_2 \rvert$. Let us consider the small and large distance behaviours of potential energy. As for the small distance, by keeping in mind that the modified Bessel functions behave like  
\begin{equation}
 K_0 (m_g \lvert r \sim -\ln(m_g \lvert r \rvert)-\gamma_{E}, \quad   K_2 (m_g \lvert r \rvert) \sim \frac{2}{m^2_g} \frac{1}{\lvert r \rvert^2}, 
\label{limitc}
 \end{equation}
then we have
\begin{equation}
\begin{aligned}
 {\cal U}  \sim \frac{\frac{\kappa}{8 \pi}  \Big ( J^{tot}_1  J^{tot}_2-\frac{m_1 m_2}{m^2_g} \Big )}{ r^2}  
-\frac{\kappa m^2_g}{16 \pi}  \Big ( J^{tot}_1 J^{tot}_2+\frac{m_1 m_2}{m^2_g} \Big ) \Big (\ln(m_g r)+\gamma_{E} \Big ).
  \end{aligned}
\end{equation}
Finally, for large distances, the Newtonian potential can be obtained as 
 \begin{equation}
 {\cal U} \sim \frac{\kappa m^2_g J^{tot}_1 J^{tot}_2}{8  \pi} \sqrt{\frac{\pi}{ 2 m_g r}} \, e^{-m_g r},
\end{equation}
where we have used the Eq.(\ref{Besselfp2}).
\section{Shock wave geometry}
\label{chp:appendixc}
In this part, we will review the shock wave geometry which is relevant for chapter 4 discussed in the thesis. Suppose $(t,x,y)$ be the coordinates in the flat space.
The shock wave metric generated by a high-energy massless particle moving in the $+x$ direction with 3 momentum as $p^\mu=\lvert p\rvert(\delta^\mu_0+\delta^\mu_x) $:
\begin{equation}
ds^2=-du dv+H(u,y) du^2+dy^2,
\label{shockwavemetric}
\end{equation}
where  $u=t-x$ and $v=t+x$ are null-cone coordinates. The corresponding energy momentum tensor produced by this particle can be given as $T_{uu}=\lvert p\rvert \delta(y)\delta(u)$. For the sake of simplicity, let us rewrite the shock-wave metric in the Kerr-Schild form as \cite{gurses_sisman}
\begin{equation}
g_{\mu \nu}=\eta_{\mu \nu}+H(u,y) \lambda_\mu \lambda_\nu,
\end{equation} 
where $\eta_{\mu \nu}$ is the Minkowski metric and the $\lambda_\mu$ is a vector which verifies the following properties
\begin{equation}
\begin{aligned}
\lambda^{\mu}\lambda_\mu=0,\hskip .6 cm\nabla_{\mu}\lambda_\nu=0,\hskip .6 cm
\lambda^\mu\partial_\mu H(u,y)=0.
\label{plane1}
\end{aligned}
\end{equation}
For the null cone coordinates in three dimensions, the only non-vanishing components of Minkowski metric $\eta_{\mu \nu}$ are  $\eta_{u v} = - \frac{1}{2}$ and  $\eta_{ yy} = 1$ and hence we have  $\det{g} = \det{\eta} = -\frac{1}{4}$.
Firstly, one needs to find the Christoffel symbols in the Kerr-Schild form. It can be given as 
\begin{equation}
2\Gamma^\sigma_{\mu \nu}= \lambda^\sigma \lambda_\mu \partial_\nu H + \lambda^\sigma \lambda_\nu \partial_\mu H - \lambda_\mu \lambda_\nu \eta^{\sigma \beta} \partial_\beta H, 
\end{equation}
with which non-vanishing components are 
\begin{equation}
 \Gamma^y_{uu}=-\frac{1}{2}\partial_y H(u,y),\hskip .6 cm \Gamma^v_{uu}=-\partial_u H(u,y),\hskip .6 cm \Gamma^v_{uy}=-\partial_y H(u,y).
\end{equation}
Note that $\lambda$ contractions of the Christoffel symbol are $\lambda_\sigma\Gamma^\sigma_{\mu \nu}=0$, $\lambda^\mu\Gamma^\sigma_{\mu \nu}=0$ and the Riemann tensor can be given in terms of the metric function $H$ as 
\begin{equation}
2 R_{\mu\alpha\nu\beta}=\lambda_{\mu}\lambda_{\beta}\partial_{\alpha}\partial_{\nu}H+\lambda_{\alpha}\lambda_{\nu}\partial_{\mu}\partial_{\beta}H
-\lambda_{\mu}\lambda_{\nu}\partial_{\alpha}\partial_{\beta}H
-\lambda_{\alpha}\lambda_{\beta}\partial_{\mu}\partial_{\nu}H,\label{Riemann_pp-wave}
\end{equation}
and followingly  Ricci tensor  and the "Box"  of the Ricci tensor (which are needed in the case of NMG as discussed in the chapter 4) become
\begin{equation}
R_{\mu\nu}=-\frac{1}{2}\lambda_{\mu}\lambda_{\nu}\partial_y^2 H(u,y),\hskip .6 cm  \Box R_{\mu \nu} = -\frac{1}{2}\lambda_{\mu}\lambda_{\nu}\partial_y^4 H(u,y).
\label{ricci_shock}
\end{equation}
On the other hand, as for the TMG case, one needs to find the Cotton tensor in terms of profile function $H$. For that reason, one can easily show that Levi-Civita symbol can be written in the shock wave geometry as \footnote{This corresponds to the sign choice $\epsilon^{t x y} =- 1$.} $ \eta^{ u v y} =  2$. Finally, Cotton tensor can be recast in the following form
\begin{equation}
C_{\mu\nu}=\frac{1}{2}\lambda_{\mu}\lambda_{\nu}\partial_y^3 H(u,y).
\label{cotton_shock}
\end{equation}
Consequently, by using all this set-ups, one can obtain the shock wave solutions in various gravity theories, which we have done in the chapter 4 above.  Now, we shall be interested the spin-2 perturbations about shock wave background as $g_{\mu\nu}=\bar{g}_{\mu\nu}+\delta g_{\mu \nu}$. 

\subsection{Perturbations about the Shock Wave}
\label{Appdx31}
In this part, we shall consider the perturbation as $ h_{\mu \nu} \equiv \delta g_{\mu \nu}$ and choosing the light-cone gauge which seems to be probably the most convenient choice for computations. So we start with 
\begin{equation}
 h_{v\mu}=0,
\end{equation}
or equivalently $\lambda^{\mu}h_{\mu\nu}=0$ and one can easily satisfies the following relations in light-cone gauge
\begin{equation}
 \Gamma^\sigma_{\mu \nu}h^\mu\,_{\sigma}=0,\hskip .6 cm \Gamma^\sigma_{\mu \nu}h_{\sigma\alpha}=-\frac{1}{2}\lambda_{\mu}\lambda_{\nu}h_{y\alpha}\partial_yH
 ,\hskip .6 cm \Gamma^\sigma_{\mu \nu}h^\mu\,_{\alpha}=\frac{1}{2}\lambda^{\sigma}\lambda_{\nu}h^y\,_{\alpha}\partial_yH.
\end{equation}
The linearized Christoffel connection can be obtained as 
\begin{equation}
\delta \Gamma^\sigma_{\mu \nu} = \frac{1}{2}\eta^{\sigma \alpha}\Big ( \partial_\mu h_{ \nu \alpha} + \partial_\nu h_{ \mu \alpha}  - \partial_\alpha h_{ \mu \nu}+  \lambda_\mu \lambda_\nu  h_{\alpha y} \partial_y H \Big ) - H \lambda^\sigma \partial\label{delayphoton1}_v h_{ \mu \nu},
\end{equation}
whose components are
\begin{equation}
\begin{aligned}
&\delta \Gamma^u_{\mu \nu} = \partial_v h_{ \mu \nu},\\
&\delta \Gamma^v_{\mu \nu} = -\partial_\mu h_{\nu u}-\partial_\nu h_{\mu u}+\partial_uh_{\mu \nu}-\lambda_\mu \lambda_\nu h_{yu}\partial_y H+2H\partial_v h_{\mu\nu}, \\
&\delta \Gamma^y_{\mu \nu} =  \frac{1}{2} \Big (\partial_\mu h_{\nu y}+\partial_\nu h_{\mu y}-\partial_yh_{\mu \nu}+\lambda_\mu \lambda_\nu h_{yy}\partial_y H \Big).
\label{SWChr}
\end{aligned}
\end{equation}
By using of Eq.(\ref{SWChr}), the linearized Ricci tensor can be computed as
\begin{equation}
\begin{aligned}
 \delta R_{\mu\nu}
 =\frac{1}{2} \Big (\nabla_\sigma\nabla_\mu h^\sigma\,_{\nu}+\nabla_\sigma\nabla_\nu h^\sigma\,_{\mu}-\square h_{\mu\nu}-\nabla_\mu\nabla_\nu h \Big ),
\end{aligned}
\end{equation}
which reduces to the following form in this gauge
\begin{equation}
 \begin{aligned}
 2\delta R_{\mu\nu}&=2\partial_{(\mu}\partial_\sigma h^\sigma\,_{\nu)}+\lambda_\mu \lambda_\nu\partial_y H\partial_\sigma h^\sigma\,_{y}
 +h\lambda_\mu \lambda_\nu\partial_y^2 H - g^{ \alpha \beta} \partial_\alpha \partial_\beta h_{\mu \nu}\\
& +4\partial_y H\partial_v\lambda_{(\mu} h_{\nu)y}-\partial_\mu\partial_\nu h+\Gamma^\sigma_{\mu \nu}\partial_\sigma h,
\end{aligned}
\end{equation}
and then the linearized curvature scalar can be found as
\begin{equation}
\delta R=\partial_{\mu}\partial_\sigma h^{\sigma\mu}-\square h.
\end{equation}
To compute linearized Cotton tensor, one needs to find linearized Cotton tensor in a arbitrary background. It can be given as  \cite{Kilicarslan:2015cla}  
\begin{equation}
\begin{aligned}
 2\delta C^{\mu\nu}=&-\frac{h}{2}\,C^{\mu\nu}+\eta^{\mu \rho \sigma}\,\nabla_{\rho} \delta G^\nu{_\sigma}+\eta^{\mu \rho \sigma}\,\delta \Gamma^\nu_{\rho \alpha}G^{\alpha}{_{\sigma}}
 +\mu \leftrightarrow \nu\\
 =&-\frac{3h}{2}\,{C^{\mu\nu}}-\frac{1}{2}\eta^{\mu \rho \sigma}\,\square \nabla_{\rho }h^{\nu}{_{\sigma}}
 +\frac{1}{2}\eta^{\mu \rho \sigma}\,\nabla^{\nu}\nabla_{\lambda}\nabla_{\rho }h_{\sigma}{^{\lambda}}
\\&+ \frac{3}{2}\eta^{\mu \rho \sigma}\, \nabla_{\rho }({\cal S}^{\lambda\nu} h_{\lambda\sigma})
 +\frac{1}{6} \eta^{\mu \rho \sigma}\, R \nabla_\rho  h^\nu{_\sigma}
 -\frac{1}{2}\eta^{\mu \rho \sigma}\,{\cal S}^\nu{_\sigma} \nabla_{\rho }h \\&-\frac{1}{2}\eta^{\mu \rho \sigma}\, h^\lambda{_\sigma} \nabla_\lambda {\cal S}^\nu{_\rho }+\eta^{\mu \rho \sigma}\,{\cal S}_{\lambda\rho }\nabla^\nu h^\lambda{_\sigma}
 +\eta^{\mu \rho \sigma}\,{\cal S}_\sigma{^\lambda} \nabla_{\lambda}h^{\nu}{_{\rho }}+\mu \leftrightarrow \nu,
 \label{cottlin}
 \end{aligned}
 \end{equation}
where traceless Ricci tensor ${\cal S}_{\mu\nu}$ is defined as ${\cal S}_{\mu\nu} = R_{\mu\nu}  - \frac{1}{3}g_{\mu\nu}R$. 
For the shock wave ansatz Eq.(\ref{shockwavemetric}) as well as imposing the light-cone gauge, Eq.(\ref{cottlin}) boils down to
 \begin{equation}
\begin{aligned}
 2\delta C^{\mu\nu}=&-\frac{3h}{4}\,\lambda^\mu\lambda^\nu \partial_y^3H-\frac{1}{2}\eta^{\mu \rho \sigma}\,\square \nabla_{\rho }h^{\nu}{_{\sigma}}
 +\frac{1}{2}\eta^{\mu \rho \sigma}\,\nabla^{\nu}\nabla_{\lambda}\nabla_{\rho }h_{\sigma}{^{\lambda}}\\
&-\frac{1}{2}\eta^{\mu \rho u}\,\delta^\nu_v  \partial_y^2H \nabla_{\rho }h -\frac{1}{2}\eta^{\mu u y}\, h^\lambda{_y}\delta^\nu_v \nabla_\lambda  \partial_y^2H
+\eta^{\mu y u}\,\partial_y^2H\nabla_v h^\nu{_y}+\mu \leftrightarrow \nu.
 \end{aligned}
 \end{equation}
In the light-cone gauge, components of the  linearized Ricci tensor can be given as
\begin{equation}
\begin{aligned}
&\delta R_{uu}= (\partial_u\partial_y+ \partial_yH\partial_v)f+(2 H
 \partial_v^2 -\frac{1}{2}\partial_y^2)g
 +\frac{1}{2}(\partial_y^2 H +\frac{1}{2}\partial_yH \partial_y-\partial_uH \partial_v-\partial_u^2)h,\\
 &\delta R_{uv}= \frac{1}{2}\partial_v\partial_yf-\partial_v^2g-\frac{1}{2}\partial_u\partial_vh,\\
 &\delta R_{uy}=(2H \partial_v^2+\partial_u\partial_v)f-\partial_v\partial_yg+\frac{1}{2} \partial_yH \partial_vh,\\
 &\delta R_{vv}= -\frac{1}{2}\partial_v^2h,\\
 &\delta R_{vy}= -\partial_v^2f,\\
 &\delta R_{yy}=-2\partial_v\partial_yf+2(H \partial_v^2+\partial_u\partial_v)h,
 \end{aligned}
\end{equation}
where $g,f$ and $h$ are functions of all coordinates and defined as $g \equiv h_{u u}$, $ f \equiv h_{u y} $, $ h  \equiv h_{y y}$. So the linearized Ricci scalar can be obtained in terms of these functions
\begin{equation}
\delta R = 4 \left( - \partial_v \partial_y  f + \partial_v^2 g  + H \partial_v^2 h + \partial_u \partial_v h \right) ~.
\end{equation}
Components of the  linearized Cotton tensor can be computed as
\begin{equation}
\begin{aligned}
&\delta C_{uu}=\frac{1}{4} \bigg((-10\partial_yH \partial_v\partial_y+4\partial_u H \partial_v^2+16 H^2
 \partial_v^3-4H\partial_v\partial_y^2+16H \partial_v^2\partial_u+4\partial_v\partial_u^2\\&-4 \partial_u\partial_y^2-6\partial_y^2H \partial_v)f
 +(-4\partial_yH \partial_v^2-8H \partial_v^2\partial_y+2\partial_y^3g-4\partial_v\partial_u\partial_yg)g
\\&+(\partial_yH (12 H \partial_v^2-\partial_y^2 +8  \partial_v\partial_u)+2\partial_u H \partial_v\partial_y+4 \partial_u\partial_yH \partial_v +4H\partial_v\partial_u\partial_y
 \\& +2\partial_u^2\partial_y-3\partial_y^2H \partial_y -3\partial_y^3H )h\bigg),\\  
 &\delta C_{uv}=\frac{1}{2}\partial_v^2\bigg(-(4H \partial_v+2 \partial_u)f+\partial_y g+(-\partial_yH +H\partial_y)h
 \bigg), \\
 &\delta C_{uy}=(-2 H\partial_v^2\partial_y -2 \partial_v\partial_u\partial_y-3\partial_yH 
\partial_v^2) f+(-2 H \partial_v^3 + \partial_v\partial_y^2 - \partial_v^2\partial_u )g\\&+(\partial_uH \partial_v^2+ \partial_v\partial_u^2
-\frac{1}{2}\partial_yH \partial_v\partial_y- \partial_y^2H \partial_v+H (2 H \partial_v^3 +3
\ \partial_v^2\partial_u)) h,\\
 &\delta C_{vv}=\partial_v^2(\partial_vf-\frac{1}{2} \partial_y h),\\
 &\delta C_{vy}=\partial_v^2 \left(\partial_v g-(H\partial_v+\partial_u) h\right),\\
 &\delta C_{yy}=\partial_v^2 \left(-4(H \partial_v+\partial_u)f+2\partial_y g-2h\partial_yH\right).
 \end{aligned}
\end{equation}
In the light of above derivations, TMG field equations can be written in terms of shock wave coordinates . But because field equations are  simply too cumbersome, for the sake of simplicity, let us make a simplification by working away from the source which leads to $\partial_y  H = - m_g H$. With this simplification, each components of TMG field equations read 
\begin{equation}
\begin{aligned}
&\partial_v^2\Big(2\partial_v f+( m_g - \partial_y)h\Big) =0 \hskip  1.3 cm vv-\mbox{component}
\end{aligned}
\end{equation}
\begin{equation}
\begin{aligned}
& \partial_v\Big((-4 H\partial_v^2 +m_g \partial_y-2\partial_v\partial_u)f+\partial_v\partial_y g -( m_g  H
  \partial_v -H\partial_v \partial_y +m_g \partial_u )h\Big)= 0\\& \hskip  8.7 cm vu-\mbox{component}
\end{aligned}
\end{equation}
\begin{equation}
\begin{aligned}
\partial_v^2\Big(-m_g f-\partial_v g+(H \partial_v+\partial_u )h\big)  = 0 \hskip  1.2 cm vy-\mbox{component}
\end{aligned}
\end{equation}
\begin{equation}
\begin{aligned}
&\partial_v^2\Big(2( H\partial_v +\partial_u)f-(m_g  
 +\partial_y) g- m_g H h\Big)=0  \hskip  1 cm yy-\mbox{component}
\end{aligned}
\end{equation}
\begin{equation}
\begin{aligned}
& (m_g H \partial_v^2-2H\partial_v^2\partial_y-m_g  \partial_v\partial_u-2 \partial_v\partial_u\partial_y)f+(-2 H \partial_v^3 +m_g \partial_v\partial_y +\partial_v\partial_y^2  \\&-\partial_v^2\partial_u )g+(2 H^2 \partial_v^3 +2H \partial_u \partial_v^2 
  + \frac{H}{2}m_g \partial_v \partial_y -\frac{H}{2}m_g^2  \partial_v+\partial_v \partial_u^2 )h
  =0 \\& \hskip  8.4 cm uy-\mbox{component}
\end{aligned}
\end{equation}
\begin{equation}
\begin{aligned}
 & (-2m_g^2H\partial_v+2m_g H\partial_v\partial_y+16 H^2\partial_v^3 
 -4 H\partial_v\partial_y^2+12 H\partial_v^2\partial_u-4 m_g \partial_u\partial_y
-4\partial_u\partial_y^2\\&+4\partial_v\partial_u^2)f+ (-2m_g^2 H  \partial_y-4 m_g H^2
 \partial_v^2 + m_g H\partial_y^2 + m_g^3 H+2m_g H \partial_u\partial_v  
+2m_g  \partial_u^2\\&+2 H \partial_u\partial_v\partial_y+2\partial_u^2\partial_y)h+(-8 H
 \partial_v^2\partial_y +2 m_g \partial_y^2 +2\partial_y^3 -4 \partial_v\partial_u\partial_y +4 m_g H \partial_v^2\label{delayphoton1})g\\& =0  \hskip  8.7 cm uu-\mbox{component}
\end{aligned}
\end{equation}
After obtaining linearized field equations of TMG, let us compute the linearized equations of NMG around the shock-wave background. For this calculations, we will impose the light-cone gauge and suppose that the perturbation is also traceless, in other words $ h=0$, otherwise the linearized field equations are complicated. Thereupon, the perturbation is simply given by   
\[ h_{\mu \nu}(u,v,y)= \left( \begin{array}{ccc}
g & 0 &  f \\
0 & 0 & 0 \\
 f & 0 & 0\end{array} \right).
\]
By using of this, NMG field equations can be calculated. But because the field equations are  somewhat lengthy, let us consider on them again by working away from $y = 0$ which yields $\partial_y  H = - m_g H$. Then the components of NMG equations are
\begin{equation}
\begin{aligned}
\partial_v^4g-\partial_v^3\partial_yf =0 \hskip  1.3 cm vv-\mbox{component}
\end{aligned}
\end{equation}
\begin{equation}
\begin{aligned}
& (-4 m_g H\partial_v^3 +m_g^2 \partial_v\partial_y+2\partial_v^2\partial_u\partial_y)f+( 
-\partial_v^2\partial_y^2+2\partial_v^3\partial_u+4H\partial_v^4)g = 0\\& \hskip  8 cm vu-\mbox{component}
\end{aligned}
\end{equation}
\begin{equation}
\begin{aligned}
(-4 H\partial_v^4 -4 \partial_v^3\partial_u-m^2\partial_v^2)f+\partial_v^3\partial_y g= 0 \hskip  1 cm vy-\mbox{component}
\end{aligned}
\end{equation}
\begin{equation}
\begin{aligned}
(-4 m_g H\partial_v^3 +2H \partial_v^3\partial_y+2\partial_v^2\partial_u\partial_y)f+( 
-\partial_v^2\partial_y^2+2\partial_v^3\partial_u
+2H\partial_v^4+m_g^2\partial_v^2)g =0\\& \hskip  7.2 cm yy-\mbox{component}
\end{aligned}
\end{equation}
\begin{equation}
\begin{aligned}
& (m_g^2\partial_v\partial_u- 2\partial_v\partial_u\partial_y^2+4 \partial_v^2\partial_u^2-4 H\partial_u\partial_v^3+5 m_g H\partial_v^2\partial_y-2 m_g^2 H\partial_v^2
-2 H\partial_v^2\partial_y^2\\&+12H\partial_v^3\partial_u+8H^2 \partial_v^4)f
+(-m_g^2\partial_v\partial_y + \partial_v \partial_y^3 
 -3 \partial_v^2 \partial_u\partial_y +3m_g H  \partial_v^3\\&-4H\partial_v^3 \partial_y )g
  =0  \hskip  5.4 cm uy-\mbox{component}
\end{aligned}
\end{equation}
\begin{equation}
\begin{aligned}
   &(-8mH\partial_u\partial_v^2-2\partial_u\partial_y^3+6 \partial_v\partial_u^2\partial_y
 +2m^3 H \partial_v-7\partial_y H \partial_v\partial_y^2
 +24 H \partial_yH \partial_v^3\\&-2H \partial_v\partial_y^3+10H \partial_v^2\partial_u\partial_y
 +8H^2 \partial_v^3\partial_y+2 m^2 \partial_u\partial_y-5m^2H \partial_v\partial_y)f \\&+(-4
\partial_v\partial_u\partial_y^2-7 \partial_yH \partial_v^2\partial_y+2\partial_v^2\partial_u^2-6 H \partial_v^2\partial_y^2+6 H \partial_v^3\partial_u\\&+8 H^2
  \partial_v^4 -m^2 \partial_y^2+\partial_y^4-3\partial_y^2H \partial_v^2)g
   =0  \hskip  2 cm uu-\mbox{component}
\end{aligned}
\end{equation}

\section{Scattering amplitudes in Massive Gravity}
\label{chp:appendixc1}
To be able to obtain eikonal scattering amplitudes in massive gravity theories which we have done in the chapter 4, the most easiest way to introduce a complete set of orthogonal six projection operators in the space of symmetric tensor fields as
\begin{eqnarray}
P^{(2)}_{\mu\nu,\rho\sigma} &=& \dfrac{1}{2}(\theta_{\mu\rho}\theta_{\nu\sigma}+\theta_{\mu\sigma}\theta_{\mu\rho}-\theta_{\mu\nu}\theta_{\rho\sigma}) ~, \qquad P^{(0,s)}_{\mu\nu,\rho\sigma} = \dfrac{1}{2}\theta_{\mu\nu}\theta_{\rho\sigma} ~, \nonumber \\ [0.5em] 
P^{(1)}_{\mu\nu,\rho\sigma} &=& \dfrac{1}{2}(\theta_{\mu\rho}\omega_{\nu\sigma}+\theta_{\mu\sigma}\omega_{\nu\rho}+\theta_{\nu\rho}\omega_{\mu\sigma}+\theta_{\nu\sigma}\omega_{\mu\rho}) ~, \\ [0.5em] \nonumber
P^{(0,w)}_{\mu\nu,\rho\sigma} &=& \omega_{\mu\nu}\omega_{\rho\sigma} ~, \qquad P^{(0,sw)}_{\mu\nu,\rho\sigma} = \dfrac{1}{\sqrt{2}}\theta_{\mu\nu}\omega_{\rho\sigma} ~, \qquad P^{(0,ws)}_{\mu\nu,\rho\sigma} = \dfrac{1}{\sqrt{2}}\omega_{\mu\nu}\theta_{\rho\sigma} ~,
\end{eqnarray}
which are obtained from the transverse and longitudinal projectors \cite{Barnes,Rivers} 
$$
\theta_{\mu\nu}=\eta_{\mu\nu}-\dfrac{\partial_{\mu}\partial_{\nu}}{\Box} ~, \qquad \omega_{\mu\nu}=\partial_{\mu}\partial_{\nu} ~.
$$
Firstly, let us expand Lagrangian density of Einstein-Hilbert term, one obtains
\begin{equation}
\mathcal{L}^{(2)}_{EH} = \sigma \sqrt{-g} R = \dfrac{\sigma}{2} h^{\mu\nu} \left[ P^{(2)}_{\mu\nu,\rho\sigma} - P^{(0,s)}_{\mu\nu,\rho\sigma} \right] \Box h^{\rho\sigma} ~.
\end{equation}
To calculate the propagator one needs to add a term in the Lagrangian fixing the de Donder gauge,
\begin{equation}
\mathcal{L}_{\rm gf}=-\dfrac{1}{2\alpha}\partial_{\mu}(\sqrt{-g}g^{\mu\nu})\partial^{\lambda}(\sqrt{-g}g_{\lambda\nu}) ~,
\end{equation}
whose second order expansion takes the form by virtue of the above projector operators 
\begin{equation}
\mathcal{L}^{(2)}_{\rm gf} = \dfrac{1}{2\alpha}h^{\mu\nu} \left[ \dfrac{1}{2}P^{(1)} + \dfrac{1}{2} P^{(0,s)} + \dfrac{1}{4} P^{(0,w)} - \dfrac{1}{2\sqrt{2}}(P^{(0,sw)} + P^{(0,ws)}) \right]_{\mu\nu,\rho\sigma} \Box h^{\rho\sigma} ~.
\end{equation}
On the other hand, to write TMG propagator in terms of projection operators, one needs find the second order expansion of the Chern-Simons term which is given in \cite{Pinheiro}
\begin{equation}
\mathcal{L}^{(2)}_{CS} = \dfrac{1}{\mu}\varepsilon^{\lambda\mu\nu}\Gamma^{\rho}_{\lambda\sigma} \left( \Gamma^{\sigma}_{\rho\nu,\mu} + \dfrac{2}{3} \Gamma^{\sigma}_{\mu\tau} \Gamma^{\tau}_{\nu\rho} \right) = \dfrac{1}{2\mu} h^{\mu\nu} \left[ S^{(1)}_{\mu\nu,\rho\sigma} + S^{(2)}_{\mu\nu,\rho\sigma} \right] \Box h^{\rho\sigma} ~,
\label{spinproCS}
\end{equation}
where we have used the following spin projection operators
\begin{eqnarray}
& & S^{(1)}_{\mu\nu,\rho\sigma} = \dfrac{1}{4}\Box(\varepsilon_{\mu\rho\lambda}\partial_{\nu}\omega^{\lambda}_{\sigma}+\varepsilon_{\mu\sigma\lambda}\partial_{\nu}\omega^{\lambda}_{\rho}+\varepsilon_{\nu\rho\lambda}\partial_{\mu}\omega^{\lambda}_{\sigma}+\varepsilon_{\nu\sigma\lambda}\partial_{\mu}\omega^{\lambda}_{\rho}) ~, \\[0.5em] \nonumber
& & S^{(2)}_{\mu\nu,\rho\sigma} = -\dfrac{1}{4}\Box(\varepsilon_{\mu\rho\lambda}\eta_{\nu\sigma}+\varepsilon_{\nu\rho\lambda}\eta_{\mu\sigma}+\varepsilon_{\mu\sigma\lambda}\eta_{\nu\rho}+\varepsilon_{\nu\sigma\lambda}\eta_{\mu\rho}) \partial^{\lambda} ~.
\end{eqnarray}
Accordingly, by using Eq.(\ref{spinproCS}), one can obtain the gauge-fixed propagator for TMG as follows (we set $\alpha = 1$):
\begin{eqnarray}
\mathcal{D}^{\rm TMG}_{\mu\nu\alpha\beta} &=& \dfrac{i4(-p^{2})}{\frac{(-p^{2})^3}{\mu} - \frac{(\sigma\mu)^2}{\mu}(-p^{2})^2} \left[ -\sigma\mu P^{(2)}_{\mu\nu,\alpha\beta} - \frac{1}{4} (\epsilon_{\mu\alpha\lambda}\theta_{\beta\nu} + \epsilon_{\mu\beta\lambda}\theta_{\alpha\nu} + \epsilon_{\nu\alpha\lambda}\theta_{\beta\mu} \right. \nonumber \\ [0.5em]
& & \left. + \epsilon_{\nu\beta\lambda}\theta_{\alpha\mu}) (ip^{\lambda}) \right] + \dfrac{4}{\sigma(-p^2)} \left[ P^{(1)}_{\mu\nu,\alpha\beta} - P^{(0,s)}_{\mu\nu,\alpha\beta} \right. \nonumber \\ [0.5em]
& &\left.\hskip 5 cm- \sqrt{2} \left( P^{(0,sw)}_{\mu\nu,\alpha\beta} + P^{(0,sw)}_{\alpha\beta,\mu\nu} \right) \right] ~.
\label{propTMG}
\end{eqnarray}
Finally, for the NMG, the second order expansion of the Lagrangian density takes the form
\begin{equation}
\mathcal{L}_{K}^{(2)} = \dfrac{1}{4m^2} h^{\mu\nu} P^{(2)}_{\mu\nu,\rho\sigma} \Box^2 h^{\rho\sigma} ~,
\end{equation}
which yields the propagator
\begin{equation}
\begin{aligned}
\mathcal{D}^{\rm NMG}_{\mu\nu\alpha\beta} = \dfrac{im^2}{(-p^2)(-p^2+\sigma m^2)}P^{(2)}_{\mu\nu,\alpha\beta} + \dfrac{2i}{\sigma(-p^2)}\bigg[& P^{(1)}_{\mu\nu,\alpha\beta} - P^{(0,s)}_{\mu\nu,\alpha\beta}\\& - \sqrt{2} \left( P^{(0,sw)}_{\mu\nu,\alpha\beta} + P^{(0,sw)}_{\alpha\beta,\mu\nu} \right) \bigg] ~.
\label{propNMG}
\end{aligned}
\end{equation}

\end{document}